\documentclass[
aps,prd,
twocolumn,
superscriptaddress,nofootinbib,longbibliography]{revtex4-2} %

\usepackage[utf8]{inputenc}
\usepackage[T1]{fontenc}
\usepackage[english]{babel}
\usepackage{graphicx}
\usepackage{tikz}
\usetikzlibrary{arrows,scopes}
\usepackage{amsmath,amsfonts,amssymb}
\usepackage{mathtools}
\usepackage{indentfirst}
\usepackage{hyperref}
\hypersetup{
	colorlinks,%
	citecolor=black,%
	linkcolor=black,%
	urlcolor=purple
}
\usepackage{enumitem}

\makeatletter
\renewcommand*{\thesection}{\arabic{section}}

\renewcommand*{\p@subsection}{}

\renewcommand*{\p@subsubsection}{}
\numberwithin{equation}{section}
\renewcommand*{\theequation}{\thesection.\arabic{equation}}
\renewcommand*{\p@equation}{}
\makeatother

\DeclareMathOperator{\tr}{tr}

\newcommand{\overright}[1]{\overset{\rightarrow}{#1}}
\newcommand{\overleft}[1]{\overset{\leftarrow}{#1}}
\newcommand{\overboth}[1]{\overset{\leftrightarrow}{#1}}
\newcommand{\la}{\langle}
\newcommand{\ra}{\rangle}
\newcommand{\eu}{E}
\newcommand{\re}{{\mathrm{re}}}
\newcommand{\bvphi}{{\boldsymbol{\varphi}}}
\newcommand{\bphi}{{\boldsymbol{\phi}}}
\newcommand{\bom}{{\boldsymbol{\omega}}}
\newcommand{\bOm}{{\boldsymbol{\varOmega}}}
\newcommand{\Bom}{{\boldsymbol{\varOmega}}}
\newcommand{\bnu}{{\boldsymbol{\nu}}}
\newcommand{\bbnu}{{\boldsymbol{\nu}}}

\newcommand{\bF}{{\boldsymbol{F}}}
\newcommand{\bW}{{\boldsymbol{W}}}
\newcommand{\bw}{{\boldsymbol{w}}}
\newcommand{\bG}{{\boldsymbol{G}}}
\newcommand{\bg}{{\boldsymbol{g}}}
\newcommand{\bDel}{{\boldsymbol{\Delta}}}
\newcommand{\bid}{{\boldsymbol{I}}}
\newcommand{\bX}{{\boldsymbol{X}}}
\newcommand{\bA}{{\boldsymbol{A}}}
\newcommand{\bj}{{\boldsymbol{j}}}
\newcommand{\bJ}{{\boldsymbol{J}}}
\newcommand{\bv}{{\boldsymbol{v}}}
\newcommand{\bS}{{\boldsymbol{S}}}
\newcommand{\calD}{{\mathcal{D}}}

\begin{document}
	\title{Nonequilibrium Schwinger-Keldysh formalism for density matrix states: \\analytic properties and implications in cosmology}
	\author{Andrei O. Barvinsky}
	\email{barvin@td.lpi.ru}
	\affiliation{Theory Department, Lebedev Physics Institute, Leninsky Prospect 53, Moscow 119991, Russia}
	\affiliation{Institute for Theoretical and Mathematical Physics, Moscow State University, 119991, Leninskie Gory, GSP-1, Moscow, Russia}
	
	\author{Nikita Kolganov}
	\email{nikita.kolganov@phystech.edu}
	\affiliation{Moscow Institute of Physics and Technology, 141700, Institutskiy pereulok, 9, Dolgoprudny, Russia}
	\affiliation{Institute for Theoretical and Mathematical Physics, Moscow State University, 119991, Leninskie Gory, GSP-1, Moscow, Russia}
	\affiliation{Theory Department, Lebedev Physics Institute, Leninsky Prospect 53, Moscow 119991, Russia}
	\begin{abstract}
Motivated by cosmological Hartle-Hawking and microcanonical density matrix prescriptions for the quantum state of the Universe we develop Schwinger-Keldysh in-in formalism for generic nonequilibrium dynamical systems with the initial density matrix. We build the generating functional of in-in Green's functions and expectation values for a generic density matrix of the Gaussian type and show that the requirement of particle interpretation selects a distinguished set of positive/negative frequency basis functions of the wave operator of the theory, which is determined by the density matrix parameters. Then we consider a special case of the density matrix determined by the Euclidean path integral of the theory, which in the cosmological context can be considered as a generalization of the no-boundary pure state to the case of the microcanonical ensemble, and show that in view of a special reflection symmetry its Wightman Green's functions satisfy Kubo-Martin-Schwinger periodicity conditions which hold despite the nonequilibrium nature of the physical setup. Rich analyticity structure in the complex plane of the time variable reveals the combined Euclidean-Lorentzian evolution of the theory, which depending on the properties of the initial density matrix can be interpreted as a decay of a classically forbidden quantum state.\\
	\begin{center}
		 {\normalsize\emph{To the memory of Jim Hartle}}
	\end{center}
	\end{abstract}
	
	\maketitle

\tableofcontents

\section{Introduction}
The purpose of this paper is to construct the Schwinger-Keldysh in-in formalism \cite{Schwinger,Keldysh} for expectation values and correlation functions in a rather generic non-equilibrium system with the initial state in the form of a special density matrix. This density matrix is itself assumed to be determined by the dynamical content of the system. The motivation for this construction comes from the scope of ideas of quantum cosmology suggesting that the initial state of the Universe should be prescribed not from some ad hoc and freely variable initial conditions like in a generic Cauchy problem, but rather intrinsically fixed by the field theory model of the Universe. The pioneering implementation of these ideas was the prescription of the Harle-Hawking no-boundary cosmological wavefunction \cite{HH,H}, \emph{no-boundary} connotation indicating the absence of the notion of the initial Cauchy (boundary) surface of spacetime. Such a pescription replaces the existence of this surface by the requirement of regularity of all fields at all spacetime points treated in the past as regular internal points of spacetime manifold.

Applied to a wide class of spatially closed cosmological models this prescription qualitatively leads to the picture of expanding Friedmann Universe with the Lorentzian signature spacetime nucleating from the domain of a Euclidean space with the topology of a 4-dimensional hemisphere, the Euclidean and Lorentzian metrics being smoothly matched by analytical continuation in the complex plane of time coordinate. This picture allows one to avoid initial singularity in the cosmological evolution and, in particular, serves as initial conditions for inflationary scenarios. This is because it implies a pure vacuum state of quantum matter perturbations on top of a quasi-exponentially expanding metric background, both the background and this vacuum state being generated by tunneling from the classically forbidden (underbarrier) state of the Universe, described by the Euclidean spacetime with the imaginary time. Correlation functions of quantum cosmological perturbations in this vacuum state have a good fit to nearly flat red-tilted primordial spectrum of the cosmic microwave background radiation (CMBR) \cite{Starobinsky,Mukhanov-Chibisov} and other features of the observable large scale structure of the Universe \cite{Planck1,*Planck2}.

Limitation of this no-boundary concept consists in the fact that it covers only the realm of pure quantum states. Moreover, it prescribes a particular quantum state which in the lowest order of the perturbation theory yields a special vacuum state. In fact, the idea of Hartle-Hawking no-boundary initial conditions came from the understanding that the vacuum state wavefunction $\varPsi[\varphi(\mathbf{x})]$ of a generic free fields model in flat spacetime can be built by the path integral over the field histories $\phi(\tau,\mathbf{x})$ on a half-space interpolating between a given 3-dimensional configuration $\varphi(\mathbf{x})$ on the boundary plane of $\tau=0$  and the vanishing value of these fields at the Euclidean time $\tau\to-\infty$. Beyond perturbation theory, in the models with a bounded from below spectrum of their Hamiltonian this procedure yields the lowest energy eigenstate. Thus, the Hartle-Hawking no-boundary wavefunction is the generalization of this distinguished state to a special case of curved spatially closed spacetime, which can be formulated even though the notion of nontrivially conserved energy does not exist for such a situation.

A natural question arises how to generalize this picture to the physical setup with the density matrix replacing this distinguished pure state. The attempt to do this encounters the problem of constructing the set of physical states $|\psi\rangle$ along with the set of their weights $w_\psi$ participating in the construction of the density matrix $\hat\rho=\sum_\psi w_\psi |\psi\rangle\langle\psi|$. This problem looks unmanageable without additional assumptions, but the simplest possible assumption ---  universal microcanonical equipartition of all physical states --- allows one to write down the density matrix in a closed form provided one has a complete set of equations which determine a full set of $|\psi\rangle$. These are the Wheeler-DeWitt equations $\hat H_\mu|\psi\rangle=0$ which are quantum Dirac constraints in gravity theory selecting the physical states \cite{DeWitt}, $\mu$ being the label enumerating the full set of Hamiltonian and diffeomorphism constraints, which includes also a continuous range of spatial coordinates. The density matrix becomes a formal operator projector on the subspace of these states, which can be written down as an operator delta functions
    \begin{equation}
    \hat\rho=\frac1Z\prod_\mu\delta(\hat H_\mu),
    \end{equation}
the factor $Z$ being a partition function which provides the normalization $\tr\hat\rho=1$ \cite{why}. Important feature of this formal projector is that a detailed construction of the delta function of \emph{noncommuting} operators $\hat H_\mu$ (which form an open algebra of first class constraints) leads to the representation of this projector in terms of the Batalin-Fradkin-Vilkovisky or Faddeev-Popov path integral of quantum gravity \cite{why,BFV} and makes it tractable within perturbation theory.

In contrast to the Hartle-Hawking prescription formulated exclusively in Euclidean spacetime this density matrix expression is built within unitary Lorentzian quantum gravity formalism \cite{Unitarity}. Euclidean quantum gravity, however, arises in this picture at the semiclssical level as a mathematical tool of perturbative loop expansion. The partition function $Z$ of the density matrix (its normalization coefficient) should be determined by the above path integral over closed periodic histories, and the dominant semiclassical  contribution comes from the saddle points --- periodic solutions of classical equations of motion. The practice of applications to concrete cosmological models shows, however, that such solutions do not exist in spacetime with the Lorentzian signature, but can be constructed in Euclidean spacetime. The deformation of the integration contour in the complex plane of both dynamical variables and their time argument suggests that these Euclidean configurations can be taken as a ground for a dominant contribution of semiclassical expansion. This gives rise to the following definition of the Euclidean path integral density matrix.

Let the classical background have at least two turning points and describe the periodic (classically forbidden or underbarrier) motion between them in imaginary Lorentzian time (or real Euclidean time $\tau$). Then the two-point kernel $\rho_\eu(\varphi_+,\varphi_-)=\la\varphi_+|\mathinner{\hat\rho_\eu}|\varphi_-\ra$ of the density matrix in question is defined by
	\begin{equation} \label{eq:density_m_eucl00}
	\rho_\eu(\varphi_+,\varphi_-) = \frac1Z\int D\phi \,
    e^{- S_\eu[\phi]}\Bigr|_{\phi(\tau_\pm)=\varphi_\pm},
	\end{equation}
where $S_\eu[\phi]$ is the Euclidean action of the field perturbations $\phi(\tau)$ on top of the given background, defined on the period of the Euclidean time, $\tau_-\leq\tau\leq\tau_+$, the functional integration runs over field histories interpolating between their values $\varphi_\pm$ --- the arguments of the density matrix kernel. $Z$ is the partition function given by the path integral over the periodic histories with the period $\beta=\tau_+-\tau_-$,
	\begin{equation} \label{Z00}
	Z=\int D\phi \,
    e^{- S_\eu[\phi]}\Bigr|_{\phi(\tau_+)=\phi(\tau_-)},
	\end{equation}
providing the normalization $\tr\hat\rho_\eu=1$. Hermiticity of this density matrix, which in view of its reality reduces to its symmetry $\rho_\eu(\varphi_+,\varphi_-)=\rho_\eu(\varphi_-,\varphi_+)$, implies that the background solution is a bounce that has a reflection symmetry with respect to the middle turning point at $\tfrac{\tau_++\tau_-}2$, and the turning points $\tau_\pm$ are in fact identified.

Up to a normalization the expression (\ref{eq:density_m_eucl00}) is the evolution operator of the Schroedinger equation in imaginary time, $t=-i\tau$, with the quantum Hamiltonian  $\hat H_S(\tau)$ calculated on top of the \emph{non-stationary} background. The Hamiltonian operator here is written down in the Schroedinger picture (which is indicated by the subscript $S$) and explicitly depends on the Euclidean time because of this non-stationarity, so that the evolution operator is the Dyson chronological $\tau$-ordered exponent
    \begin{equation} \label{Euclid_evolution}
	\rho_\eu(\varphi_+,\varphi_-) =
    {\rm const} \times\la\varphi_+|\mathrm{T}
    e^{- \int_{\tau_-}^{\tau_+}d\tau\,\hat H_S(\tau)}|\varphi_-\ra.
	\end{equation}

\begin{figure}
\centering
\includegraphics[scale=1]{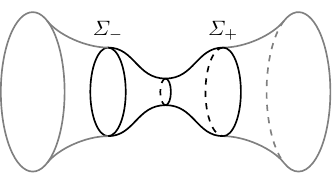}
\caption{Picture of instanton representing the density matrix. Gray lines
depict the Lorentzian Universe nucleating from the instanton at the
minimal surfaces $\varSigma_-$ and $\varSigma_+$.}
\label{Fig.1}
\end{figure}

Because of the properties of the turning points (zero derivatives of the background field) the Euclidean background can be smoothly matched at $\tau_\pm$ with the classically allowed and \emph{real} background solution of equations of motion parameterized by real Lorentzian time $t$. The evolution of quantum perturbations on this Lorentzian branch of the background is then driven by the unitary version of the $t$-ordered exponent (\ref{Euclid_evolution})
    \begin{equation}
    \hat{U}(t_+, t_-)=\mathrm{T}
    e^{-i \int^{t_+}_{t_-} dt \,\hat H_S(t)}  \label{Unitary0}
	\end{equation}
with the Hermitian time-dependent Hamiltonian which is evaluated on this Lorentzian background. In the cosmological context, when the spatial sections of spacetime of $S^3$-topology are represented by circles of a variable scale factor, the graphical image of the combined Euclidean-Lorentzian evolution operator $\hat U(T,0)\hat\rho_\eu\hat U^\dagger(T,0)$ is depicted on Fig.~\ref{Fig.1}. It shows the Euclidean spacetime instanton with the topology $R^1 \times S^3$, $R^1=[\tau_-,\tau_+]$, bounded at the turning points $\tau_\pm$ by two minimal surfaces $\varSigma_\pm$ with a vanishing extrinsic curvature. This instanton represents the density matrix $\hat\rho_\eu$ and connects the Lorenzian spacetime branches. These branches correspond to the unitary and anti-unitary evolution from $\varSigma_\pm$ in some finite interval of the Lorentzian time $0\leq t\leq T$.\footnote{Of course, the second Lorentzian branch could have been attached to the middle turning point $\tfrac{\tau_++\tau_-}2$ of the total period, but this reflection asymmetric setup would correspond to the calculation of the in-out amplitude of underbarrier tunneling through the Euclidean domain, which is not the goal of this paper.}

\begin{figure}
\centering
\includegraphics[scale=1]{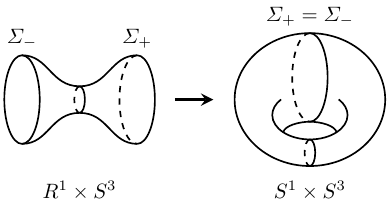}
\caption{\small Origin of the partition function instanton from the density matrix instanton by the procedure of gluing the boundaries $\varSigma_+$ and $\varSigma_-$ --- tracing the density matrix.}
 \label{Fig.2}
\end{figure}

The pictorial representation of the cosmological partition function $Z$ in view of cancellation of unitary evolution factors, $\tr\bigl(\hat U(T,0)\hat\rho_\eu\hat U^\dagger(T,0)\bigr)=\tr \hat\rho_\eu=1$, contains only the Euclidean part of Fig.~\ref{Fig.1}. It is represented by the closed cosmological instanton with the identified surfaces $\varSigma_+=\varSigma_-$ and their 3-dimensional field configurations $\varphi_+=\varphi_-$ (following from the identification of the arguments in $\tr\hat\rho_\eu=\int d\varphi\,\rho_\eu(\varphi,\varphi)$). The origin of this instanton having a donut topology $S^1\times S^3$ is shown on Fig.~\ref{Fig.2}.

\begin{figure}
\centering
\includegraphics[scale=1]{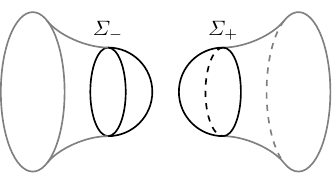}
\caption{Density matrix of the pure Hartle-Hawking state represented by the
union of two no-boundary instantons.}
\label{Fig.3}
\end{figure}

\begin{figure}
	\includegraphics[scale=1]{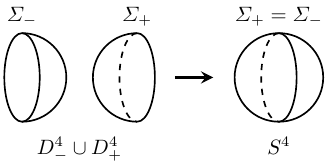}
\caption{\small Origin of the partition function instanton from the density matrix instanton by the procedure of gluing the boundaries $\varSigma_+$ and $\varSigma_-$ --- tracing the density matrix.}
 \label{Fig.4}
\end{figure}

The Euclidean space bridge incorporates the density matrix correlations between the fields on opposite Lorentzian branches, which only vanish for the density matrix of the pure state factorizable in the product of the wavefunction $\varPsi(\varphi_+)$ and its complex conjugated counterpart $\varPsi^*(\varphi_-)$, $\rho_\eu(\varphi_+,\varphi_-)=\varPsi(\varphi_+)\,\varPsi^*(\varphi_-)$. In the cosmological context this situation is depicted on Fig.~\ref{Fig.3}  with two disconnected Euclidean-Lorentzian manifolds corresponding to these factors. Each of them corresponds to the Hartle-Hawking state, and the partition function is based on the instanton with $S^4$-topology of Fig.~\ref{Fig.4}. The latter originates by glueing together two 4-dimensional hemispheres (discs $D^4_\pm$) along their common equatorial boundary.

So the goal of this paper is to construct the generating functional of expectation values and correlation functions of Heisenberg operators defined with respect to such a density matrix. Motivated by applications of quantum cosmology, this is essentially non-equilibrium physical setup, because the cosmological inflationary background is very non-stationary. Because of this it raises many questions which for the impure density matrix case go essentially beyond what is known about the Hartle-Hawking state. In particular, despite non-equilibrium nature this pure state selects a distinguished set of positive/negative frequency basis functions of the so-called Euclidean vacuum which for the de Sitter metric background turns out to be a special case of the de Sitter invariant vacuum \cite{Laflamme,Mottola,Allen}. But for a density matrix case this distinguished choice is unknown and, moreover, its reasonable particle interpretation is not granted at all to be possible.

The notion of the Euclidean quantum gravity density matrix was pioneered in \cite{Page}. Then, within the concept of the above type, it was built in a concrete inflationary model driven by the trace anomaly of Weyl invariant fields \cite{slih}. Interpreted as a microcanonical density matrix of spatially closed cosmology \cite{why}\footnote{This interpretation follows from the analogy with the microcanonical ensemble whose density matrix is a projector on the subspace of fixed conserved energy. As mentioned above, in the absence of the notion of conserved energy the role of this projection in closed cosmology is played by the delta function of Hamiltonian and momentum constraints --- the projector on their conserved zero value.} it was later shown to be a very promising candidate for the initial quantum state of the Universe. In particular, it includes the existence of the quasi-thermal stage preceding the inflation \cite{slih}, provides the origin of the Higgs-type or $R^2$-type inflationary scenario \cite{Starobinsky_R^2} with subplanckian Hubble scale \cite{slih_R^2} and suppresses the contribution of Hartle-Hawking instantons to zero. Thus, this model allows one to circumvent the  main difficulty of the Hartle-Hawking prescription --- insufficient amount of inflation in the Hartle-Hawking ensemble of universes dominated by vanishingly small values of the effective cosmological constant.  Elimination of this infrared catastrophe is, on the one hand, the quantum effect of the trace anomaly which flips the sign of the Euclidean effective action and sends it to $+\infty$ \cite{slih,suppression}. On the other hand, this is the hill-top nature of inflation starting from the maximum of inflaton potential rather than from its minimum \cite{hill-top}. Finally, this model suggests that quantum origin of the Universe is the subplanckian phenomenon subject to semiclassical $1/N$-perturbation theory in the number of numerous higher-spin conformal fields \cite{CHS}. Thus, it sounds reliable even in the absence of currently unavailable non-perturbative methods of quantum gravity.

All these conclusions have been recently reviewed in \cite{SLIH_review} including certain preliminary results on the primordial CMBR spectra, which might even bear potentially observable thermal imprint of the pre-inflation stage of this model \cite{thermal}. However, detailed calculation of this spectrum and of higher order correlation functions requires the construction of the in-in Schwinger-Keldysh formalism extended to the setup with the initial density matrix of the above type.

Schwinger-Keldysh formalism \cite{Schwinger,Keldysh} was intensively applied in quantum gravity and cosmology, and the number of publications on this subject is overwhelmingly high, so that we briefly mention only their minor part. Together with early applications \cite{Jordan,Calzetta-Hu,Onemli-Woodard} and the pioneering calculation of non-gaussianities in cosmological perturbation spectra \cite{Maldacena} these works include the calculation of cosmological correlation functions \cite{Weinberg,Ford-Woodard}, the results on cosmological singularity avoidance due to nonlocal effects \cite{Donoghue-El-Menoufi}, equivalence of the Euclidean and in-in formalisms in de Sitter QFT \cite{Higuchi,Korai} and even the analysis of initial conditions within Schwinger-Keldysh formalism \cite{Adshead-Easther-Lim}. Among recent results one should mention the development of a special effective field theory method based on analyticity and unitarity features of in-in formalism \cite{Gorbenko_etal}, its applications to four-point correlators in inflationary cosmology \cite{Heckelbacher-Sachs-Skvortsov-Vanhove} and numerous conformal field theory and holography ramifications of Schwinger-Keldysh technique (see, for example \cite{Arkani-Hamed,Gorbenko_etal} and references therein). However, the success of these works essentially relies on working with the model of spatially flat Universe --- extension to spatially closed cosmology with $S^3$-sections is likely to invalidate many of these exact analytical results. At the same time, despite a general belief that inflation completely washes out details of initial quantum state, learning its imprint on the Universe requires to go beyond $K=0$ FRW model. Moreover, recent analysis of the large scale Planck 2018 data, associated with the Hubble tension problem in modern precision cosmology \cite{Hubble_tension}, testifies at more that 99\% confidence level in favor of the closed Universe preferring a positive curvature with $K=+1$  \cite{Silk1,Silk2}. Remarkably, the model of microcanonical initial conditions in early quantum cosmology of \cite{slih,why} exists only for $K=1$. Therefore, robust observational evidence in favour of a positive spatial curvature serves as an additional motivation for this model and justifies our goals.

Having said enough about the motivation coming from cosmology for the density matrix modification of the in-in formalism coming from cosmology, let us emphasize that the usefulness of this modification extends to a much wider area. Note that the expression (\ref{Euclid_evolution}) for the case of a static background is nothing but a well-known density matrix of the equilibrium canonical ensemble at the inverse temperature $\beta=\tau_+-\tau_-$,
    \begin{equation} \label{Hibbs}
	\hat\rho=
    \frac1Z e^{- \beta\hat H}.
	\end{equation}
Its evolution in time gives rise to Matsubara technique of thermal Green's functions \cite{Matsubara} and thermofield dynamics \cite{Umezawa} which satisfies nontrivial analyticity properties in the complex plane of time including periodicity in the direction of imaginary axis --- Kubo-Martin-Schwinger (KMS) condition \cite{KMS1,KMS2}. Many of these properties depend on the condition of equilibrium and associated with the conservation of energy. What we suggest here is the generalization of this technique to non-equilibrium situation with the Hamiltonian explicitly depending on time, which would be important in many areas of quantum field theory, high energy and condensed matter physics. To cover as wide scope of models and problems as possible we will try being maximally generic and use condensed notations applicable in generic dynamical systems.

In this paper we will basically consider the elements of the diagrammatic technique for the density matrix in-in formalism. Therefore we restrict ourselves with the systems having a quadratic action on top of the non-stationary background subject to reflection symmetry discussed above. The one-loop preexponential factors of this formalism will be considered elswhere.

The paper is organized as follows. Section~\ref{sec:summary} contains the summary of notations and main results. It includes the formulation of in-in generating functional in the generic non-equilibrium system with a Gaussian type initial density matrix, the selection of distinguished set of positive/negative frequency basis functions of the wave operator, determined by the density matrix parameters, and application of this formalism to a special density matrix based on the Euclidean path integral, this case demonstrating special reflection symmetry, analyticity and KMS periodicity properties. Section~\ref{sec:Preliminaries} presents preliminary material of canonical quantization and the technique of boundary value problems and relevant Green's functions in a generic dynamical system. Section~\ref{sec:Gen_Funct} contains detailed derivation of all the results. Section~\ref{Examples} is devoted to the demonstration of the formalism on concrete examples, while Section~\ref{sec:Discussion} contains a concluding discussion along with the prospects of future research. Several appendices give technical details of derivations and present certain nontrivial properties of Green's functions and Gaussian type density matrices.

%


\section{Summary of main results} \label{sec:summary}
\subsection{Schwinger-Keldysh technique for models with density matrix state}
We consider a generic system with the action $S[\phi]$ quadratic in dynamical variables $\phi=\phi^I(t)$, the index $I$ including both the discrete tensor labels and in field-theoretical context also the spatial coordinates,
\begin{align}
		S[\phi] =
		\frac12 \int dt\Bigl(\dot\phi^T A \dot\phi + \dot \phi^T B  \phi + \phi^T B^T \dot \phi + \phi^T C \phi\Bigr). \label{eq:action_lor0}
	\end{align}
Here $A=A^T\equiv A_{IJ}$, $B\equiv B_{IJ}$ and $C=C^T\equiv C_{IJ}$ are the matrices acting in the vector space of $\phi^J$, the superscript ${}^T$ denoting the transposition, $\phi$ being a column and $\phi^T$ --- a row  (the use of these canonical condensed notations including also spatial integration over contracted indices $I$ will be discussed in much detail in Section~\ref{sec:Preliminaries}). What is most important throughout the paper, all these matrices are generic functions of time $A=A(t)$, $B=B(t)$, $C=C(t)$, reflecting non-equilibrium and non-stationary physical setup. This action will be considered as a quadratic part of the full nonlinear action in field perturbations $\phi$ on a certain background whose possible symmetries will be inherited by these coefficients as certain restrictions on their time dependence. These restrictions will be very important for the results of the paper and will be discussed below, but otherwise this time dependence is supposed to be rather generic.

The prime object of our interest will be the Schwinger-Keldysh generating functional of the in-in expectation values and correlation functions of Heisenberg operators in the physical state described by the initial density matrix $\hat\rho$. This is the functional of two sources
    \begin{equation} \label{eq:gen_fun_def0}
		Z[J_1,J_2] = \tr \left[ \hat{U}_{J_1}(T,0) \mathinner{\hat{\rho}} \hat{U}^\dagger_{-J_2}(T,0) \right].
	\end{equation}
Here the trace is taken over the Hilbert space of the canonically quantized field $\hat\phi$ and $\hat{U}_J(T, 0)$ is the operator of unitary evolution from $t=0$ to $t=T$ with the time dependent Hamiltonian corresponding to the action (\ref{eq:action_lor0}) and modified by the source term $-J^T(t)\phi(t) \equiv-J_I(t)\phi^I(t)$ with the source $J^T(t)=J_I(t)$. In the Schroedinger picture (labelled by $S$) it reads as the chronologically ordered operator $\mathbb{\mathrm{T}}$-exponent
    \begin{equation}
    \hat{U}_J(T, 0)=\mathbb{\mathrm{T}}
    e^{-i \int\limits^{T}_{0} dt \,
    \bigl(\hat H_S(t)-J(t)\hat\phi_S\bigr)}.  \label{Unitary}
	\end{equation}

We will consider the class of density matrices whose kernel in the coordinate representation $\la\varphi_+|\mathinner{\hat\rho}|\varphi_-\ra =\rho(\varphi_+,\varphi_-)$ has the following Gaussian form --- exponentiated quadratic and linear forms in $\varphi_\pm$,
	\begin{align} \label{eq:dm_coord0}
    &\rho(\bvphi) = \text{const} \times \exp\left\{-\frac12 \bvphi^T \Bom  \, \bvphi + {\bj}^T \! \bvphi  \right\}, \\
    &\bvphi =
    \begin{bmatrix}\,\varphi_+
    \\ \,\varphi_-
	\end{bmatrix},\quad
    {\bj} = \begin{bmatrix*}[l]
		\,{j}_+ \\
    	\,{j}_-
	\end{bmatrix*},
	\end{align}
where we assembled $\varphi_\pm$ into the two-component column multiplets (denoted by boldfaced letters) $\bvphi$, did the same with the coefficients $\bj$ of the linear form and introduced the $2\times2$ block-matrix $\Bom$ acting in the space of such two-component multiplets
    \begin{equation}
    \Bom =
		\begin{bmatrix*}[l]
			\;R & S \\
    		\;S^* & R^*
		\end{bmatrix*}, \quad R = R^T, \quad S = S^\dagger. \label{Omega}
	\end{equation}
The blocks of this matrix $R=R_{IJ}$, $S=S_{IJ}$ and their complex and Hermitian conjugated versions, $S^\dagger\equiv S^{T*}$, should satisfy these transposition and conjugation properties in order to provide Hermiticity of the density matrix. The same concerns the ``sources'' $j_\pm$ in the definition of $\bj$, $j_+=j_-^*\equiv j$. Transposition operation above applies also to two-component objects, so that $\bvphi^T=[\varphi_+^T\,\;\varphi_-^T]$. Below we will denote $2\times2$ block matrices and relevant 2-block component columns and rows by boldfaced letters.

Such a choice of the density matrix is motivated by the fact that for a block-diagonal $\Bom$ it reduces to a pure quasi-vacuum state, its ``source'' $\bj$ allows one to induce nonzero mean value of the quantum field and by the differentiation with respect to $\bj$ one can generate a much wider class of density matrices with ``interaction'' terms in the exponential. Normalizability of the density matrix of course implies that the real part of $\Bom$ should be a positive definite matrix.

The path integral representation for the coordinate kernels of the unitary evolution operator (\ref{eq:gen_fun_def0}) allows one to rewrite the generating functional $Z[J_1,J_2]$ as the double path integral. For this purpose it is useful to introduce the two-component notations for the histories $\phi_1(t)$ and $\phi_2(t)$ as well as for their sources,
	\begin{gather}
		\phi_1,\phi_2\mapsto\bphi = \begin{bmatrix*}[l]
			\,\phi_1\\ \,\phi_2
		\end{bmatrix*}, \quad
		J_1,J_2\mapsto\bJ = \begin{bmatrix*}[l]
			\,J_1\\ \,J_2
		\end{bmatrix*},   \label{double_objects}
	\end{gather}
In terms of these notations the generating functional reads
	\begin{multline}
		Z[\bJ] =
    \int D[\bphi,\bvphi]\, \exp \Biggl\{i\bS[\bphi] + i\int_0^T dt \, \bJ^T \bphi\\
    - \frac12 \bvphi^T \Bom \, \bvphi + \bj^T\! \bvphi \Biggr\}, \label{gen_fun_pm0}
	\end{multline}
where the total action is obviously
	\begin{equation}
    \bS[\bphi]=S[\phi_1]-S[\phi_2]
	\end{equation}
with the actions $S[\phi_{1,2}]$ given by (\ref{eq:action_lor0}) in the integration range from $t=0$ to $t=T$ and the total integration measure over $\bphi$ and $\bvphi$
	\begin{equation}
    D[\bphi,\bvphi]=\int d\varphi_+\,d\varphi_-\!\!\!\!\!\!\!\!
    \int\limits_{\substack{\bphi(0)=\bvphi\\ \phi_1(T)=\phi_2(T)}}\!\!\!\!\!\!\!\! D \phi_1\,D\phi_2. \label{Measure}
	\end{equation}
Here $d\varphi$ and $D\phi$ denote respectively the integration measures over variables at a given moment of time and the integration measures $D\phi=\prod_t d\phi(t)$ over time histories subject to indicated boundary conditions.

Calculation of this Gaussian path integral leads to the expression
	\begin{multline}  \label{gen_fun_dir0}
	Z[\bJ]=\text{const} \times
    \exp\Biggl\{ -\frac{i}2 \int_0^T dt\,dt' \, \bJ^T\!(t) \bG(t,t') \bJ(t)\\
    - \int_0^T dt \,\bJ^T\!(t) \bG(t,0) \, \bj + \frac{i}2\bj^T \bG(0,0) \, \bj \Biggr\},
	\end{multline}
where we disregard the source-independent prefactor. The bilinear in the full set of sources exponential is the total action in the integrand of (\ref{gen_fun_pm0}) at its saddle point --- the point of stationarity of the action with respect to variations of both the histories $\bphi(t)$ and their boundary data $\bvphi$ at $t=0$. The condition of stationarity generates the boundary value problem for this saddle point including the linear second order equation of motion for $\bphi(t)$ and the full set of boundary conditions at $t=0$ and $t=T$. This problem is posed and solved in Section~\ref{sec:Gen_Funct} in terms of its Green's function $\bG(t,t')$ subject to homogeneous version of these boundary conditions. The Green's function has a block-matrix form typical of Schwinger-Keldysh in-in formalism composed of the Feynman $G_{\mathrm{T}}(t, t')$, anti-Feynman $G_{\bar{\mathrm{T}}}(t, t')$ and off-diagonal Wightman Green's functions blocks (\ref{eq:gf_comp}),
	\begin{equation}  \label{G_final}
		\bG(t, t') = \begin{bmatrix*}[r]
			G_{\mathrm{T}}(t, t') & G_{<}(t, t') \\
			G_{>}(t, t') & G_{\bar{\mathrm{T}}}(t, t')
		\end{bmatrix*},
	\end{equation}
which are related to one another by the equalities $G_{\bar{\mathrm{T}}}(t, t')=[G_{\mathrm{T}}(t, t')]^*$ and $G_{>}(t, t')=G_{<}^T(t', t)$ and satisfy respectively inhomogeneous and homogeneous wave equations
	\begin{equation}  \label{}
	F G_{\mathrm{T},\bar{\mathrm{T}}}(t, t')=\delta(t-t'),\quad
    F G_{\gtrless}(t, t')=0
	\end{equation}
with the operator $F$ --- the Hessian of the action (\ref{eq:action_lor0}), $F\delta(t-t')=\delta^2 S[\phi]/\delta\phi(t)\,\delta\phi(t')$,
    \begin{equation} \label{eq:phi_eq0}
	F=-\frac{d}{dt} A \frac{d}{dt}-\frac{d}{dt} B
    +B^T \frac{d}{dt} + C.
	\end{equation}

The block-matrix Green's function $\bG(t,t')$, as is usually done in boundary value problems, can be built in terms of the full set of basis functions $\bv_\pm$ of this operator, satisfying the boundary conditions of the variational problem for the action in (\ref{gen_fun_pm0}). This will explicitly be done in Section~\ref{sec:Gen_Funct}, but in view of the complexity of these boundary conditions intertwining the $\phi_{1,2}$-branches of the field space these basis functions do not have a clear particle interpretation, that is separation into positive and negative frequency parts. This difficulty is caused by the convolution of problems associated, on the one hand, with the non-equilibrium nature of a generic background (rather generic dependence of the operator coefficients $A(t)$, $B(t)$ and $C(t)$ on time) and, on the other hand, with the in-in physical setup involving a nontrivial density matrix.

Despite these difficulties, there exists a distinguished set of basis functions for the wave operator which have a clear particle interpretation, and this is one of the main results of the paper. This set is related by Bogoliubov transformations to $\bv_\pm(t)$ and is uniquely prescribed by the full set of complex conjugated positive and negative frequency basis functions of the operator (\ref{eq:phi_eq0}) $v(t)$ and $v^*(t)$ which satisfy the intial value problem at $t=0$,
	\begin{align}
    \begin{aligned}\label{eq:green_lor_neum_bdy0}
	&Fv(t) = 0,\\ 
    &(i W - \omega) v(t)\bigr|_{t=0} = 0, \;\,
    (i W + \omega^*) v^*(t)\bigr|_{t=0} = 0,
    \end{aligned}
	\end{align}
where $W$ is what we call the \emph{Wronskian} operator
	\begin{equation} \label{Wronskian}
		W = A \frac{d}{dt} + B,
	\end{equation}
which participates in the Wronskian relation for the operator $F$, which is valid for arbitrary two complex fields $\phi_{1,2}(t)$,
	\begin{equation}
	\phi_2^T F \phi_1 - (F\phi_2)^T \phi_1 =
	- \frac{d}{dt} \left[ \phi_2^T W \phi_1
    - (W \phi_2)^T\phi_1\right]              \label{Wrelation}
	\end{equation}
and, moreover, serves as the definition of the conserved (not positive-definite) inner product in the space of solutions of the homogeneous wave equation, $F\phi_{1,2}=0$,
    \begin{equation} \label{eq:kg_inner_prod0}
		(\phi_1, \phi_2) = i\phi_1^\dagger (W\phi_2)
    - i(W\phi_1)^\dagger \phi_2.
	\end{equation}
We will call the boundary conditions (\ref{eq:green_lor_neum_bdy0}) and associated with them Green's functions the Neumann ones\footnote{Strictly speaking, these are the analogue of Robin boundary conditions, because they contain together with the derivative transversal to the boundary also the tangential terms composed of the coefficient $B$ and $\omega$.}.

Important point of the definition (\ref{eq:green_lor_neum_bdy0}) is that the frequency matrix $\omega$ (remember that in the generic setup this is a matrix $\omega_{IJ}$ acting in the vector space of $\phi^J$) is not directly contained in the blocks of the matrix (\ref{Omega}), but follows from the requirement of the particle interpretation of the basis functions $v(t)$. This requirement can be formulated as follows. One defines the creation-annihilation operators $\hat a^\dagger$ and $\hat a$ in terms of which the Heisenberg operator $\hat\phi(t)$ is decomposed as the sum of positive-negative basis functions $v(t)$ and $v^*(t)$, $\hat\phi(t)=v(t)\,\hat a+v^*(t)\,\hat a^\dagger$. Then there exist non-anomalous and anomalous particle averages with respect to the density matrix,
    \begin{align}
	\nu = \tr\bigl[\hat \rho \, \hat{a}^\dagger
    \hat{a}\bigr], \quad
	\kappa = \tr\bigl[\hat \rho \,
    \hat{a} \, \hat{a}\bigr],      \label{kappa}
	\end{align}
and the requirement of vanishing anomalous average $\kappa=0$ allows one to assign the average $\nu$ the interpretation of the set of occupation numbers associated with $\hat\rho$. This requirement serves as the equation for the frequency matrix $\omega$ which, as it is shown in Section~\ref{sec:Gen_Funct}, can be explicitly solved for a special case of the real matrix $\Bom$. This solution reads
	\begin{equation}
		\omega=R^{1 / 2} \sqrt{I-\sigma^{2}} R^{1 / 2},
    \quad \sigma \equiv R^{-1 / 2} S R^{-1 / 2} \label{eq:omega_bod}
	\end{equation}
and gives the expression for the occupation number matrix in terms of the single symmetric matrix $\sigma$ after the orthogonal rotation by the orthogonal matrix $\varkappa$,
    \begin{align}
		&\nu=\frac12 \varkappa
    \left( \sqrt{\frac{I-\sigma}{I+\sigma}}
        - 1\right) \varkappa^T,                   \label{nu}\\
        &\varkappa
        \equiv \big[\omega^{1/2} R^{-1}\omega^{1/2}\big]^{1/2} \omega^{-1/2} R^{1/2}= \bigl(\varkappa^T\bigr)^{-1}.                       \label{nu_bar_bod0}
	\end{align}
As shown in Appendix~\ref{sec:dm_def}, the existence of this particle interpretation with a positive definite matrix $\nu$ fully matches with conditions of normalizability, boundedness and positivity of the density matrix incorporating positive definiteness of matrices $I\pm\sigma$ and negative definiteness of $\sigma$.

With the normalization of these distinguished basis functions $v$ to unity
    \begin{equation}
    (v_A,v_B)=-(v^*_A,v^*_B)=\delta_{AB},
    \end{equation}
where $A$ is the index enumerating the full set of basis functions, the blocks of the in-in Green's function (\ref{G_final}) take the form
	\begin{align}
			&iG_{\mathrm{T}}(t,t') = v(t) \,  v^\dagger(t') \, \theta(t-t')
    + v^*(t) \, v^T(t') \, \theta(t'-t) \nonumber\\
    &\qquad\qquad+ \, v(t) \, \nu \, v^\dagger(t') + v^*(t) \, \nu \, v^T(t'),
\\
		&iG_>(t, t') = v(t) \, \bigl(\nu + I\bigr) \, v^\dagger(t')
    + v^*(t) \, \nu \, v^T(t'). \label{lor_green_wightmann0}
	\end{align}
Here the terms of the type $ v(t)\,v^\dagger(t')$ should be understood as the matrix products $\sum_A v^I_A(t)\,v^{*J}_A(t')$ (one should bear in mind that the basis function $v(t)=v^I_A(t)$ represents the square (but asymmetric) matrix whose upper indices label the field $\phi^I$ components, whereas the subscript indices $A$ enumerate the basis functions in their full linear independent set). Correspondingly, $v(t)\,\nu\,v^\dagger(t')= \sum_{A,B}v^I_A(t)\,\nu^{AB}v^{*J}_B(t')$, etc.

This form of the Green's functions is very familiar from thermofield dynamics for simple equilibrium condensed matter systems, when all the matrices of the above type become diagonal in the momentum space of field modes labeled by $A=\mathbf{p}$, $\sum_A=\int d^3\mathbf{p}/(2\pi)^{3/2}$ and $\nu_{AB}=\nu_{\mathbf{p},\mathbf{p}'}=(\exp(\beta\omega_{\mathbf{p}})-1)^{-1} \delta(\mathbf{p}-\mathbf{p}')$ represents expected occupation number for Bose-Einstein statistics at inverse temperature $\beta$ (detailed consideration of this example is presented in Section~\ref{Examples}). Remarkably, the occupation number picture generalizes to nonequilibrium systems of a very general type --- the function of the single symmetric matrix in the parentheses of Eq.~(\ref{nu}) can be diagonalized by extra orthogonal rotation (additional to that of $\varkappa$), and its eigenvalues would serve as occupation numbers in the generic nonequilibrium state with the initial density matrix.

\subsection{Euclidean density matrix}

As discussed in Introduction, in quantum cosmology context the density matrix itself can be given in terms of the Euclidean path integral and thus dynamically determined by individual properties of the system including its action functional. So we consider the path integral expression for the Euclidean density matrix $\la\varphi_+|\mathinner{\hat{\rho}_\eu[J_\eu]}|\varphi_-\ra\equiv\rho_\eu(\varphi_+,\varphi_-; J_\eu]$,
	\begin{multline} \label{eq:density_m_eucl0}
	\rho_\eu(\varphi_+,\varphi_-; J_\eu] =
    \frac1Z\!\!\!\!\!\!\!\int\limits_{\phi(\tau_\pm)=\varphi_\pm}
    \!\!\!\!\!\!\!D\phi\,
    \exp \biggl\{- S_\eu[\phi]\\
    - \int_{\tau_-}^{\tau_+} d\tau \, J_\eu(\tau) \phi(\tau) \biggr\},
	\end{multline}
where integration runs over histories $\phi(\tau)$ in Euclidean time $\tau$ on the segment $[\tau_-,\tau_+]$, interpolating between the arguments of the density matrix $\varphi_\pm$. In what follows we will assume that $\tau_-=0$ and $\tau_+=\beta$. The Euclidean action, supplied by the Euclidean source $J_\eu(\tau)$ probing the interior of the Euclidean spacetime, has a structure similar to the Lorentzian action and can be obtained from (\ref{eq:action_lor0}) by the replacement
	\begin{equation}
    t, A(t),B(t),C(t)\,\mapsto\,\tau,A_\eu(\tau),B_\eu(\tau),C_\eu(\tau). \label{ABC_\eu }
	\end{equation}

These functions of the Euclidean time are rather generic, except that they should not contradict the basic property of the density matrix (with the source $J_\eu$ switched off) --- its Hermiticity. Sufficient conditions providing this property, $\rho_\eu(\varphi_+,\varphi_-; 0]=\rho^*_\eu(\varphi_-,\varphi_+; 0]$, read
		\begin{align} \label{eq:eucl_action_herm_cond0}
        \begin{aligned}
		&A_\eu(\beta-\tau)=A_\eu^*(\tau),
        \quad B_\eu(\beta-\tau)=-B_\eu^*(\tau),\\
        &C_\eu(\beta-\tau)=C_\eu^*(\tau).
        \end{aligned}
        \end{align}
For real values of these coefficients these relations reduce to the reflection symmetry of the action and the whole formalism relative to inversions with respect to the center of the Euclidean segment $[0,\beta]$. Here we consider this property as given, but it can be derived from the assumption that the quadratic Euclidean action is built on top of the Euclidean spacetime background --- the bounce which solves full nonlinear equations of motion of the theory and represents the periodic (underbarrier) motion of the system between two turning points. One of these points is associated with the center of the above Euclidean segment $\tfrac{\tau_++\tau_-}2=\tfrac\beta2$, and the other one corresponds to the (identified) points of nucleation $\tau_\pm$ from the classically forbidden Euclidean regime to the Lorentzian regime, the latter being described by the two branches of the Schwinger-Keldysh formalism (labelled above by 1 and 2). For an equilibrium situation with constant coefficients (\ref{ABC_\eu }) at the inverse temperature $\beta=\tau_+-\tau_-$ this setup is even simpler and corresponds to the density matrix of the thermal canonical ensemble.

The Gaussian integration in (\ref{eq:density_m_eucl0}) allows one to express the result in terms of the Green's function of the Hessian of the Euclidean action $F_{E}$, $F_{E}\delta(\tau-\tau')=\delta^2 S_\eu[\phi]/\delta\phi(\tau)\,\delta\phi(\tau')$, which can be obtained from (\ref{eq:phi_eq0}) by the replacement (\ref{ABC_\eu }), subject to Dirichlet boundary conditions,
	\begin{equation}
	F_{E}\,G_D(\tau, \tau') = \delta(\tau-\tau'),
    \quad G_D(\tau_\pm, \tau') = 0.
	\end{equation}
The resulting density matrix looks like the expression (\ref{eq:dm_coord0}) amended by the quadratic form in the Euclidean source,
	\begin{multline} \label{eq:dm_eucl0}
	\rho_\eu(\varphi_+,\varphi_-; J_\eu] =
    {\rm const}\times\exp \biggl\{ -\frac12 \bvphi^T \Bom_\eu \, \bvphi + \bj^T_\eu \, \bvphi\\
    + \frac12 \int_0^\beta d\tau\,d\tau' J_\eu(\tau)\, G_D(\tau,\tau') J_\eu(\tau')\biggr\}
	\end{multline}
with the special expressions for the matrix $\Bom_\eu$ and the source $\bj_\eu$. The matrix $\Bom_\eu$ reads
	\begin{align}   \label{Bomega}
	\Bom_\eu =
	\begin{bmatrix*}[r]
	-{W}_{\!\eu} \, G_D(\beta,\beta)
    \overleft{W}_{\!\eu}  & \overright{W}_{\!\eu} \, G_D(\beta,0) \overleft{W}_{\!\eu} \\
	\overright{W}_{\!\eu} \, G_D(0,\beta)
    \overleft{W}_{\!\eu}  & -\overright{W}_{\!\eu} \, G_D(0,0) \overleft{W}_{\!\eu} 
	\end{bmatrix*}
    \end{align}
where we use the arrow to indicate the direction in which the Wronskian operator is acting on the corresponding first or second time argument of the Green's function, so that for the left action the following rule holds $\phi^T\!(\tau)\overleft{W}_{\!\eu} \equiv(W_\eu\phi(\tau))^T$, and $\overright{W}_{\!\eu} \, G_D(\beta,0) \overleft{W}_{\!\eu} $ of course implies
$\overright{W}_{\!\eu} \, G_D(\tau,\tau') \overleft{W}_{\!\eu} |_{\tau=\beta,\tau'=0}$, etc. The column $\bj_\eu$ is given by the following integral
    \begin{align}
	\bj_\eu  &= \int_0^\beta d\tau'
    \begin{bmatrix*}[r]
	W_\eu \, G_D(\beta, \tau')\,  \\
    -W_\eu \, G_D(0,\tau') \,
	\end{bmatrix*} J_\eu(\tau').
		\end{align}
Using (\ref{gen_fun_dir0}) in the generating functional with the Euclidean density matrix (\ref{eq:dm_eucl0})
    \begin{equation}
    	Z[\bJ, J_\eu] \equiv
		\tr \left[ \hat{U}_{J_1}(T,0) \mathinner{
        \hat{\rho}_\eu[J_\eu]} \hat{U}^\dagger_{-J_2}(T,0) \right]
	\end{equation}
one directly finds the total Schwinger-Keldysh generating functional with the full set of sources probing the two Lorentzian branches and the Euclidean branch of the in-in formalism
	\begin{align}               \label{Zfinal}
		Z[\bJ, &J_\eu ] = {\rm const}\times
    \exp \Biggl\{-\frac{i}2\int_0^T dt\,dt' \,
    \bJ^T\!(t) \bG(t,t') \bJ(t')\nonumber\\
      &-\int_0^T dt \,\bJ^T\!(t)\,\bG(t,0)\,\bj_\eu
      + \frac{i}2\,\bj^T_\eu \,\bG(0,0)\,\bj_\eu \nonumber\\
			& +\frac12\int_0^\beta d\tau\,d\tau'
    \, J_\eu (\tau)\,G_D(\tau,\tau')\,J_\eu (\tau')
       \Biggr\}.
	\end{align}

\subsection{Reflection symmetry and analyticity properties}

The obtained expression for $Z[\bJ, J_\eu]$ features a nontrivial mixup of the Neumann and Dirichlet Green's functions of different Lorentzian $F$ and Euclidean $F_\eu$ wave operators, but it becomes essentially unified if we assume that the Lorentzian and Euclidean actions are related by the analytic continuation of the form
	\begin{equation}
		i S[\phi(t)]\bigr|_{t=-i\tau} = -S_\eu[\phi_\eu(\tau)],
	\end{equation}
where the Lorentzian and Euclidean histories are also related by the same continuation rule $\phi(t)|_{t=-i\tau}=\phi_\eu(\tau)$. This, in particular, implies that the coefficients of the operators $F_\eu$ and $F$ are related by
    \begin{align}
	\begin{aligned}
		&A_\eu(\tau) = A(-i\tau), \quad
		B_\eu(\tau) = -i B(-i\tau),\\
		&C_\eu(\tau) = -C(-i\tau),
	\end{aligned}
    \end{align}
so that $F|_{t=-i\tau}=-F_\eu$. The origin of these relations, especially in connection with reality condition for the coefficients of both Lorentzian and Euclidean operators at their respective \emph{real} $t$ and $\tau$ arguments can be traced back to the properties of the full nonlinear action which gives rise to its quadratic part on top of a special background solution of full equations of motion. It is assumed that the Euclidean background solution has a turning point at $\tau=0$ where all real field variables have zero $\tau$-derivatives and can be smoothly continued to the imaginary axis of $\tau=it$ where they become again real functions of real $t$. This leads to the above continuation rule with real $A(t),B(t),C(t)$ and $A_\eu(\tau),B_\eu(\tau),C_\eu(\tau)$ at real $t$ and $\tau$.

With this analytic continuation rule, the expression (\ref{Zfinal}) for $Z[\bJ, J_\eu]$ can indeed be uniformly rewritten in terms of the Lorentzian, Euclidean and mixed Lorentzian-Euclidean Green's functions, all of them subject to one and the same set of Neumann type boundary conditions which select in the Lorentzian branch of the in-in formalism a distinguished set of positive and negative frequency basis functions. This expression reads
	\begin{align} \label{Z_FINAL}
		Z[\mathbb{J}] = {\rm const}\times
    \exp \Biggl\{\frac12\int_{\mathbb{C}} dz\,dz'\,\mathbb{J}^T(z)\,\mathbb{G}(z,z')\;
    \mathbb{J}(z')
       \Biggr\},
	\end{align}
where the $z$-integration runs respectively over $t$ or $\tau$ in the domain $\mathbb{C}=[0\leq t\leq T]\cup[0\leq\tau\leq\beta]$ depending on which of these Lorentzian or Euclidean time variables is in the argument of the following block matrix Green's function $\mathbb{G}(z,z')$ and the corresponding source $\mathbb{J}(z)$,
	\begin{align} \label{totalG}
	\mathbb{G}(z,z')\!=\!\!
	\begin{bmatrix*}
	-i\bG(t,t') & \;\bG_{LE}^<(t,\tau')\,\\
	\;\bG^>_{LE}(\tau,t') & \;G_\eu(\tau,\tau')\,
	\end{bmatrix*}\!,\;
    \mathbb{J}(z)\!=\!	
    \begin{bmatrix*}[c]
	\,\bJ(t)\, \\
	\,J_\eu(\tau)\,
	\end{bmatrix*}\!.
    \end{align}
Here the Euclidean and Lorentzian-Euclidean blocks of the total Green's function
	\begin{align}
		&G_\eu(\tau, \tau') = G_\eu^>(\tau, \tau')
        \, \theta(\tau-\tau')\nonumber\\
        &\qquad\qquad\qquad\qquad+ G_\eu^<(\tau,\tau') \, \theta(\tau' - \tau), \label{EuclidG}\\
		&\bG_{LE}^<(t,\tau)=\begin{bmatrix}
			\,G^1_{LE}(t,\tau)\; \\ G^2_{LE}(t,\tau)
		\end{bmatrix}
        =\begin{bmatrix}
			&I& \\ &I&
		\end{bmatrix} G_{LE}^<(t,\tau), \label{GLE0}\\
        &\bG^>_{LE}(\tau,t)
        =\big[\bG^<_{LE}(t,\tau)\big]^T,   \label{GLE}
	\end{align}
express in terms of the relevant Euclidean and Lorentzian-Euclidean Wightman functions
   \begin{align}
	\hspace{-0.2cm}G_\eu^>(\tau, \tau') &= u_+(\tau)(\nu + I) u_-^T(\tau')
    \!+\! u_-(\tau)\,\nu\, u_+^T(\tau'),\\
    \hspace{-0.2cm}G_\eu^<(\tau,\tau') &= \bigl[G_\eu^>(\tau', \tau)\bigl]^T,\\
    \hspace{-0.2cm}G_{LE}^<(t,\tau)&=v(t) \, (\nu + I)
        \, u_-^T(\tau)+v^*(t)\,\nu\,u_+^T(\tau).  \label{GLE1}
	\end{align}
In their turn these Green's functions, as one can see, are built according to one and the same universal pattern out of the full set of Lorentzian $v(t)$ and $v^*(t)$ and Euclidean $u_\pm(\tau)$ basis functions. All these functions are subject to Neumann boundary conditions (\ref{eq:green_lor_neum_bdy0}) and
	\begin{equation} \label{green_eucl_neum_bdy0}
		(W_\eu + \omega)u_+|_{\tau=\beta} = 0, \qquad
		(W_\eu - \omega)u_-|_{\tau=0} = 0.
	\end{equation}

For $\omega$ fixed by the above condition of particle interpretation, leading to the expressions (\ref{eq:omega_bod})--(\ref{nu_bar_bod0}), the Euclidean basis functions $u_\pm$ have a remarkable property. They satisfy at opposite ends of the Euclidean segment $\tau_\mp$ the same boundary conditions\footnote{In fact, the requirement of $\kappa=0$ in (\ref{kappa}) turns out to be the necessary and sufficient condition for this property of Euclidean basis functions, that is the coincidence of boundary condition for $u_\pm(\tau)$ at both ends of the time segment leads to $\kappa=0$.}
    \begin{equation} \label{eq:basis_fun_bdy_add0}
		(W_\eu + \omega)u_+|_{\tau=0} = 0, \quad
		(W_\eu - \omega)u_-|_{\tau=\beta} = 0.
	\end{equation}
If one smoothly continues the operator $F_\eu$ beyond the segment $\tau_-\leq\tau\leq\tau_+$, then it becomes periodic with the period $\beta$ (which is possible because $\tau_\pm$ are assumed to be the turning points of the background solution on top of which the Hessian of the nonlinear action of the theory is built). This means that the basis functions $u_\pm$ of this operator become quasi-periodic --- $u_\pm(\tau+\beta)$ expresses as a linear combination of the same basis functions $u_\pm(\tau)$ (no mixing between $u_-$ and $u_+$ occurs in their monodromy matrix). As shown in Section~\ref{sec:Gen_Funct}, with the normalization $u_\pm(0)=1/\sqrt{2\omega}$ this quasi-periodicity property reads in terms of the occupation number matrix (\ref{nu})
\begin{equation} \label{eq:bf_monodromy}
	\begin{aligned}
		u_-(\tau+\beta)&= u_-(\tau)\frac{\nu+I}\nu, \\
		u_+(\tau+\beta)&=u_+(\tau)\frac\nu{\nu+I}. 
	\end{aligned}
\end{equation}

Together with the reflection symmetry relative to the middle point of the Euclidean time segment (\ref{eq:eucl_action_herm_cond0}) the periodicity of the operator $F_\eu$ implies its reflection symmetry with respect to the point $\tau=0$
    \begin{align} \label{eq:coef_fun_eucl_refl0}
    \begin{aligned}
	&A_\eu(\tau) = A_\eu(-\tau), \quad B_\eu(\tau) = -B_\eu(-\tau),\\
    &C_\eu(\tau) = C_\eu(-\tau).
    \end{aligned}
	\end{align}
Therefore, similarly to quasi-periodicity the basis functions $u_\pm(\pm\tau)$ are also related by the analogue of the anti-diagonal monodromy matrix $L$, $u_+(\tau)=u_-(-\tau)\,L$, which is trivial in view of the normalization $u_\pm(0)=1/\sqrt{2\omega}$,
	\begin{equation}
	u_+(\tau) = u_-(-\tau).
	\end{equation}

The above relations introduce the analytic structure which allows one to express all basis and Green's functions on the Euclidean-Lorentzian domain $\mathbb{C}$ in terms of one analytic function $V(z)$ of the complexified time variable $z=t-i\tau$. This follows from the fact, mentioned above, that the Lorentzian wave operator $F$ can be regarded as the analytic continuation of the Euclidean operator $F_\eu$ into the complex plane of time at the point $z=0$, $F\equiv F(t,d/dt)=-F_\eu |_{\tau=it}$. As a consequence its basis function $v(t)$ in view of its boundary conditions and boundary conditions (\ref{eq:basis_fun_bdy_add0}) for the Euclidean function $u_+(\tau)$ also turns out to be the analytic continuation of the latter,
	\begin{equation}
	v(t)=u_+(it).   \label{v_vs_u_+}
	\end{equation}
Therefore, the operators $F$ and $-F_\eu$ as well as the full set of their basis functions $v(t)$ and $u_\pm(\tau)$ can be represented respectively as the boundary values at the real and imaginary axes of the complex $z$-plane of the complex operator $F_{\mathbb{C}}$ and the solution $V(z)$ of its homogeneous wave equation,
	\begin{align}
	&\hspace{-0.22cm}F_{\mathbb{C}}V(z)\equiv\biggl[-\frac{d}{dz}A(z)
    \frac{d}{dz}-\frac{d}{dz} B(z)+B^T(z)\frac{d}{dz}\nonumber\\
    &\qquad\qquad\quad+ C(z)\biggr] V(z) = 0,\quad z=t-i\tau,\\
	&\hspace{-0.22cm}\bigl(i W_{\mathbb{C}}
    -\omega\bigr) V(z)\bigr|_{z=0}=0, \;\,
    W_{\mathbb{C}} \equiv A(z)\frac{d}{dz} + B(z).    \label{eq:basis_V_in_cond0}
	\end{align}
The function $V(z)$ gives rise to basis functions as
    \begin{align}
    v(t)=V(z)\bigr|_{z=t},\quad u_\pm(\tau)=V(z)\bigr|_{z=\mp i\tau},
    \end{align}
and thus can be used in (\ref{totalG}) for the construction of all Green's functions of the Schwinger-Keldysh in-in formalism. Conversely $V(z)$ can be obtained by analytic continuation of the single Euclidean function $u_+(\tau)$ from the imaginary axes $z=-i\tau$,
    \begin{align}
    V(z)=u_+(iz)=u_+(\tau+it),
    \end{align}
and in view of reality of $u_+(\tau)$ for real $\tau$ it has the property $[V(z)]^*=V(-z^*)$.

Important corollary of these analiticity properties is that in view of the monodromy relations for Euclidean basis functions (\ref{eq:bf_monodromy}) the Lorenzian basis functions become quasi-periodic in the imaginary time
    \begin{align}
		\hspace{-0.18cm}v(t-i\beta)\!=\! v(t)\frac{\nu+I}\nu,\;\,
		v^*(t-i\beta)\!=\!v^*(t)\frac\nu{\nu+I}.
	\end{align}
Due to inverse matrix factors of positive and negative basis functions here the Lorentzian Wightman functions $G_>(t,t')$ (given by the expression (\ref{lor_green_wightmann0})) and $G_<(t,t')=G^T_>(t',t)$ satisfy the relation
	\begin{equation}
		G_>(t-i\beta, t') = G_<(t, t'),
	\end{equation}
which is nothing but Kubo-Martin-Schwinger condition \cite{KMS1,KMS2}. It is important that this condition is satisfied in the generic non-equilibrium system with the special Euclidean density matrix (\ref{eq:density_m_eucl0}) even despite the fact that no notion of conserved energy can be formulated in such a physical setup.

\begin{figure}
	\includegraphics[scale=1]{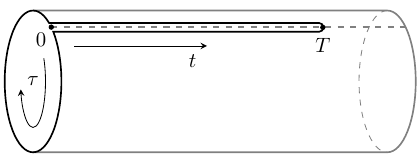}
	\caption{\small Euclidean-Lorentzian contour $\mathbb{C}$ on the Riemann surface of complex time $z=t-i\tau$. Wightman functions are periodic in imaginary (Euclidean) time direction with a period $\beta$, whereas the basis function $v(z)$ suffers a jump at the cut denoted by the horizontal dashed line, the two Lorentzian time branches running along the shores of this cut. }
	\label{Fig.5}
\end{figure}

The tubular Riemann surface of complex time $z=t-i\tau$ whose main sheet is compactified in $\tau$ to the circle of curcumference $\beta$ is shown on Fig.~\ref{Fig.5}. The boundaries of the main sheet of this surface form two shores of the cut depicted by dashed line, along which two branches of Lorentzian evolution are running. This rich analytic structure of Euclidean-Lorentzian evolution  suggests that the equivalence of the Euclidean and Lorentzian formalisms proven beyond tree level for interacting QFT on top of the de Sitter spacetime  \cite{Higuchi,Korai} might be extended to a generic reflection-symmetry background underlying our definition of the Euclidean density matrix.

\section{Preliminaries}\label{sec:Preliminaries}
To derive the aforementioned results we dwell here in more detail on the introduced above notations and develop a canonical formalism and quantization of the underlying theory. In particular, we pose rather generic initial value and boundary value problems for equations of motion and discuss the properties of the related Green's functions.
	
\subsection{Condensed notations}
	The elements of the field space will be denoted as $\phi^I(t)$, where the index $I$ is, in fact, a multi-index, and contains both the dependence on spatial coordinates denoted as $\mathbf{x}$ and discrete spin-tensor labels $i$, $I = (\mathbf{x}, i) $. Thus, we can equivalently write the fields in the form, emphasizing its dependence on the spatial coordinates $\phi^I(t) = \phi^i(t, \mathbf{x})$.
	
Assuming that equations of motion are of the second order in time derivatives one has the most general quadratic action of the theory of the form (\ref{eq:action_lor0}) where we explicitly specify the initial and final moments of time range $t_\pm$,	
	\begin{align}
		S[\phi] =
		\frac12 \int_{t_-}^{t_+} dt \, \Bigl(\dot\phi^T  A \dot\phi
        + \dot \phi^T  B  \phi + \phi^T B^T \dot \phi + \phi^T C \phi \Bigr). \label{eq:action_lor}
	\end{align}
Here dots denote the derivatives with respect to time $t$, and $A$, $B$ and $C$ are the time-dependent real bilinear forms in the space of fields. Moreover, $A$ and $C$ are assumed to be symmetric. The explicit action of these bilinear forms on the fields, e.g.\ for $A$ reads
	\begin{equation}
		(A \phi)_I(t) = A_{IJ}(t) \phi^J(t) \equiv \sum_{j} \int d \mathbf{x}' \, A_{ij}(t, \mathbf{x}, \mathbf{x}') \, \phi^j(t, \mathbf{x}'),
	\end{equation}
where $A_{ij}(t, \mathbf{x}, \mathbf{x}')$ is the kernel of the operator. Thus, the first term in (\ref{eq:action_lor}) has the following explicit structure
	\begin{align}
		\dot\phi^T  A \dot\phi &= \dot{\phi}^I (A \dot{\phi})_I
    \nonumber\\
    &=\sum_{ij} \int d\mathbf{x} \, d\mathbf{x}' \, \dot\phi^i(t, \mathbf{x}) A_{ij}(t, \mathbf{x}, \mathbf{x}') \dot\phi^j(t, \mathbf{x}).
	\end{align}
The superscript $T$ applied to the bilinear form denotes the functional matrix transposition operation which implies the transposition of discrete and spatial labels of the corresponding kernel, but does not touch the time variable
	\begin{equation}
		(B^T)_{ij}(t, \mathbf{x}, \mathbf{x}') = 	B_{ji}(t, \mathbf{x}', \mathbf{x}).
	\end{equation}
Consequently, the second and the third terms in (\ref{eq:action_lor}) are the same. However, we will keep them separate for symmetry reasons.

In local non-gauge theories the kernels of the above coefficients are represented by delta functions of spatial coordinates and their finite order derivatives. For local gauge theories treated within reduction to the physical sector in certain gauges these coefficients can become nonlocal in space, but locality in time derivatives within canonical quantization should be strictly observed.
	
The equations of motion, obtained by varying the action (\ref{eq:action_lor}) with respect to $\phi$ have the form
	\begin{equation} \label{eq:phi_eq}
		F \phi(t) = 0, \quad F \equiv  -\frac{d}{dt} A \frac{d}{dt} - \frac{d}{dt} B + B^T \frac{d}{dt} + C,
	\end{equation}
where the wave operator $F$, or the Hessian of the action (\ref{eq:action_lor}), has already been defined above by Eq.~(\ref{eq:phi_eq0}). Another form of this operator, obtained by integration by parts and involving both left and right time derivatives, the direction of their action being indicated by arrows
	\begin{equation}
		\overboth{F} \equiv  \overleft{\frac{d}{dt}} A \overright{\frac{d}{dt}} + \overleft{\frac{d}{dt}} B + B^T \overright{\frac{d}{dt}} + C,
	\end{equation}
allows one to rewrite the quadratic action (\ref{eq:action_lor}) in even more condensed form
	\begin{align}
		S[\phi] &= \frac12\int_{t_-}^{t_+} dt \,
        \phi^T \overboth{F} \, \phi\nonumber\\
        &= \frac12\int_{t_-}^{t_+} dt \, \phi^T (F \phi) + \frac12 \phi^T (W\phi)\Bigr|_{t_-}^{t_+}.
	\end{align}
Here the Wronskian operator $W$ is defined by (\ref{Wronskian}) and the origin of the boundary term at $t_\pm$ is the result of integration by parts, which is also associated with the Wronskian relation (\ref{Wrelation}).

\subsection{Canonical formalism}
The Hamiltonian formalism of the theory with the action (\ref{eq:action_lor}), which is the first step to the canonical quantization begins with the determination of the momentum $\pi$ canonically conjugated to the field $\phi$
	\begin{equation} \label{eq:field_mom_expr}
		\pi = \frac{\partial L}{\partial \dot\phi} =
    A \dot \phi + B \phi = W \phi, \quad W = A \frac{d}{dt} + B,
	\end{equation}
where $L$ is the Lagrangian of the action (\ref{eq:action_lor}). The corresponding Hamiltonian has the form
	\begin{align} \label{eq:ham_cl}
		H = \pi^T \dot \phi - L
		=\frac12 (\pi - B \phi)^T  A^{-1}(\pi - B\phi)-\frac12\phi^T C \phi.
	\end{align}
Together with the Poisson bracket $\{ \phi^I, \pi_J\} = \delta^I_J$ it defines the dynamics of the system. The Hamiltonian equations of motion read
	\begin{subequations} \label{eq:ham_eq}
	\begin{align}
		\dot\phi &= \{\phi, H\} = A^{-1} (\pi - B \phi) \label{eq:ham_eq_1}\\
		\dot \pi &= \{\pi, H\} = B^T A^{-1} (\pi - B\phi) + C \phi. \label{eq:ham_eq_2}
	\end{align}
	\end{subequations}
Transition to the Lagrangian formalism by expressing $\pi$ in terms of $\phi$ and $\dot\phi$ obviously leads to equations of motion (\ref{eq:phi_eq}) following from the variation of the action (\ref{eq:action_lor}).
	
Let us denote the basis of the independent solutions to (\ref{eq:phi_eq}) as $v_\pm{}^I_A(t)$, where the multi-index $A$ enumerates the number of the particular solution and has the same range as the index $I$. The general solution in terms of basis functions reads
	\begin{equation}
		\phi^I(t) = v_+{}_A^I(t) \, \alpha^{+A} +v_-{}_A^I(t) \, \alpha^{-A}
	\end{equation}
	and can be rewritten in shortened notations as
	\begin{equation}
		\phi(t) = v_+(t) \, \alpha^+ +v_-(t) \, \alpha^-.
	\end{equation}
	Here $\alpha^{\pm A}$ constitute a set of constants, specifying particular initial conditions. Using (\ref{eq:field_mom_expr}), we find the corresponding solution for the momentum
	\begin{equation}
		\pi(t) = Wv_+(t) \, \alpha^+ + Wv_-(t) \, \alpha^-,
	\end{equation}
	so, the evolution of phase space variables can be rewritten in the joint form as
	\begin{equation}
		\begin{bmatrix}
			\,\phi(t)\, \\ \,\pi(t)
		\end{bmatrix} =
		\mathcal{M}(t)
		\begin{bmatrix*}[l]
			\,\alpha^+ \\ \,\alpha^-
		\end{bmatrix*}, \;
		\mathcal{M}(t) =
		\begin{bmatrix}
			\,v_+(t) & v_-(t)\, \\
			\,Wv_+(t) & Wv_-(t)\,
		\end{bmatrix}.
	 \label{eq:phi_pi_evol_pm}
	\end{equation}
	
Now, we can equip the space of initial conditions, consisting of $\alpha^\pm$, with the Poisson bracket structure inherited from the Poisson brackets of $\phi$ and $\pi$. Substituting (\ref{eq:phi_pi_evol_pm}) into the left hand side of
	\begin{align}
		&\left\{ \begin{pmatrix}
			\phi^I \\ \pi_{J}
		\end{pmatrix},
		\begin{pmatrix}
			\phi^{I'} & \pi_{J'}
		\end{pmatrix}
		\right\} =
		\begin{bmatrix}
			\,0 & \delta^I_{J'}\; \\
			\;-\delta_{J}^{I'} & 0\;
		\end{bmatrix},
\end{align}
we have in condensed notations
	\begin{align} \label{eq:symp_tr_pre}
		\mathcal{M}(t)
		\begin{bmatrix}
			\;\{ \alpha^{+}, \alpha^{+}\} & \{ \alpha^{+}, \alpha^{-}\}\; \\
			\;\{ \alpha^{-}, \alpha^{+}\} & \{ \alpha^{-}, \alpha^{-}\}\;
		\end{bmatrix}\mathcal{M}^T(t)
        =
		\begin{bmatrix*}[r]
			\,0 & & I & \\
			\,-I & & 0 &
		\end{bmatrix*},
	\end{align}
where $I$ denotes the identity matrix. The identity above fixes the pairwise Poisson brackets of $\alpha^\pm$. Let us denote the right hand side of this equality, playing the role of the Poisson bivector in the Darboux coordinates, as
	\begin{equation}
		\mathcal{P} \equiv \begin{bmatrix*}[r]
			0 & & I &\\
			-I & & 0 &
		\end{bmatrix*}.
	\end{equation}
Introducing also the matrix $\mathcal D$ as inverse to the matrix of the pairwise Poisson brackets
	\begin{equation} \label{eq:Delta_def}
		\mathcal{D} = \begin{bmatrix}
			\;\Delta_{++} && \Delta_{+-} \\
			\;\Delta_{-+} && \Delta_{--}
		\end{bmatrix} \equiv -
		\begin{bmatrix}
			\,\{ \alpha^{+}, \alpha^{+}\} & \{ \alpha^{+}, \alpha^{-}\}\, \\
			\,\{ \alpha^{-}, \alpha^{+}\} & \{ \alpha^{-}, \alpha^{-}\}\,
		\end{bmatrix}^{-1}
	\end{equation}
where the matrices $\Delta$ denote the corresponding block-elements of $\mathcal{D}$, we can invert the equality (\ref{eq:symp_tr_pre}) as
	\begin{equation} \label{eq:poiss_ev}
		\mathcal{M}^T(t) \, \mathcal{P} \, \mathcal{M}(t) = \mathcal{D}.
	\end{equation}
Thus, one can express the inverse of $\mathcal{M}(t)$ in terms of its transpose, namely
	\begin{equation} \label{eq:M_ev_inv}
		\mathcal{M}^{-1}(t) = \mathcal{D}^{-1} \,
        \mathcal{M}^T(t) \, \mathcal{P}.
	\end{equation}
Before proceeding further, let us show explicitly that the right hand side (\ref{eq:poiss_ev}) is indeed independent of time $t$. To demonstrate this, we contract l.h.s\ of the equation (\ref{eq:phi_eq}) where field $\phi = \phi_1$, with another field $\phi_2$, and subtract the same quantity, but with $F$, acting on $\phi_2$ ($\phi_{1,2}$ are not necessarily solve e.o.m.). The result can be written as
	\begin{equation} \label{eq:phi_wronski_rel}
		 \phi_2^T F \phi_1 - (F\phi_2)^T \phi_1 =
		  - \frac{d}{dt} \left[ \phi_2^T W \phi_1 - (W \phi_2)^T\phi_1\right].
	\end{equation}
	Thus, for $\phi_{1,2}$ --- solutions of (\ref{eq:phi_eq}) l.h.s.\ vanishes, so we have
	\begin{equation}  \label{eq:W_const}
		\phi_2^T W \phi_1 - (W \phi_2)^T\phi_1  = \text{const}.
	\end{equation}
	It is easy to see, that each element of (\ref{eq:Delta_def}) has the form  (\ref{eq:W_const}) as above, where the role of solutions $\phi_1$,~$\phi_2$ is played by the basis functions $v^+$,~$v^-$. Applying the matrix transposition operator to both sides of (\ref{eq:symp_tr_pre}), we obtain that the matrix $\mathcal{D}$ is skew-symmetric, since $\mathcal{P}^T=-\mathcal{P}$. In terms of the block elements of $\mathcal{D}$ this means that
	\begin{equation}
		\Delta_{+-}^T = - \Delta_{-+},\; \Delta_{++}^T = -\Delta_{++},
    \; \Delta_{--}^T = - \Delta_{--}.
	\end{equation}
	Moreover, using the fact that the coefficient matrices $A$,~$B$,~and~$C$ in (\ref{eq:action_lor}) are real, we conclude that basis functions $v_+$,~$v_-$ can also be chosen to be real. Thus, the matrix $\mathcal{D}$ is real skew-symmetric, so there is a time-independent linear transformation $\mathcal{S}$, bringing it to the canonical form, i.e.\ $\mathcal{S}^T \mathcal{D} \mathcal{S} = \mathcal{P}$. Without the loss of generality one set $\mathcal{D} = \mathcal{P}$ by default\footnote{This choice allows to give an additional interpretation for the equation (\ref{eq:poiss_ev}), which becomes $\mathcal{M}^T(t) \mathcal{P} \mathcal{M}(t) = \mathcal{P}$. Namely, the matrix $\mathcal{M}(t)$ performs a time-dependent symplectomorphism of the Poisson bivector $\mathcal{P}$.}. However, for the reasons which will become clear soon (see equation (\ref{eq:Green_full}) below), we will assume that $\mathcal{D}$ has the following more general form
	\begin{equation} \label{eq:D_def_0}
		\mathcal{D} = \begin{bmatrix*}[l]
			\;\;\,0 && \Delta_{+-} \\
			\;\;\Delta_{-+} && \,0
		\end{bmatrix*},
	\end{equation}
	where
	\begin{equation}
		\Delta_{+-} = -\Delta_{-+}^T = v_+^T W v_- - (W v_+)^T v_-.
	\end{equation}
In terms of the basis functions, the vanishing of the diagonal blocks of $\mathcal{D}$ implies that $v_+$,~$v_-$ are chosen such that
    \begin{align}
	\begin{aligned}
		\Delta_{++} &= v_+^T W v_+ - (W v_+)^T v_+ = 0, \\
		\Delta_{--} &= v_-^T W v_- - (W v_-)^T v_- = 0.
	\end{aligned}
    \end{align}
This can always be done by an appropriate transformation of the basis functions, possibly mixing $v_+$ and~$v_-$. Consequently, the pairwise Poisson brackets of $\alpha^+$ and $\alpha^-$ take the form
    \begin{subequations}
	\begin{align} \label{eq:alpha_pb_pm}
		&\{\alpha^+, \alpha^-\} =
    -\{\alpha^+, \alpha^-\} = -\Delta_{-+}^{-1}, \\
		&\{\alpha^+, \alpha^+\} = \{\alpha^-, \alpha^-\} = 0.
	\end{align}
    \end{subequations}
As noted above, one can go further, and set $\Delta_{+-} = -\Delta_{-+}^T = I$.

	Now, let us modify the Hamiltonian by introducing time-dependent sources $J_\phi$,~$J_\pi$ for the field and its conjugate momentum
	\begin{equation}
		H \; \mapsto \; H + J_\phi^T \phi + J_\pi^T \pi.
	\end{equation}
The modified equations of motion can be written as
	\begin{align} \label{eq:ham_eq_src}
		&\frac{d}{dt}
		\begin{bmatrix}
			\phi_J(t) \\ \pi_J(t)
		\end{bmatrix} =
		\mathcal{A}(t)
		\begin{bmatrix}
			\phi_J(t) \\ \pi_J(t)
		\end{bmatrix} + \mathcal{P}
		\begin{bmatrix*}[r]
			J_\phi(t) \\ J_\pi(t)
		\end{bmatrix*},\\
		&\mathcal{A}(t) \equiv
		\begin{bmatrix}
			-A^{-1}B &&& A^{-1} \\
			-B^T A^{-1} B + C &&& B^T A^{-1} \nonumber
		\end{bmatrix},
	\end{align}
where the subscript $J$ of $\phi$,~$\pi$ emphasizes the presence of the sources in equations of motion. We will find a solution to modified equations of motion using the constant variation method. Namely, we start with the solution (\ref{eq:phi_pi_evol_pm}) to equations of motion with vanishing sources, but make the integration constants $\alpha^+$,~$\alpha^-$ in its definition time-dependent
	\begin{equation} \label{eq:phi_pi_evol_var}
		\begin{bmatrix}
			\phi_J(t) \\ \pi_J(t)
		\end{bmatrix} =
		\mathcal{M}(t)
		\begin{bmatrix}
			\alpha^+(t) \\ \alpha^-(t)
		\end{bmatrix},
	\end{equation}
	Then, we substitute the result to the modified e.o.m.\ and obtain
	\begin{equation}
		\mathcal{M}(t) \frac{d}{dt} \!
		\begin{bmatrix}
			\alpha^+(t) \\ \alpha^-(t)
		\end{bmatrix} =
		\mathcal{P}
		\begin{bmatrix*}[r]
			J_\phi(t) \\ J_\pi(t)
		\end{bmatrix*},
	\end{equation}
	where we exploit the fact that $\mathcal M(t)$ satisfies the system (\ref{eq:ham_eq}). Using the equality (\ref{eq:M_ev_inv}) for the inverse of the matrix $\mathcal{M}(t)$ and integrating the equation on $\alpha^+(t)$ and $\alpha^-(t)$ we obtain
	\begin{equation}
		\begin{bmatrix}
			\alpha^+(t) \\ \alpha^-(t)
		\end{bmatrix} =
		\begin{bmatrix}
			\alpha^+_0 \\ \alpha^-_0
		\end{bmatrix}
		-
		\int_{t_-}^t dt' \; \mathcal{D}^{-1} \mathcal{M}^T(t')
		\begin{bmatrix*}[r]
			J_\phi(t') \\ J_\pi(t')
		\end{bmatrix*},
	\end{equation}
where $\alpha^+_0$ and $\alpha^-_0$ are integration constants. Substitution back to (\ref{eq:phi_pi_evol_var}) gives the solution to the equations (\ref{eq:ham_eq_src})
	\begin{equation} \label{eq:pi_phi_inhom_sol}
		\begin{bmatrix}
			\phi_J(t) \\ \pi_J(t)
		\end{bmatrix} =
		\begin{bmatrix}
			\phi_0(t) \\ \pi_0(t)
		\end{bmatrix} -
		\int_{t_-}^t dt' \;\mathcal{M}(t)\, \mathcal{D}^{-1} \mathcal{M}^T(t')
		\begin{bmatrix*}[r]
			J_\phi(t') \\ J_\pi(t')
		\end{bmatrix*},
	\end{equation}
where the initial conditions $\phi_0(t)$,~$\pi_0(t)$ are related to constants of integration by
	\begin{equation}
		\begin{bmatrix}
			\phi_0(t) \\ \pi_0(t)
		\end{bmatrix}=
		\mathcal{M}(t)
		\begin{bmatrix}
			\alpha^+_0 \\ \alpha^-_0
		\end{bmatrix},
	\end{equation}
and represent the solution to homogeneous equation, i.e. for vanishing sources $J_\phi$ and $J_\pi$.
	
	Now, let us focus on the case of vanishing momentum source and also redefine the field source for the convenience
	\begin{equation}
		J_\pi(t) = 0, \qquad J(t) \equiv -J_\phi(t).
	\end{equation}
	The corresponding e.o.m. in the Lagrange form reads
	\begin{equation}\label{eq:phi_eq_inhom}
		F\phi_J(t) + J(t) = 0.
	\end{equation}
	From (\ref{eq:pi_phi_inhom_sol}), one obtains the explicit form of the solution for $\phi(t)$, which is
	\begin{equation} \label{eq:phi_sol_ret}
		\phi_J(t) = \phi_0(t) - \int_{t_-}^{t_+} dt' \, G_R(t, t') J(t'),
	\end{equation}
	where $G_R(t, t')$ is called the retarded Green's function and expressed through the top-left block of the matrix $\mathcal{M}(t)\, \mathcal{D}^{-1} \mathcal{M}^T(t')$, specifically
	\begin{equation} \label{eq:gf_ret_def}
		G_R(t, t')\! =\! -\Bigl(v_+(t) \Delta_{-+}^{-1} v_-^T(t') + v_-(t) \Delta_{+-}^{-1} v_+^T(t') \Bigr) \, \theta(t-t').
	\end{equation}
	The fact that $\Delta_{++}=\Delta_{--}=0$ is crucial in obtaining this simple expression for $G_R$. From (\ref{eq:phi_sol_ret}) we find that $G_R$ satisfies the equation
	\begin{equation}
		F G_R(t, t') = I \,\delta(t-t')
	\end{equation}
	and is uniquely determined by the condition
	\begin{equation} \label{eq:ret_cond}
		G_R(t,t') = 0, \qquad t < t'.
	\end{equation}
	The latter fact follows, in particular, from the fact that any two Green's functions of the same differential operator differ by the solution of the homogeneous equation. Once some Green's function, satisfying the condition (\ref{eq:ret_cond}) is found, a shift by a solution to homogeneous equation will violate this condition.
	Alternatively, $G_R$ can be defined via initial value problem
	\begin{equation}
		G_R(t, t')\bigr|_{t'=t+0} = 0, \; WG_R(t, t')\bigr|_{t'=t+0} = - I.
	\end{equation}
	The fact that solution (\ref{eq:phi_sol_ret}) is expressed through the retarded Green's function means that $\phi(t)$ is subject to the following initial (rather than boundary) value problem
	\begin{equation}
		\phi_J(t_-) = \phi_0(t_-), \; W\phi_J(t_-) = W\phi_0(t_-) \equiv \pi_0(t_-).
	\end{equation}

\subsection{The solution of Dirichlet and Neumann boundary value problems}
	The Green's functions, solving the boundary problems, can be obtained from the retarded Green's function by shifting it by the solution of the homogeneous equation (\ref{eq:phi_eq}). In particular, one constructs the so-called symmetric Green's function as
	\begin{align} \label{eq:gf_sym}
		G_S(t, t')&= G_R(t, t') + v_+(t) \Delta_{-+}^{-1} v_-^T(t') \nonumber \\
    &= -v_+(t) \Delta_{-+}^{-1} v_-^T(t') \, \theta(t-t') \nonumber\\
    &\quad+ v_-(t) \Delta_{+-}^{-1} v_+^T(t') \, \theta(t'-t).
	\end{align}
	It is symmetric under the simultaneous transposition and exchange of the time arguments, i.e. $G_S^T(t,t') = G_S(t', t)$. Unlike the retarded Green's function it is defined non-uniquely and the concrete boundary conditions should be specified. These are in one-to-one correspondence to the boundary conditions, satisfied by the basis functions $v_+$ and $v_-$ at the higher and lower time limits $t=t_+$ and $t=t_-$, respectively.
	
	In particular, to solve the inhomogeneous equation (\ref{eq:phi_eq_inhom}) supplemented with the vanishing Dirichlet boundary conditions
	\begin{equation} \label{eq:bdy_cond_dir}
		\phi_J(t_\pm) = 0,
	\end{equation}
	one can use the Dirichlet Green's function subject to the same boudary conditions
	\begin{equation} \label{eq:gf_dir_bdy}
		G_D(t_\pm, t') = 0 \qquad \leftrightarrow \qquad v_\pm(t_\pm)  = 0,
	\end{equation}
so that the solution reads
	\begin{equation}
		\phi_J(t) = -\int_{t_-}^{t_+} dt' \, G_D(t, t') \, J(t').
	\end{equation}
	Similarly, in solving Neumann boundary problem
	\begin{equation} \label{eq:bdy_cond_neum}
		(iW\mp\omega_\pm)\phi_J(t_\pm) = 0,
	\end{equation}
	one defines the corresponding Neumann Green's function demanding
	\begin{equation} \label{eq:gf_neumann_bdy}
		(iW\mp\omega_\pm)G_N(t_\pm,t') = 0
		\leftrightarrow
		(iW\mp\omega_\pm)v_\pm(t_\pm) = 0,
	\end{equation}
	and obtains the solution as
	\begin{equation} \label{eq:phi_sol_neum}
		\phi_J(t) = -\int_{t_-}^{t_+} dt' \, G_N(t, t') \, J(t').
	\end{equation}

Notably, the Dirichlet and Neumann Green's functions, which are subject to homogeneous boundary conditions, allows one to solve the modified boundary problems, namely with inhomogeneous boundary conditions. Namely, the solutions can be obtained as follow. First, we exploit the equality (\ref{eq:phi_wronski_rel}) and perform in it the substitutions $\phi_2 \mapsto \phi(t')$, $\phi_1 \mapsto G(t', t)$, where $\phi(t')$ solves (\ref{eq:phi_eq_inhom}) and $G(t',t)$ is some Green's function, solving $F G(t',t) = \delta(t-t')$. Next, integrating both sides of the equality over $t'$ from $t_-$ to $t_+$, we obtain
	\begin{align} \label{eq:phi_wronski_int}
		\phi_J(t) = {}& -\int_{t_-}^{t_+} dt' \, G(t, t') \, J(t') \nonumber\\&+ (WG(t_+,t))^T \phi(t_+) - (WG(t_+,t))^T \phi_J(t_+)  \nonumber\\
		&- G^T(t_+,t) \, W \phi_J(t_+) + G^T(t_-,t) \, W \phi_J(t_-)
	\end{align}
	Now, suppose we are to solve (\ref{eq:phi_eq_inhom}) supplemented by inhomogeneous boundary conditions (in contrast to homogeneous ones (\ref{eq:bdy_cond_dir}))
	\begin{equation}
		\phi_J(t_\pm) = \varphi_\pm,
	\end{equation}
	for some constants $\varphi_+$,~$\varphi_-$. Substituting these conditions to (\ref{eq:phi_wronski_int}) together with Dirichlet Green's function $G \mapsto G_D$, satisfying (\ref{eq:gf_dir_bdy}), we observe that the third line vanishes, so we get	
	\begin{equation}  \label{eq:phi_sol_dir_mod}
		\phi_J(t) = - \bw^T(t)
        \begin{bmatrix}
			\;\varphi_+ \\
        \;\varphi_-
		\end{bmatrix} -\int_{t_-}^{t_+} dt' \, G_D(t, t') \, J(t'),
	\end{equation}
where we introduce the notation for the two-component row as the transposition of the newly introduced column
	\begin{align}
		 \bw^T(t) &\equiv\begin{bmatrix}
		 \;G_D(t, t_+) \overleft{W} & & -G_D(t, t_-) \overleft{W}\;
		 \end{bmatrix}\nonumber\\
            &=[\bw(t)]^T,     \label{WGrow}\\
        \bw(t)& \equiv
        \begin{bmatrix*}[r]
	   \overright{W} G_D(t_+,t)  \\
        -\overright{W} G_D(t_-,t)        \label{GWcolumn}
	   \end{bmatrix*},
	\end{align}
and	$\overleft{W}$ denotes the Wronskian operator (\ref{Wronskian}) acting from the right on the second argument of $G_D(t,t')$ at the total boundary of the time domain at $t_\pm$ (the sign taking into account the outward pointing time derivative in $W$) --- the notation used above in (\ref{Bomega}). The transposition law here, of course, takes into account the symmetry of Dirichlet Green's function,
    \begin{align}
    [G_D(t, t_+) \overleft{W}]^T&=
    \bigl(A(t_+)\frac{d}{dt_+}+B(t_+)\bigr)G_D^T(t,t_+)\nonumber\\
    &=\overright{W}G_D(t_+,t).
    \end{align}
The quantity $\bw(t)$ introduced above has the following important property. Namely, evaluating both sides of (\ref{eq:phi_sol_dir_mod}) at the boundary points $t=t_\pm$, and using (\ref{eq:gf_dir_bdy}) we observe that
\begin{align} \label{eq:Wgd_bdy_subs}
	\bw^T(t_+) = \begin{bmatrix}
		\,-I && 0\,
	\end{bmatrix}, \qquad
	\bw^T(t_-) & = \begin{bmatrix}
		\,0 && -I\,
	\end{bmatrix}.
\end{align}
	
Similarly, one can consider inhomogeneous Neumann boundary conditions
	\begin{equation}
		\bigl(\pm iW-\omega_\pm\bigr)\phi_J(t_\pm) = j_\pm,
	\end{equation}
with some boundary sources $j_+$ and $j_-$. Substitution of this condition and Neumann Green's function $G \mapsto G_N$, satisfying (\ref{eq:gf_neumann_bdy}), to (\ref{eq:phi_wronski_int}) gives the solution to (\ref{eq:phi_eq_inhom}) with the boundary conditions above
	\begin{equation} \label{eq:phi_sol_neum_mod}
		\phi_J(t) = -i\, \bg_N^T(t)\begin{bmatrix} \;j_+ \\
    \;j_- \end{bmatrix} - \int_{t_-}^{t_+} dt' \, G_N(t, t') J(t').
	\end{equation}
Here $\bg_N^T(t)$ is the notation analogous to (\ref{WGrow}) --- the row built in terms of the Neumann Green's function kernels with the second  argument located at the total 2-point boundary of the time domain (points $t_-$ and $t_+$),
	\begin{equation}
		\bg_N^T(t) \equiv \begin{bmatrix}
			\;G_N(t, t_+) & & G_N(t, t_-)\;    \label{GNrow}
		\end{bmatrix}.
	\end{equation}

\subsection{The relation between Dirichlet and Neumann Green's functions}	
	There is important explicit connection between Dirichlet and Neumann Green's functions, which can be derived in the following way. The idea is to consider the problem with homogeneous Neumann boundary conditions (\ref{eq:bdy_cond_neum}) as the Dirichlet problem with some nontrivial boundary values $\varphi_\pm$. Substituting the solution of this problem (\ref{eq:phi_sol_dir_mod}) into (\ref{eq:bdy_cond_neum}) one obtains a linear equation on $\varphi_\pm$, which can be solved as
	\begin{equation}
		\begin{bmatrix}
			\;\varphi_+ \\ \;\varphi_-
		\end{bmatrix} =
		(i\bom + \Bom)^{-1} \int_{t_-}^{t_+} dt \, \bw(t) \, J(t),
	\end{equation}
where the matrices $\bom$ and $\Bom$ read
	\begin{align}
		\hspace{-0.2cm}&\qquad\qquad\qquad
        \bom \equiv \begin{bmatrix}
			\;\omega_+ & 0 \\
			0 & \;\omega_-
		\end{bmatrix}, \\
		\hspace{-0.2cm}&
        \Bom\equiv
        \begin{bmatrix*}[r]
		-\overright{W} G_D(t_+,t_+)\overleft{W} &
        \overright{W} G_D(t_+,t_-) \overleft{W}\; \\
			\overright{W} G_D(t_-,t_+) \overleft{W} &
        -\overright{W} G_D(t_-,t_-) \overleft{W}\;
		\end{bmatrix*}.
	\end{align}
Substituting these $\varphi_\pm$ back into (\ref{eq:phi_sol_dir_mod}) gives
	\begin{align}
		\phi_J(t)& = -\int_{t_-}^{t_+} dt' \, \Bigl[G_D(t,t')\nonumber\\
    &+\bw^T(t) \, (i\bom + \Bom)^{-1} \bw(t')
    \Bigr] J(t'),
	\end{align}
which implies, after comparing with (\ref{eq:phi_sol_neum}), the following expression for the Neumann Green's functrion
	\begin{equation} \label{green_neum_dir}
	G_N(t, t') = G_D(t,t') + \bw^T(t)\,
    (i\bom + \Bom)^{-1} \bw(t').
	\end{equation}
Here we use the notations (\ref{WGrow})--(\ref{GWcolumn}) introduced above. Substituting $t=t_\pm$ to the both sides of the equality and using (\ref{eq:Wgd_bdy_subs}), we get the equality
\begin{equation} \label{green_dir_neum_1}
	\bg_N^T(t)  =
	- \bw^T(t)\,  (i\bom + \Bom)^{-1},
\end{equation}
that allows us to express the Dirichlet Green's function from (\ref{green_neum_dir}) via the Neumann one as
	\begin{equation} \label{green_dir_neum_2}
	G_D(t, t') = G_N(t, t') - \bg_N^T(t) \,
    (i\bom + \Bom) \, \bg_N(t'),
	\end{equation}
where we use the notation (\ref{GNrow}) for the row $\bg_N(t)=[G_N(t,t_+)\;\,G_N(t,t_-)]$ and its transpose. Using (\ref{eq:Wgd_bdy_subs}) once again, we can write down the expression for the block matrix of boundary values of the Neumann function $\bg_N$ at both ends of the time segment (double bar denoting the restriction of both arguments to $t_\pm$)
	\begin{equation} \label{green_neum}
		\hspace{-0.2cm}\bG_N\|=\!\begin{bmatrix*}[r]
		\;G_N(t_+,t_+) & G_N(t_+, t_-)\; \\
		\;G_N(t_-,t_+) & G_N(t_-, t_-)\;
		\end{bmatrix*}\!= \!(i\bom + \Bom)^{-1}.
	\end{equation}

\subsection{Canonical quantization}
	Before proceeding to the canonical quantization of the theory (\ref{eq:action_lor}), whose Hamiltonian formalism was constructed in the previous subsection, let us make a more specific choice of basis functions, which is more convenient for the quantization purposes. We first choose the basis functions $v_\pm(t)$ real, and such that the matrix $\mathcal{D}$ defined by (\ref{eq:D_def_0}) has a canonical form, $\mathcal{D}=\mathcal{P}$. Together with the reality of $\phi(t)$ this implies also the reality of the corresponding integration constants $\alpha^\pm$. Next, we combine these basis functions and integration constants into the following complex conjugated pairs
	\begin{align}
		&\hspace{-0.3cm}\begin{bmatrix}
			\,v_+ && v_-
		\end{bmatrix}\mapsto
		\begin{bmatrix}
			\,v && v^*
		\end{bmatrix} =
		\frac1{\sqrt{2}}\begin{bmatrix}
			\,v_+ && v_-
		\end{bmatrix}
		\begin{bmatrix*}[c]
			I & I & \\
			-iI & iI &
		\end{bmatrix*}, \label{eq:bf_conj_def}\\
		&\hspace{-0.3cm}\begin{bmatrix}
			\,\alpha^+ \\ \,\alpha^-
		\end{bmatrix} \;\mapsto\;
		\begin{bmatrix*}[l]
			& \alpha \\ &\alpha^*
		\end{bmatrix*} =
		\frac1{\sqrt{2}}
		\begin{bmatrix*}[r]
			\,I & iI \,  \\
			\,I & -iI\,
		\end{bmatrix*}
		\begin{bmatrix}
			\,\alpha^+ \\ \,\alpha^-
		\end{bmatrix}.
	\end{align}
After this change of basis, the matrix $\mathcal{D}$ becomes
	\begin{equation}
		\mathcal{D} \mapsto i \mathcal{P} =  \begin{bmatrix*}[c]
			\phantom{-}0 && iI &\\
			-iI && 0 &
		\end{bmatrix*}.
	\end{equation}
	According to (\ref{eq:alpha_pb_pm}), this implies the following pairwise Poisson brackets of $\alpha$,~$\alpha^*$
	\begin{equation} \label{eq:alpha_pb}
		\{\alpha, \alpha^*\} = -\{\alpha^*, \alpha\} = -iI, \;
		\{\alpha, \alpha\} = \{\alpha^*, \alpha^*\} = 0.
	\end{equation}
	In terms of the new basis functions, the evolution law (\ref{eq:phi_pi_evol_pm}) of the field and the canonical momentum becomes
	\begin{equation} \label{eq:phi_pi_evol}
		\begin{bmatrix}
			\phi(t) \\ \pi(t)
		\end{bmatrix} =
		\begin{bmatrix}
			v(t) & v^*(t) \\
			Wv(t) & Wv^*(t)
		\end{bmatrix}
		\begin{bmatrix*}[l]
			&\alpha \\ &\alpha^*
		\end{bmatrix*}.
	\end{equation}
	The equation (\ref{eq:M_ev_inv}) takes the form
	\begin{equation} \label{eq:symp_tr_s}
		\mathcal{M}^{-1}(t) = i \, \mathcal{P} \, \mathcal{M}^T\!(t) \, \mathcal{P},
		\quad \mathcal{M}(t) = \begin{bmatrix}
			v(t) & v^*(t) \\
			Wv(t) & Wv^*(t)
		\end{bmatrix}
	\end{equation}
	and allows to invert the equality (\ref{eq:phi_pi_evol}) as
	\begin{equation} \label{eq:phi_pi_evul_inv}
		\begin{bmatrix*}[l]
			\;\alpha \\ \;\alpha^*
		\end{bmatrix*} =
		i \, \mathcal{P} \, \mathcal{M}^T\!(t) \, \mathcal{P}
		\begin{bmatrix}
			\,\phi(t)\, \\ \,\pi(t)\,
		\end{bmatrix}.
	\end{equation}
Evaluating at $t=t_-$ and substituting back to (\ref{eq:phi_pi_evol}), one obtains evolving phase space variables in terms of the basis functions $v(t)$, $v^*(t)$ and initial data,
	\begin{equation} \label{eq:phi_pi_sol}
		\begin{bmatrix}
			\,\phi(t)\, \\ \,\pi(t)\,
		\end{bmatrix} =
		i \, \mathcal{M}(t) \, \mathcal{P}
        \, \mathcal{M}^T\!(t_-) \, \mathcal{P}
		\begin{bmatrix}
			\phi(t_-) \\ \pi(t_-)
		\end{bmatrix}.
	\end{equation}
	
Now, we are ready perform the canonical quantization of the system under consideration, whose Hamiltonian form was obtained in the previous subsection. We will quantize it in the Heisenberg picture. Thus, we map the solutions of the Hamiltonian equations to the corresponding Heisenberg operators, i.e. $\phi(t), \pi(t)\mapsto\hat\phi(t),\hat\pi(t)$, whereas the Poisson bracket is replaced by the commutator times the factor $i$, so that we obtain the equal-time quantum commutation relations $[\hat\phi(t), \hat\pi(t)] = i \hat{I}$, where $\hat I$ is the identity operator in the Hilbert space. Thus, the Hamiltonian equations (\ref{eq:ham_eq}) are mapped to the corresponding Heisenberg equations, defining the evolution of the operators
	\begin{subequations} \label{eq:heis_eq}
		\begin{align}
			\frac{d}{dt} \hat \phi(t) &= -i [\hat \phi(t), \hat H(t)] = A^{-1} (\hat\pi(t) - B \hat\phi(t)) \label{eq:heis_eq_1}
			\\ \frac{d}{dt} \hat \pi(t) &= -i [\hat \pi(t), \hat H(t)]\nonumber\\ &= B^T A^{-1} \bigl(\hat\pi(t) - B\hat\phi(t)\bigr) + C \hat\phi(t). \label{eq:heis_eq_2}
		\end{align}
	\end{subequations}
Here $\hat H(t)$ is the classical Hamiltonian (\ref{eq:ham_cl}) where the field and the momentum are replaced by the corresponding Heisenberg operators.

Linearity of the system obviously implies that the classical Hamiltonian and the Heisenberg equations formally coincide and their solutions are in one-to-one correspondence. In particular, the relation (\ref{eq:field_mom_expr}) between the field $\phi$ and its conjugate momentum $\pi$ is literally the same at classical and quantum levels $\hat\pi(t) = W \hat\phi(t)$. Formal coincidence and linearity of the Hamiltonian and Heisenberg equations allow one to obtain the solution of the latter ones from classical equations (\ref{eq:phi_pi_sol})
	\begin{equation} \label{eq:heis_eq_sol}
		\begin{bmatrix}
			\,\hat{\phi}(t)\, \\ \,\hat{\pi}(t)\,
		\end{bmatrix} =
		i \, \mathcal{M}(t) \, \mathcal{P} \,
        \mathcal{M}^T\!(t_-) \, \mathcal{P}
		\begin{bmatrix}
			\,\hat{\phi}(t_-)\, \\ \hat{\pi}(t_-)
		\end{bmatrix}.
	\end{equation}
Similarly, our quantization procedure implies that the integration constants $\alpha$, $\alpha^*$ are in one-to-one correspondence to the creation/annihilation operators $\hat{a}$,~$\hat{a}^\dagger$. According to (\ref{eq:phi_pi_evol}) the operators $\hat{\phi}(t)$,~$\hat{\pi}(t)$ are decomposed in the creation/annihilation operators as
	\begin{equation} \label{eq:heis_eq_sol_a}
		\begin{bmatrix}
			\,\hat{\phi}(t)\, \\ \hat{\pi}(t)
		\end{bmatrix} =
		\begin{bmatrix}
			v(t) & v^*(t) \\
			\,Wv(t) & Wv^*(t)\,
		\end{bmatrix}
		\begin{bmatrix*}[l]
			\;\hat{a} \\ \;\hat{a}^\dagger
		\end{bmatrix*},
	\end{equation}
	that can be inverted similar to (\ref{eq:phi_pi_evul_inv}) as
	\begin{equation} \label{eq:phi_pi_inv_a}
		\begin{bmatrix*}[l]
			\;\hat{a} \\ \;\hat{a}^\dagger
		\end{bmatrix*} =
		i \, \mathcal{P} \, \mathcal{M}^T\!(t) \, \mathcal{P}
		\begin{bmatrix}
			\,\hat{\phi}(t)\, \\ \hat{\pi}(t)
		\end{bmatrix}.
	\end{equation}	
	The fact that $\hat{a}$ and $\hat{a}^\dagger$ are indeed Hermitian conjugate to each other immediately follows from the Hermicity of $\hat{\phi}(t)$. Indeed, comparing $\hat{\phi}(t)$ to it's conjugate
	\begin{align}
    \begin{aligned}
		\hat{\phi}(t) &= v(t)\, \hat{a} + v^*(t) \, \hat{a}^\dagger, \\
		\hat{\phi}^\dagger(t) &= \bigl(v(t) \, \hat{a} + v^*(t)
        \, \hat{a}^\dagger\bigr)^\dagger = v^*(t) \, \hat{a}^\dagger + v(t) \, \hat{a},
    \end{aligned}
	\end{align}
we find the coincidence, for which the choice (\ref{eq:bf_conj_def}) of two complex conjugated basis functions is crucial. The commutation relations of the creation/annihilation operators are inherited from the Poisson brackets (\ref{eq:alpha_pb}), namely
	\begin{align}\label{eq:a_comm_rel}
    \begin{aligned}
	&[\hat{a}^A, \hat{a}^{\dagger B}] =
    - [\hat{a}^{\dagger B}, \hat{a}^B] = \delta^{AB} \hat I,\\ &[\hat{a}^A, \hat{a}^B] = [\hat{a}^{\dagger A}, \hat{a}^{\dagger B}] = 0.
    \end{aligned}
	\end{align}

	Though the explicit solution to the Heisenberg equations (\ref{eq:heis_eq_sol}) is obtained, we still have no the expression for the evolution operator in a closed form. The latter solves the Schroedinger equation in the form of the chronological ordering,
	\begin{equation}
	i \frac{d}{dt} \hat{U}(t, t') = \hat H_S(t) \hat{U}(t, t'), \;
    \hat{U}(t_+, t_-)=\mathrm{T} e^{-i \int_{t_-}^{t_+} dt \,\hat H_S(t)},
	\end{equation}
where $\hat{H}_S(t)$ is the Hamiltonian in the Schroedinger picture, so that its time dependence is only due to time-dependent coefficients $A$,~$B$, and~$C$. The operators $\hat{\phi}$,~$\hat{\pi}$ in the Schroedinger picture are identified with the Heisenberg ones, evaluated at the initial time
	\begin{equation} \label{eq:phi_pi_schr}
	\hat{\phi}=\hat{\phi}(t_-), \qquad \hat{\pi}
    =\hat{\pi}(t_-).
	\end{equation}	
In the presence of the source, $H \mapsto H - J^T \phi$, the solution (\ref{eq:heis_eq_sol_a}) to the Heisenberg equation generalizes to
	\begin{equation}
		\hat{\phi}_J(t) = \hat{\phi}(t) - \int_{t_-}^{t_+} dt' \, G_R(t, t') J(t'),
	\end{equation}
that can be easily derived from (\ref{eq:phi_sol_ret}). Here $\hat{\phi}(t)$ is the solution (\ref{eq:heis_eq_sol_a}) to the sourceless Heisenberg equation. The Schroedinger equation for the evolution operator in the presence of the source
	\begin{equation} \label{eq:ev_op_eq}
		i \frac{d}{dt} \hat{U}_J(t, t') = \bigl(\hat H_S(t)
    - J^T(t) \hat{\phi} \bigr)\, \hat{U}_J(t, t'), \; \hat{U}_J(t,t) = \hat{I}
	\end{equation}
can be solved by the chronological Dyson $t$-exponent (cf. Eq.~(\ref{Unitary})),
    \begin{equation}
    \hat{U}_J(t_+, t_-)=\mathbb{\mathrm{T}}\,
    e^{-i \int^{t_+}_{t_-} dt \,
    \bigl(\hat H_S(t)-J(t)\hat\phi_S\bigr)}.  \label{Unitary1}
	\end{equation}

Another representation for the evolution operator follows from functional integral formalism. If one introduces the coordinate representation, associated with the Scroedinger operators (\ref{eq:phi_pi_schr}),
	\begin{equation} \label{eq:coord_repr}
		\hat{\phi}|\varphi\ra = \varphi\, |\varphi\ra, \;
		\hat{\pi}|\varphi\ra =
    i\frac{\partial}{\partial\varphi} |\varphi\ra,
    \; \hat{I} = \int d\varphi \, |\varphi\ra\la\varphi|,
	\end{equation}
then the matrix elements of $\hat{U}_J$ in the coordinate representation express in terms of the following functional integral
	\begin{align} \label{eq:ev_op_pi}
		&\la\varphi_+|\hat{U}_J(t_+, t_-)\,
        |\varphi_-\ra\nonumber\\
        &\quad=\!\!\!\!\!\!\int
		\limits_{\phi(t_\pm)=\varphi_\pm}\!\!\!\!\!\!\calD\phi \;
        \exp\Bigl\{ iS[\phi]
         + i \int_{t_-}^{t_+}dt \,J^T(t) \phi(t)\Bigr\}.
	\end{align}
	Since the action (\ref{eq:action_lor}) is quadratic in the field $\varphi$, the latter integral is Gaussian, so it can be calculated explicitly. We will do this in the next section with the use of the saddle point method.
%

\subsection{Bogoliubov transformations}
	In the previous subsection we made the choice (\ref{eq:bf_conj_def}) of basis functions which implies a simple form of the commutation relations (\ref{eq:a_comm_rel}) for the creation/annihilation operators. It will be useful to study the transformations, preserving these commutation relations. For this purpose, let us define a new set of creation/annihilation operators $\hat{b}$,~$\hat{b}^\dagger$ as a linear combination of the initial ones,
	\begin{align}
		\begin{bmatrix*}[l]
			\, \hat b \\ \,\hat b^\dagger
		\end{bmatrix*}
		= \begin{bmatrix*}[l]
			&U & V \\
			&V^* & U ^*
		\end{bmatrix*}
		\begin{bmatrix*}[l]
			\, \hat a \\ \,\hat a^\dagger
		\end{bmatrix*},
	\end{align}
where $U$, $V$ are referred to as the matrices of Bogoliubov transformations. Demanding that the commutation relations of the new cration/annihilation operators coincides with those of the initial ones (\ref{eq:a_comm_rel}), one obtains the equality
	\begin{equation}
		\begin{bmatrix*}[l]
			&U & V \\
			&V^* & U ^*
		\end{bmatrix*}
		\begin{bmatrix*}[r]
			0 && I &\\
			-I && 0 &
		\end{bmatrix*}
		\begin{bmatrix}
			&U^T & V^\dagger\, \\
			&V^T & U^\dagger
		\end{bmatrix} =
		\begin{bmatrix*}[r]
			0 && I &\\
			-I && 0 &
		\end{bmatrix*} \label{eq:bogol_cond}
	\end{equation}	
so that $U$ and $V$ should satisfy
	\begin{align}
		U U^\dagger - V V^\dagger = I,  \qquad
		U V^T - V U^T = 0.
	\end{align}
Thus, the field operator $\hat{\phi}(t)$ has two equivalent decompositions
	\begin{align}
		\hat \phi(t) = v(t) \, \hat a + v^*(t) \,  \hat a^\dagger = \tilde v(t) \, \hat b + \tilde v^*(t) \, \hat b^\dagger,	
	\end{align}
where the new set of the basis functions $\tilde{v}(t)$,~$\tilde{v}^*(t)$ is related to the initial one via the following relation
	\begin{equation}
		\begin{bmatrix}
			\,v && v^*\,
		\end{bmatrix} =
		\begin{bmatrix}
			\,\tilde v && \tilde v^*\,
		\end{bmatrix}
		\begin{bmatrix*}[l]
			&U & V \\
			&V^* & U ^*
		\end{bmatrix*}
	\end{equation}
	or, in more explicit form
	\begin{equation}
		v = \tilde v \, U + \tilde v^* V^*. \label{eq:bogol_mode}
	\end{equation}
Equality (\ref{eq:bogol_cond}) leads to the following formula for the inverse matrix of the Bogoliubov transformation coefficients
	\begin{align}
		\begin{bmatrix*}[l]
			&U & V \\
			&V^* & U ^*
		\end{bmatrix*}^{-1}
 =
		\begin{bmatrix*}[r]
			U^\dagger & -V^T \\
			-V^\dagger & U^T
		\end{bmatrix*}
	\end{align}
so that (\ref{eq:bogol_mode}) can be inverted as
	\begin{equation} \label{eq:bogol_mode_inv}
		\tilde v = v \, U^\dagger - v^* V^\dagger.
	\end{equation}

	Now, let us solve the inverse problem. Namely, suppose we have two sets of the basis functions $v(t)$,~$v^*(t)$ and $\tilde{v}(t)$,~$\tilde{v}^*(t)$, such that the commutation relation of the corresponding creation/annihilation operators are of the canonical form (\ref{eq:a_comm_rel}). The question is what are the Bogoliubov coefficients relating these two sets? To find the explicit form of the coefficients, let us introduce the following inner product in the space of solutions of the equation (\ref{eq:phi_eq})
	\begin{equation} \label{eq:kg_inner_prod}
		(\phi_1, \phi_2) = i\,\phi_1^\dagger (W\phi_2) - i(W\phi_1)^\dagger \phi_2.
	\end{equation}
	This is constant if $\phi_1$,~$\phi_2$ solve (\ref{eq:phi_eq}) due to the Wronskian property (\ref{eq:W_const}), together with the fact that the operator $F$, defining equations of motion (and the Wronskian $W$) is real. The inner product (\ref{eq:kg_inner_prod}) is usually referred to as the Klein-Gordon type inner product. The choice (\ref{eq:bf_conj_def}) of the basis functions implies the following normalization with respect to this inner product
	\begin{equation} \label{eq:bogol_norm}
		(v, v)= -(v^*, v^*) = I, \quad
		(v^*, v) = 0,
	\end{equation}
and the same for $\tilde{v}(t)$,~$\tilde{v}^*(t)$. Projecting the equality (\ref{eq:bogol_mode}) onto $\tilde v$, and using the property $(v_1, v_2)^* = -(v_1^*, v_2^*)$, one obtains the explicit expressions for the Bogoliubov coefficients
	\begin{equation} \label{eq:bogol_dot}
		U = (\tilde v, v), \qquad V = (\tilde v, v^*).
	\end{equation}
	
If both the old and the new sets of basis functions satisfy the Neumann conditions with different frequency matrices $\omega$ and $\tilde\omega$ at the initial moment of time
	\begin{equation}
		(i W - \omega) v(t_-) = 0, \quad (i W - \tilde{\omega}) \tilde{v}(t_-) = 0,
	\end{equation}
one can find the explicit expressions for $U$, $V$ in terms of $\omega$ and $\tilde \omega$. Let us first write down the normalization conditions (\ref{eq:bogol_norm}) explicitly
	\begin{equation}
		(v, v) = v^\dagger (\omega + \omega^\dagger) v = I, \; (v^*, v) = v^T (\omega - \omega^T) v = 0,
	\end{equation}
where all quantities are evaluated at $t=t_-$. The same equations hold for $\tilde{v}(t)$ and $\tilde{v}^*(t)$. The second equation implies that the matrices $\omega$ and $\tilde{\omega}$ are symmetric (in view of invertibility of the matrix $v(t)$ at a generic moment of time), whereas the first equation allows one to fix the initial value of the basis functions as
	\begin{equation}
		v(t_-) = \frac1{\sqrt{2 \omega_\re}}, \quad \tilde v(t_-)
        = \frac1{\sqrt{2\tilde{\omega}_\re}},
	\end{equation}
where $\omega_\re$ and $\tilde{\omega}_\re$ are the real parts of $\omega$ and $\tilde{\omega}$, respectively. Using (\ref{eq:bogol_dot}) with the inner product defined in (\ref{eq:kg_inner_prod}) one finds the following expressions for Bogoliubov coefficients relating two sets of Neumann basis functions with different frequency matrices,
    \begin{subequations}\label{eq:bogol_coef}
	\begin{align}
		&U = \frac1{\sqrt{2\tilde{\omega}_\re}} (\omega
    + \tilde \omega^\dagger) \frac1{\sqrt{2\omega_\re}}, \\
		&V = \frac1{\sqrt{2\tilde{\omega}_\re}} (\tilde\omega^\dagger
    - \omega^\dagger) \frac1{\sqrt{2\omega_\re}}.
	\end{align}
    \end{subequations}

\subsection{Fock space and the coherent states}
	Once the basis functions $v(t)$,~$v^*(t)$ are chosen, we can define the Fock space, associated to the corresponding creation/annihilation operators. Namely, introducing the vacuum state~$|0\ra$ as
	\begin{equation}\label{eq:fock_vac_def}
		\hat{a}|0\ra = 0,
	\end{equation}
	one defines the Fock space as a linear space spanned by
	\begin{equation} \label{eq:fock_state}
		|A_1, A_2, \ldots A_n\ra \equiv 	\hat{a}^{\dagger A_1} \hat{a}^{\dagger A_2} \ldots \hat{a}^{\dagger A_n}|0\ra.
	\end{equation}

Next, let us obtain the coordinate representation of the Fock states. For this purpose, we rewrite (\ref{eq:phi_pi_inv_a}) explicitly for $t=t_-$ as
	\begin{align}
		&\begin{bmatrix*}[l]
			&\hat{a} \\ &\hat{a}^\dagger
		\end{bmatrix*} =
		\frac1{\sqrt{2\omega_\re}}
		\begin{bmatrix*}[r]
			&\omega^* & iI&\\
			&\omega^{\phantom{*}} & -iI&
		\end{bmatrix*}
		\begin{bmatrix}
			&\hat{\phi}& \\ &\hat{\pi}&
		\end{bmatrix},
	   \end{align}
where $\omega$ is given in terms of the positive frequency basis function
    	\begin{align} \label{omega_in_vac}
        &\omega =  \bigl(iWv\bigr)v^{-1}\bigr|_{t=t_-},
	   \end{align}
and rewrite the definition (\ref{eq:fock_vac_def}) of the vacuum state in the coordinate representation (\ref{eq:coord_repr}) as
	\begin{equation}
		\frac1{\sqrt{2\omega_\re}} \left(\frac\partial{\partial\varphi} + \omega^* \varphi\right) \la \varphi| 0 \ra= 0.
	\end{equation}
Therefore, up to $\pi$-dependent normalization the wavefunction of a vacuum reads
	\begin{equation} \label{eq:vac_coord_repr}
		\la \varphi| 0\ra = \bigl(\det \omega_\re\bigr)^{\frac14}  \, \exp \biggl\{-\frac12 \varphi^T \omega^* \varphi \biggr\}.
	\end{equation}
Coordinate representation of the excited states can be found by using the definition (\ref{eq:fock_state}) and the expression for $\hat{a}^\dagger$ in the coordinate representation.
	
	Similarly, one can define the coherent states $|\alpha\ra$ as eigenstates of the annihilation operator
	\begin{equation} \label{eq:coh_state}
		\hat{a}|\alpha\ra = \alpha |\alpha\ra.
	\end{equation}
	Projecting the definition on the coordinate representation basis vector $|\varphi\ra$, one obtains the equation
	\begin{equation}
		\left(\frac\partial{\partial\varphi} + \omega^* \varphi\right) \la \varphi| \alpha\ra = \sqrt{2\omega_\re} \la \varphi|\alpha \ra,
	\end{equation}
	whose integration gives the (unnormalized) solution
	\begin{equation}
		\la \varphi| \alpha \ra = \exp \biggl\{-\frac12 \varphi^T \omega^* \varphi + \alpha^T\! \sqrt{2\omega_\re} \,\varphi - \frac12 \alpha^T\alpha \biggr\}.
	\end{equation}
	For this normalization we have the following expression for the Fock states in terms of coherent state
	\begin{equation} \label{eq:fock_gen}
		|A_1, A_2, \ldots A_n\ra = \left.	\frac{\partial^n}{\partial\alpha^{A_1}\, \partial\alpha^{A_2} \ldots \partial\alpha^{A_n}}|\alpha\ra\right|_{\alpha=0}.
	\end{equation}
	Coherent states allows to perform a partition of unity as
	\begin{equation} \label{eq:coh_unity_part}
		\hat I = \int d \alpha^* \, d\alpha \, e^{-\alpha^\dagger \alpha} |\alpha \ra\la \alpha|.
	\end{equation}

\section{Generating functional in the path integral formalism}\label{sec:Gen_Funct}
	We begin our derivation of the in-in Green's function generating functional for the theory, defined in the previous section, by the physical motivation and the definition of an arbitrary Gaussian initial state. After that, we derive the corresponding two-component Green's functions. As we will observe, there is an ambiguity in the definition of these Green's function, parameterized by a matrix, defining initial conditions for the modes, employed in the mode expansion of the field operators. There is no any a priory preferred choice, fixing this ambiguity. However, being motivated by the simple harmonic oscillator in a thermal state, we make a choice of the modes such that the resulting Green's function has the form and the properties, very close to that of the Green's functions for the equilibrium system in a thermal state. Further, we introduce the notion of the quasi-thermal state, which is a very particular case of the Gaussian state, in which the properties of the Green's functions become even more closer to those of the thermal ones, in particular, satisfying the Kubo-Martin-Schwinger (KMS) condition.
	
\subsection{Gaussian states}
	Our goal is to find the explicit and useful form of the generating functional
	\begin{equation} \label{eq:gen_fun_def}
		Z[J_1,J_2] = \tr \left[ \hat{U}_{J_1}(T,0) \, \hat{\rho} \,\, \hat{U}^\dagger_{-J_2}(T,0) \right]
	\end{equation}
	of in-in correlation functions
	\begin{align}
		&\tr \left[ \hat{\rho} \, \bar{\mathrm{T}}\bigl(\hat \phi(t'_1) \ldots \phi(t'_m) \bigr) \, \mathrm{T}\bigl(\hat \phi(t_1) \ldots \phi(t_n) \bigr) \right] \nonumber\\ &=
		\frac{i^{m-n}}{Z} \left.\frac{\delta^{n+m}Z[J_1,J_2]}{\delta J_1(t_1) \ldots \delta J_1(t_n) \, \delta J_2(t'_1) \ldots \delta J_2(t'_m)}\right|_{J_1=J_2=0}. \label{eq:corr_fun_def}
	\end{align}
where $\hat{U}_{J}$ are the evolution operators subject to equation (\ref{eq:ev_op_eq}) with different sources $J_1$ and $-J_2$, whereas $\mathrm{T}$ and~$\bar{\mathrm{T}}$ denote chronological and anti-chronological ordering, respectively. The relation between (\ref{eq:gen_fun_def}) and the correlation functions (\ref{eq:corr_fun_def}) obviously follows from (\ref{Unitary1}). The basic elements are the two-point correlation functions, namely
	\begin{subequations} \label{eq:gf_comp}
		\begin{align}
			iG_{\mathrm{T}}(t, t') &\equiv \tr \left[\hat \rho \,
        \mathrm{T} \bigl( \hat \phi(t) \hat\phi(t') \bigr)\right], \\
			iG_{\bar{\mathrm{T}}}(t, t') &\equiv \tr \left[\hat \rho
        \, \bar{\mathrm{T}} \bigl( \hat \phi(t) \hat\phi(t') \bigr)\right], \\
			iG_{<}(t, t') &\equiv \tr \left[\hat \rho \,
        \hat \phi(t) \hat\phi(t')\right],
		\end{align}
	\end{subequations}
where $G_{\mathrm{T}}$,~$G_{\bar{\mathrm{T}}}$, and~$G_{<}$ are Feynman, anti-Feynman and Wightman Green's functions, respectively.
		
	The density matrix $\hat \rho$ is assumed to be the Hermitian positive-definite operator of unit trace.
	Inserting the partition of unity in the coordinate representation to the definition (\ref{eq:gen_fun_def}) of the generating functional three times, and using the path integral representation (\ref{eq:ev_op_pi}) of the evolution operator, one obtains the following expression for the generating functional
	\begin{align} \label{eq:gen_fun_pi_gen}
	&\hspace{-0.3cm}Z[J_1,J_2] = \int d\varphi_+\, d\varphi_-\;
    \rho(\varphi_+, \varphi_-) \nonumber\\
    &\;\;\times\int\calD \phi_1 \, \calD\phi_2 \; \exp \Biggl\{iS[\phi_1] - iS[\phi_2]\nonumber\\
    &\;\;+ i\int_{0}^{T} dt \, \left.\bigl(J_1^T \phi_1+ J_2^T \phi_2\bigr) \Biggr\}\,
    \right|_{\substack{\phi_1(T)=\phi_2(T),\\
    \phi_1(0)=\varphi_+,\,
    \phi_2(0)=\varphi_- }}\!\!\!\!,
	\end{align}
where the integration over $\phi_{1,2}(t)$ runs with the indicated boundary conditions and we introduce the notation for the coordinate representation of the density matrix $\rho(\varphi_+,\varphi_-)=\la \varphi_+| \mathinner{\hat \rho} | \varphi_- \ra$.

Now, we restrict ourselves with the Gaussian density matrices, i.e.\ those whose coordinate representation has the form of the Gaussian exponent
	\begin{equation} \label{eq:dm_coord}
		\hspace{-0.1cm}\rho(\bvphi) =
    \frac1Z \exp\left\{ -\frac12 \bvphi^T \Bom \, \bvphi + \bj^T \! \bvphi  \right\}, \; \bvphi = \begin{bmatrix}
			\,\varphi_+ \\ \,\varphi_-
		\end{bmatrix},
	\end{equation}
	where the matrix $\Bom$, and the vector $\bj$ play the role of the parameters of $\hat{\rho}$, and normalization constant $1/Z$ is independent of $\bvphi$. The Hermitian property of the density matrix, $\la\varphi_+| \mathinner{\hat{\rho}} |\varphi_- \ra = \la \varphi_-|\mathinner{\hat \rho}| \varphi_+\ra^*$, which in the coordinate representation reads
	\begin{equation}
    \rho(\varphi_+, \varphi_-) = \rho^*(\varphi_-, \varphi_+),
	\end{equation}
implies the following conditions on $\Bom$ and $\bj$
	\begin{equation}
		\bX \, \Bom \, \bX = \Bom^*, \quad
		\bX \,\bj = \bj^*,
        \quad \bX \equiv \begin{bmatrix}
			\;\,0 && I\; \\ \;I && 0\,
		\end{bmatrix}, \label{eq:bom_conj}
	\end{equation}
or, in a more explicit block-matrix form
	\begin{equation}
		\bj = \begin{bmatrix*}[l]
			\;\,j \\
\;\,j^*
		\end{bmatrix*}, \;
\Bom =
		\begin{bmatrix*}[l]
			\;\,R\, & S \\
 \;\,S^* & R^*
		\end{bmatrix*}, \; R = R^T, \; S = S^\dagger.
	\end{equation}
	Normalizability of $\hat{\rho}$ implies that the real part of the sum $R + S$ is positive-definite.
	The case in which the matrix $S$ is non-vanishing corresponds to the mixed states, i.e.\ such that $\hat{\rho}^2 \ne \hat{\rho}$. The role of the linear term in the exponential in (\ref{eq:dm_coord}) is two-fold. Firstly, $\bj$ defines non-vanishing mean value of the field operator. Secondly, it can also be used to introduce non-linearities to the density matrix, namely, by differentiating it with respect to \ $\bj$. The typical example of the (pure) Gaussian state is the vacuum state (\ref{eq:fock_vac_def}), i.e. $\hat{\rho} = |0\ra\la0|$, associated with some choice of the annihilation operator, for which $R = \omega^*$, $S = 0$, and $\bj$ = 0. Another example of the pure Gaussian state is the coherent state (\ref{eq:coh_state}) whose density matrix reads $\hat{\rho} = |\alpha\ra\la\alpha|$, and $R = \omega^*$,~$S=0$ again, but $j = \sqrt{2\omega_\re} \alpha^*$.

\subsection{In-in boundary value problem} \label{subs:in-in}
Substituting the general Gaussian density matrix to (\ref{eq:gen_fun_pi_gen}), one obtains
	\begin{multline} \label{eq:gen_fun_pi_gaus}
		Z[J_1,J_2] =
    \!\!\!\!\!\!\int\limits_{\substack{\phi_1(T)=\phi_2(T)\\\bphi(0)=\bvphi}}
    \!\!\!\!\!\! \calD \phi_1 \, \calD\phi_2 \, \exp \Biggl\{iS[\phi_1] - iS[\phi_2]\\
    \,\,\,\,+ i\int_{0}^{T}\! dt\bigl(J_1^T \phi_1+J_2^T \phi_2\bigr)\! -\! \frac12 \bvphi^T \Bom \, \bvphi\!+\! \bj^T \! \bvphi \Biggr\},
	\end{multline}
	where we put the boundary points $\varphi_\pm$,~$\varphi'$, appearing in (\ref{eq:gen_fun_pi_gen}) to the functional integration measure, and also omit the constant normalization factor of the density matrix.

	We will compute the integral (\ref{eq:gen_fun_pi_gaus}) representing the generating functional, with the use of saddle point method. The latter turns out to be exact since the integral has the Gaussian form. First of all, we introduce the notations for the block-matrix operators acting on columns of fields and sources (\ref{double_objects}) introduced in Section~\ref{sec:summary},
	\begin{gather}
		\bF = \begin{bmatrix}
			\;F & 0 \\ 0 & - F\,
		\end{bmatrix}, \quad
		\bW = \begin{bmatrix}
			\;W & 0 \\ 0 & - W\,
		\end{bmatrix},
	\end{gather}
so that the sum of the actions for $\phi_1$ and $\phi_2$ in (\ref{eq:gen_fun_pi_gaus}) can be rewritten in the joint form
	\begin{align}
	\bS[\bphi] &= \frac12 \int_{0}^{T} dt \,
    \bphi^T \overboth{\bF} \bphi\nonumber\\
     &=\frac12 \int_{0}^{T} dt \, \bphi^T \bF \bphi + \frac12 \bphi^T  \bW \bphi\,\biggr|_{0}^{T}.
	\end{align}
This allows us to treat the underlying equations of motion, Green's functions, etc. in exactly the same way as the original theory with the action (\ref{eq:action_lor}), except that now the field content is doubled. In terms of the new notations the expression for the generating functional is given by Eqs.(\ref{gen_fun_pm0})--(\ref{Measure}) of Section~\ref{sec:summary}.

The saddle point equation obtained by varying the exponential of this double-field action (\ref{gen_fun_pm0}) with respect to all fields including the boundary values at $t=0$ and $t=T$ reads
	\begin{align}
		&\delta\left\{i\bS[\bphi] + i\int_0^T dt \, \bJ^T \bphi
    - \frac12 \bvphi^T \Bom \, \bvphi + \bj^T\! \bvphi \right\}\nonumber\\
    &\qquad\quad=\int_0^T dt \, \delta\bphi^T (\bF \bphi+ \bJ)+i\,\delta\bphi^T\,\bW \bphi \bigr|_{t=T}\nonumber\\
    &\qquad\quad-\delta\bvphi^T\Bigl[(i \bW + \Bom) \bphi\bigr|_{t=0}-\bj\,\Bigr]=0.
	\end{align}
Independent variation of the fields $\delta\bphi(t)$ in the interior of the time interval gives equations of motion
	\begin{equation} \label{eq:bulk_eq_lor}
		\bF \bphi(t) + \bJ(t) = 0,
	\end{equation}
whereas the variation of the boundary values $\delta\bphi(T)$ and $\delta\bphi(0)=\delta\bvphi$ supply these equations with the boundary conditions. They read as the following matrix relations
		\begin{align}
			&(i \bW + \Bom)\,
            \bphi\bigr|_{t=0} = \bj,      \label{eq:bdy_eq_lor_1}\\
			&
			\begin{bmatrix}
				\,I && I\,
			\end{bmatrix}\,\bW \bphi \bigr|_{t=T} = 0, \;
			\begin{bmatrix}
				\,I & -I\,
			\end{bmatrix}\bphi \bigr|_{t=T} = 0, \label{eq:bdy_eq_lor_2}
		\end{align}
where we took into account that in view of $\phi_1(T)=\phi_2(T)$ the variation $\delta\bphi^T(T)=\delta\phi_1^T(T)\begin{bmatrix}\,I & I\,\end{bmatrix}$, and the boundary conditions at $t=T$ reduce to the equality of both the fields and their time derivatives of both $\phi_1$ and $\phi_2$.
	
To solve the boundary value problem above, we first find the Green's function subject to the homogeneous version of the above boundary conditions, i.e. those of vanishing $\bj$
	\begin{align}
		&\bF \bG(t,t') = \boldsymbol{I}\, \delta(t-t'), \\
		&(i \bW + \Bom)\bG(t,t')\bigr|_{t=0} = 0 , \\
		&\begin{aligned}
		&\begin{bmatrix}
			\,I && I\,
		\end{bmatrix}\bW \bG(t,t')\bigr|_{t=T} = 0, \\
		&\begin{bmatrix}
			\,I & -I\,
		\end{bmatrix} \bG(t,t') \bigr|_{t=T} = 0.
    \end{aligned}
	\end{align}
	We can construct the Green's function $\bG$ solving the problem above, out of the basis functions $\bv_\pm$. These basis function should solve the homogeneous equation, and satisfy the same boundary conditions as those of the Green's function,
	\begin{align}
		&\bF \bv_\pm(t) = 0,  \quad
		(i \bW + \Bom)\,\bv_-(t)\bigr|_{t=0} = 0 , \\
        &\begin{aligned}
		  &\begin{bmatrix}
			\,I && I\,
		  \end{bmatrix}\bW \bv_+(t) \bigr|_{t=T} = 0, \\
		  &\begin{bmatrix}
			\,I & -I\,
		  \end{bmatrix} \bv_+(t) \bigr|_{t=T} = 0.
        \end{aligned}
	\end{align}
Applying the generic Green's function expression (\ref{eq:gf_sym}) to the case of the doubled field content, we obtain the Green's function $\bG$ in terms of these basis functions
	\begin{align}
		\bG(t, t') &= -\bv_+(t) \, \bDel_{-+}^{-1} \, \bv^T_-(t') \,
    \theta(t - t') \nonumber\\
    &\quad+ \bv_-(t) \, \bDel_{+-}^{-1} \,
    \bv^T_+(t') \, \theta(t' - t)           \label{eq:Green_full} \\
	\bDel_{-+} &= \bv_-^T \,\bW \bv_+
    - (\bW\bv_-)^T \,\bv_+ = -\bDel_{+-}^T. \label{eq:Delta_Green_full}
	\end{align}

\subsection{Neumann type basis functions and Green's function representation}

However, we do not have the explicit form of basis functions $\bv_\pm$. We will construct $\bv_\pm$ with the help of another set basis functions $\bv$, $\bv^*$ subject to much simpler boundary conditions
	\begin{align}
		&\bF \bv(t) = 0, \quad
		(i \bW - \bom)\bv(t)\bigr|_{t=0} = 0, \label{eq:bv_eq} \\
		&\bom = \begin{bmatrix}
			\;\,\omega\; & 0 \phantom{{}^*} \\ \,\;0\; &\omega^*
		\end{bmatrix}. \label{bom_diag}
	\end{align}
Since $\bW$ and $\bom$ are block-diagonal, the basis functions $\bv$,~$\bv^*$ can be chosen block-diagonal too, namely
	\begin{equation}
		\bv = \begin{bmatrix*}[l]
			&v && 0 \\ &0 && v^*
		\end{bmatrix*}, \qquad
		\bv^* = \begin{bmatrix*}[l]
			&v^* & 0 &\\ &0 & v&
		\end{bmatrix*}.
	\end{equation}
With a real operator $F$ the blocks of these matrices solve the equations $Fv(t) = 0$ and $Fv^*(t) = 0$, subject to the complex conjugated boundary conditions
	\begin{gather}
	(i W - \omega) v(t)\bigr|_{t=0} = 0,
    \quad (i W + \omega^*) v^*(t)\bigr|_{t=0} = 0. \label{eq:green_lor_neum_bdy}
	\end{gather}
Thus, $v$ and $v^*$ are simply the basis functions for the single field $\phi_+$ or $\phi_-$ subject to the Neumann boundary conditions introduced above. We assume that $\omega$ is the symmetric matrix with a positive-definite real part.

The answer for the basis function $\bv_+$ in terms of $v$ and $v^*$ can be easily constructed as
	\begin{align} \label{v+_vs_v}
		\bv_+ = \bv + \bv^* \bX = \begin{bmatrix}
			\;v && v^* \\ \;v && v^*
		\end{bmatrix}, \; \bX = \begin{bmatrix}
			\,0 && I\, \\ \,I && 0\,
		\end{bmatrix},
	\end{align}
while the calculation of $\bv_-$ requires more efforts. We will obtain the answer for $\bv_-$ with the use of Bogoliubov coefficients relating two sets of different Neumann basis functions (\ref{eq:bogol_coef}) by treating $\bv_-$ as the negative frequency basis function complex conjugated to its positive frequency counterpart $\bv_-^*$ satisfying at $t=0$ the boundary condition $(i \bW  - \Bom^*)\,\bv_-^*|_{t=0}=0$. Thus, in accordance with (\ref{eq:bogol_coef}) the answer for $\bv_-$ reads
	\begin{equation} \label{eq:bv_bog_tr}
		\bv_- = \bv^* \boldsymbol{U}^T - \bv \boldsymbol{V}^T.
	\end{equation}
where $\boldsymbol{U}$, $\boldsymbol{V}$ are the corresponding Bogoliubov coefficients
    \begin{subequations}	
	\begin{align} \label{eq:bv_bog_coef}
		&\boldsymbol{U} = \frac1{\sqrt{2\Bom_\re}} (\Bom + \bom) \frac1{\sqrt{2\bom_\re}}, \\
		&\boldsymbol{V} = \frac1{\sqrt{2\Bom_\re}} (\Bom - \bom^*) \frac1{\sqrt{2\bom_\re}}.
	\end{align}
    \end{subequations}
Here we assume the normalization $\bv(0) = 1/\sqrt{2\bom_\re}$, $\bv_-(0) = 1/\sqrt{2\Bom_\re}$, and denote the real parts of $\bom$ and~$\Bom$ respectively as $\bom_\re$ and~$\Bom_\re$.

	Finally, let us consider the details of the Green's function $\bG(t, t')$ defined by (\ref{eq:Green_full}) for a particular form of $\bv_\pm$ we have just built. Its matrix $\bDel_{-+}$ given by (\ref{eq:Delta_Green_full}) reads
	\begin{align} \label{eq:Delta_Green_full_expl}
		i \bDel_{-+} =
		\frac{1}{\sqrt{2\Bom_\re}} \Bigl[  (\bid - \bX) \, \bom    + \Bom \, (\bid + \bX)\Bigr] \frac{1}{\sqrt{2{\bom_\re}}}.
	\end{align}
Next, let us consider separately the first term in (\ref{eq:Green_full}). After the calculation presented in Appendix~\ref{B} one obtains for it the following form
	\begin{equation}
		\bv_+(t) \, (i\bDel_{-+})^{-1} \, \bv^T_-(t')  = \bv_+(t) \, \bv^\dagger(t')  + \bv_+(t) \, \bnu\, \bv_+^T(t'), \label{Green_1st_part}
	\end{equation}
where we introduce the following symmetric matrix
	\begin{equation} \label{nu_bar_def}
			\bnu= \Bigl[\bid\! +\! \bX\!
            -\! \sqrt{2\bom_\re} \,\bX \,
        (\bom\! +\!\Bom)^{-1} \bX \sqrt{2\bom_\re}\Bigr]^{-1}\!\! - \bX.
	\end{equation}

Recalling that the second term of the expression (\ref{eq:Green_full}) can be obtained from the first one by the simultaneous transposition and exchange of time arguments, and observing that the second term in (\ref{Green_1st_part}) is symmetric under this transformation, we find that two theta functions sum up to identity, so that the final expression for the Green's function reads
	\begin{equation}
		i\bG(t,t') = i\bG_0(t, t') +
		\bv_+(t) \, \bbnu \, \bv_+^T(t'), \label{lor_green}
	\end{equation}
	where $\bG_0$ is defined as
	\begin{align}
		i\bG_0(t,t')&= \bv_+(t) \, \bv^\dagger(t') \, \theta(t-t') +
		\bv^*(t) \, \bv_+^T(t') \, \theta(t'-t) \nonumber\\
		&= \bv(t) \, \bv^\dagger(t') \, \theta(t-t') +
		\bv^*(t) \, \bv^T(t') \, \theta(t'-t)\nonumber\\
        &\quad+ \bv^* (t) \, \bX \, \bv^\dagger(t'),
	\end{align}
and interpreted as the Green function, corresponding to the vacuum state, having the density matrix $\hat{\rho}_0 = |0\ra\la0|$, associated with the basis functions $v(t)$,~$v^*(t)$. Indeed, from (\ref{eq:vac_coord_repr}) one observes that the matrix $\Bom$, defining the vacuum density matrix $\hat{\rho}_0$, coincides with $\bom^*$, i.e. $\Bom = \bom^*$. In this case, $\bnu$ vanishes due to its definition (\ref{nu_bar_def}), so from (\ref{lor_green}) we find that~$\bG = \bG_0$.
	
	Substituting the generating functional obtained to (\ref{eq:corr_fun_def}), we observe that for vanishing $\bj$ the block-matrix components of $\bG$ are composed of the Feynman, anti-Feynman and Wightman Green's functions (\ref{eq:gf_comp}), namely
	\begin{equation}
		\bG(t, t') = \begin{bmatrix*}[l]
			G_{\mathrm{T}}(t, t') & G_{<}(t, t') \\
			G_{>}(t, t') & G_{\bar{\mathrm{T}}}(t, t')
		\end{bmatrix*},
	\end{equation}
	where $G_{>}(t, t') \equiv G_{<}^T(t', t)$, and the explicit form of the block components can be read off form (\ref{lor_green}).

Now we have to find the solution $\bphi(t)$ of the boundary value problem (\ref{eq:bulk_eq_lor})--(\ref{eq:bdy_eq_lor_2}) in order to substitute it into the exponential of (\ref{gen_fun_pm0}). The only inhomogeneous boundary conditions in this problem are the Neumann conditions (\ref{eq:bdy_eq_lor_1}), so that the solution is given by the double-field version of (\ref{eq:phi_sol_neum_mod}) with the substitutions $j_+ \mapsto 0$ (remember that there is no $j_+$ at the point $t=T$) and $j_- \mapsto-\bj$. Thus it reads
	\begin{equation}
		\bphi(t) =  i\bG(t,0) \, \bj - \int_0^T dt'\, \bG(t,t') \bJ(t').
	\end{equation}
Substituting it to the exponential of (\ref{gen_fun_pm0}) then gives Eq.~(\ref{gen_fun_dir0}) advocated in Section~\ref{sec:summary}.
	\begin{align}  \label{gen_fun_dir}
		&Z[\bJ] = \text{const} \times
    \exp \Biggl\{ -\frac{i}2 \int dt\,dt' \, \bJ^T\!(t) \bG(t,t') \bJ(t) \nonumber\\
    &\qquad\quad- \int dt \,\bJ^T\!(t) \bG(t,0) \, \bj + \frac{i}2\bj^T \bG(0,0) \, \bj \Biggr\},
	\end{align}
where all time integrations run from $t=0$ to $t=T$. Here the restriction of $\bG(t,t')$ to $\bG(t,0)$ does not lead to essential simplification whereas $\bG(0,0)$ has, as shown in Appendix~\ref{sec:app_G00_der}, the following explicit and simple form in terms of the parameters of the density matrix
	\begin{align} \label{G00}
		i\bG(0,0) = \frac{\bid + \bX}{2 \Bom_\re},
	\end{align}
where the ``ratio'' of matrices $\bid+ \bX$ and $\Bom_\re$ is unambiguous because these matrices are commuting in view of the special form of $\Bom$ subject to the relation $\bX\Bom\bX=\Bom^*$.
	

\subsection{Keldysh rotation}
For further convenience it is useful to perform the change of the basis in the doubled field space $\phi_+$,~$\phi_-$ and introduce the so-called classical and quantum fields $\phi_c$ and $\phi_q$ \cite{Leonidov-Radovskaya,Radovskaya-Semenov},
	\begin{equation}
		\hspace{-0.3cm}\bphi_K(t)=\begin{bmatrix}
			\;\phi_c(t)\, \\ \phi_q(t)
		\end{bmatrix} =\boldsymbol{C}\bphi(t),\;
\boldsymbol{C} \equiv \begin{bmatrix*}[c]
			\;\frac12 I && \, \frac 12 I\; \\
			I && - I
		\end{bmatrix*}.
	\end{equation}
This transformation is called Keldysh rotation. In the new basis, the Green's function $\bG$ takes the form
	\begin{equation} \label{eq:gf_keld_rot}
		\hspace{-0.2cm}\bG_K(t, t')\!=\!
    \boldsymbol{C} \bG(t, t')\, \boldsymbol{C}^T\!\!
    =\! \begin{bmatrix}
			\,G_K(t, t') & G_R(t, t')\, \\
			\,G_A(t, t') & 0
		\end{bmatrix}\!.
	\end{equation}
Here $G_R$ and $G_A$ are the retarded and advanced Green's functions, respectively, having the following operator form
	\begin{align}
		G_R(t, t') &= -i\tr \Bigl( \hat \rho \, \bigl[\hat\phi(t), \hat\phi(t')\bigr] \Bigr) \, \theta(t-t')\nonumber\\
 &= -i\bigl[v(t) v^\dagger(t') - v^*(t) v^T(t')\bigr] \, \theta(t-t'),\\
		 G_A(t, t') &= G_R^T(t',t).
	\end{align}
They are consistent with the classical definition (\ref{eq:gf_ret_def}), in particular, because of independence of the commutator average of the state $\hat{\rho}$. The block $G_K$ is called Keldysh Green's function and contains the information about the state. In view of operator averages (\ref{eq:gf_comp}) it expresses as the mean value of the anti-commutator of fields and, due to (\ref{eq:gf_keld_rot}), explicitly reads in terms basis functions as
	\begin{align} \label{eq:gf_keld_def}
		iG_K&(t, t') = \tfrac12 \tr \Bigl( \hat{\rho} \,
        \bigl\{\hat\phi(t), \hat\phi(t')\bigr\}\Bigr)\nonumber\\
        &=
		\begin{bmatrix}
			\;v(t) && v^*(t)\,
		\end{bmatrix}
		\bigl(\bnu + \tfrac12\bX\bigr) \begin{bmatrix}
			\;v^T(t')\, \\
        v^\dagger(t')
		\end{bmatrix}.
	\end{align}
	
\subsection{Special choice of basis functions and particle interpretation}\label{subs:omega_choice}
	
	Thus far, the matrix $\omega$, which defines the Neumann boundary conditions for the basis functions $v$,~$v^*$, is not fixed except the requirement of symmetry under transposition and positive definiteness of its real part. In this section we make a convenient choice of $\omega$ which leads to the expressions for the Green's functions admitting particle interpretation with well-defined notion of average occupation number.
	
For this purpose, it is useful to rewrite the Keldysh Green's function in terms of non-anomalous and anomalous particle averages
	\begin{align}
		\nu = \tr\bigl[\hat \rho \, \hat{a}^\dagger \hat{a}\bigr], \qquad
		\kappa = \tr\bigl[\hat \rho \, \hat{a} \, \hat{a}\bigr],
	\end{align}
so that from (\ref{eq:gf_keld_def}) $G_K$ becomes
	\begin{align}
		i\,G_K(t, t')& = \begin{bmatrix}
			\;v(t) && v^*(t)\;
		\end{bmatrix}\nonumber\\
		&\times\begin{bmatrix}
			\;\kappa & \nu^* + \frac12 I\;\\
			\;\nu + \frac12 I & \kappa^*\;
		\end{bmatrix}
		\begin{bmatrix}
			\;v^T(t')\; \\ v^\dagger(t')
		\end{bmatrix}.
	\end{align}
	Note that the matrix $\kappa$ is symmetric, whereas $\nu$ is Hermitian.
	Comparing with (\ref{eq:gf_keld_def}) we find the connection between particle averages and the matrix $\bnu$
	\begin{equation}
		\bnu = \begin{bmatrix}
			\;\kappa && \nu^*\\
			\;\nu && \kappa^*
		\end{bmatrix}.
	\end{equation}
Thus, we see that block-diagonal components of $\bnu$ are responsible for anomalous averages. To ascribe the particle interpretation to the creation/annihilation operators, we will try to choose the matrix $\omega$, defining the corresponding basis functions $v(t)$ and $v^*(t)$ so that the diagonal blocks of $\bnu$, defining the anomalous averages $\kappa$, vanish. Moreover, this choice will simplify the expressions for the Green's functions, since they contain the terms, containing $\kappa$. For example, with a nonzero $\kappa$ the Wightman function reads
\begin{multline} \label{eq:gf_wightman_kappa}
	G_>(t,t') = v(t) \, (\nu^* + I) \, v^\dagger(t') + v^*(t) \, \nu \, v^T(t') \\ + v(t) \, \kappa \, v^T(t') + v^*(t) \, \kappa^* \, v^\dagger(t').
\end{multline}
	
To make the matrix $\bbnu$ block off-diagonal consider the expression (\ref{nu_bar_def}) and note that the only block diagonal contribution is contained in the identity matrix $\bid$ and, possibly, in the term involving $(\bom+\Bom)^{-1}$. Thus, we want to choose $\omega$ such that block-diagonal contribution of the latter exactly cancels those of $\bid$. Using the block matrix inversion formula\footnote{The useful form of the block matrix inversion formula is
			\begin{align*}
					\begin{bmatrix}
							\; A \; & B\; \\ \;C\; & D\;
						\end{bmatrix}^{-1} =&
					\begin{bmatrix}
							\;(A-B D^{-1}C)^{-1} & 0 \\
							0 & (D-C A^{-1} B)^{-1}
						\end{bmatrix}\nonumber\\
					&\times\begin{bmatrix}
							I & - B D^{-1} \\
							- C A^{-1} & I
						\end{bmatrix}
				\end{align*}
with $A=R+\omega$, $B=S$, $C=S^*$, $D=R^*+\omega^*$.} we have the condition of the vanishing block-diagonal part of $\bnu$,
	\begin{equation} \label{eq:omega_eq}
		R + \omega - S (R^* + \omega^*)^{-1} S^* -2 \omega_\re=0.
	\end{equation}
We will focus on the case in which $R$ and $S$ are real. The formalism described below can be easily extended to the complex $R$, but it seems that there is no straightforward extension to general complex (Hermitian) $S$. Introducing the dimensionless quantities
	\begin{equation}
		r = \omega^{-1/2} \, R \, \omega^{-1/2}, \qquad
		s = \omega^{-1/2} \, S \, \omega^{-1/2}, \label{dimless_q}
	\end{equation}
the equation (\ref{eq:omega_eq}) can be rewritten as $r + I - s (r + I)^{-1} s = 2$ and further simplified by introducing the new variable $\tilde s = (r+I)^{-1/2}s\,(r+I)^{-1/2}$ and solving for $\tilde s$, so that it takes the following form
	\begin{equation}
		r^2 = s^2+I.
	\end{equation}
This is implicit equation on $\omega$, due to the above definition of $r$ and $s$. Its explicit form reads
	\begin{equation}
		R \omega^{-1} R = S \omega^{-1} S + \omega,
	\end{equation}
which can be solved in the form advocated in Section~\ref{sec:summary}
	\begin{equation}
		\omega=R^{1 / 2} \sqrt{I-\sigma^{2}} R^{1 / 2}, \quad \sigma \equiv R^{-1 / 2} S R^{-1 / 2}. \label{omega_bod}
	\end{equation}

Note that the assumption of positive definiteness of $\omega$ implies that $I-\sigma^2=(I-\sigma)(I+\sigma)$ is positive definite. Recalling that $R+S = R^{1/2}(I+\sigma)R^{1/2}$ should be positive definite for normalizability of the density matrix, it is easy to see that $I-\sigma = R^{-1/2}(R-S)R^{-1/2}$, or equivalently $R-S$ should be positive definite too. Then, the substitution of the obtained expression for $\omega$ to (\ref{nu_bar_def}) gives the desired block-diagonal matrix form of $\bbnu$ advocated in Section~\ref{sec:summary}
	\begin{align}
		&\bbnu = \begin{bmatrix} \;0 && \nu\; \\ \;\nu && 0\; \end{bmatrix},
    \quad \nu \equiv \frac12 \varkappa \left( \sqrt{\frac{I-\sigma}{I+\sigma}} - I\right) \varkappa^T,\label{nu_bar_bod}\\
    &\varkappa
    \equiv \big[\omega^{1/2} R^{-1}
    \omega^{1/2}\big]^{1/2} \omega^{-1/2} R^{1/2}= \bigl(\varkappa^T\bigr)^{-1}
	\end{align}
where the matrix $\varkappa$ introduced above is orthogonal. Therefore, as a consequence of positive definiteness of $I+\sigma$ and $I-\sigma$, the matrix $\nu$ is necessarily real. As shown in Appendix~\ref{sec:dm_def}, for the density matrix to be positive definite, the matrix $\sigma$ should be negative definite, so $\nu$ is positive definite.

Substituting it to (\ref{lor_green}), one immediately obtains simple expressions for the Green's functions. In particular, for Wightmann and Feynman functions one has
	\begin{align}
	iG_{\mathrm{T}}(t,t')=&
    v(t) \,  v^\dagger(t') \, \theta(t-t') + v^*(t) \, v^T(t') \, \theta(t'-t) \nonumber\\
    &+ \, v(t) \, \nu \, v^\dagger(t') + v^*(t) \, \nu \, v^T(t'),\\
	iG_>(t, t')=& v(t) \, \bigl(\nu + I\bigr) \, v^\dagger(t')
    + v^*(t) \, \nu \, v^T(t'), \label{lor_green_wightmann}
	\end{align}
while the others can be expressed through them in a straightforward way.

It will be useful to express $\bOm$ in terms of $\nu$. Disentangling $\bOm$ from (\ref{nu_bar_def}), and then using the explicit form (\ref{nu_bar_bod}) of $\bnu$, corresponding to the special choice of Neumann basis functions with (\ref{omega_bod}), we obtain the following expression
	\begin{equation} \label{eq:bom_nu}
		\bOm = \omega^{1/2} \begin{bmatrix}
			\;\frac{2\nu^2+2\nu+I}{2\nu+I} & \;-\frac{2\nu(\nu+I)}{2\nu+I} & \\
			\;-\frac{2\nu(\nu+I)}{2\nu+I} & \;\frac{2\nu^2+2\nu+I}{2\nu+I}
		\end{bmatrix} \omega^{1/2}.
	\end{equation}

\subsection{Euclidean density matrix state} \label{subs:eucl_dm}

Now, let us focus on the particular Gaussian state, which is obtained from the Euclidean path integral, namely
	\begin{align} \label{eq:density_m_eucl}
		\rho_\eu(\varphi_+,\varphi_-; J_\eu] =
    \frac1Z\!\!\!& \int\limits_{\phi(\tau_\pm) = \varphi_\pm}
    D\phi \, \exp \biggl\{- S_\eu[\phi]\nonumber\\
    &- \int_0^\beta d\tau \, J_\eu(\tau) \phi(\tau) \biggr\},
	\end{align}
Here $S_\eu$ is the quadratic action of Euclidean field theory within time limits $\tau_\pm$ which we will chose to be $\tau_+=\beta$ and $\tau_-=0$,
	\begin{align}
		&S_\eu[\phi] = \frac12 \int_{\tau_-}^{\tau_+} d\tau \,
        \phi^T \overboth{F}_\eu\, \phi\nonumber\\
        &\qquad\quad= \frac12 \int_{\tau_-}^{\tau_+} d\tau \,\phi^T F_\eu \, \phi + \frac12 \phi^T W_\eu \phi\,\Bigr|_{\tau_-}^{\,\tau_+}, \label{eq:action_eucl} \\
        &\begin{aligned}
		&F_\eu \equiv -\frac{d}{d\tau} A_\eu \frac{d}{d\tau}
        - \frac{d}{d\tau} B_\eu + B_\eu^T \frac{d}{d\tau} + C_\eu, \\
        &W_\eu  \equiv A_\eu \frac{d}{d\tau} + B_\eu.
        \end{aligned}
	\end{align}
The partition function $Z$ in the normalization factor is such that $\tr \rho_\eu = 1$ for vanishing source $J_\eu=0$. Hermiticity of the density matrix implies the following (sufficient) condition on the  coefficient matrices $A_\eu$,~$B_\eu$, and $C_\eu$ as the functions of $\tau$
	\begin{align} \label{eq:eucl_action_herm_cond}
    \begin{aligned}
	&A_\eu(\beta-\tau)=A_\eu^*(\tau),
    \quad B_\eu(\beta-\tau)=-B_\eu^*(\tau),\\
    &C_\eu(\beta-\tau)=C_\eu^*(\tau),
    \end{aligned}
	\end{align}
which are not necessary real. Nevertheless, we restrict ourselves to the real case below. The source $J_\eu$ is included in the path integral to be able to introduce nonlinear terms of the Euclidean action, leading to the non-Gaussinities of the resulting density matrix.
		
We take the path integral (\ref{eq:density_m_eucl}) over the Euclidean fields $\phi$ by using the saddle point method. The boundary conditions of the integral fix the endpoints $\phi(\beta) = \varphi_+$, $\phi(0) = \varphi_-$, so we have the boundary problem with the Dirichlet boundary conditions
	\begin{subequations} \label{eq:eucl_saddle}
	\begin{align}
		&F_\eu \phi + J_\eu = 0, \\
		&\phi(\tau_\pm) = \varphi_\pm,
	\end{align}
	\end{subequations}
Using the Dirichlet Green's function $G_D$ for vanishing boundary conditions
	\begin{equation}
		F_\eu  G_D(\tau, \tau') = \delta(\tau-\tau'), \quad G_D(\tau_\pm, \tau') = 0,
	\end{equation}
and substituting to  (\ref{eq:phi_sol_dir_mod}), one expresses the solution of (\ref{eq:eucl_saddle}) as follows
	\begin{equation}
		\phi(\tau) = 
		- \bw_\eu^T (\tau)\, \bvphi - \int_0^\beta d\tau' \, G_D(\tau,\tau') J_\eu(\tau'),
	\end{equation}
where we introduce the notations, similarly to those of the Lorentzian context (\ref{WGrow})--(\ref{GWcolumn}), for the row $\bw_\eu^T(\tau)$ obtained by the transposition of the column $\bw_\eu(\tau)$ in
	\begin{align}
		\bw_\eu^T(\tau)=&
        \bigl[\bw_\eu(\tau)\bigr]^T=
    \begin{bmatrix*}[r]
			W_\eu  G_D(\beta, \tau)\phantom{'}  \\
        -W_\eu  G_D(0,\tau') 
		\end{bmatrix*}^T\nonumber\\
		=&
		\begin{bmatrix}
		\;G_D(\tau, \beta)
        \overleft{W}_{\!\eu}  & -G_D(\tau, 0) \overleft{W}_{\!\eu} \, \label{GW}
		\end{bmatrix}
	\end{align}
(the last equality implies the symmetry of the Dirichlet Green's function, $G_D^T(\tau,\tau')=G_D(\tau',\tau)$). Substitution back to (\ref{eq:density_m_eucl}) gives
	\begin{align} \label{eq:dm_eucl}
	\rho_\eu(\varphi_+,&\varphi_-;J_\eu] ={\rm const}\times
    \exp \biggl\{ -\frac12 \bvphi^T \Bom \, \bvphi + \bj^T \bvphi\nonumber\\ &+ \frac12 \int d\tau\,d\tau' J_\eu(\tau) G_D(\tau,\tau') J_\eu(\tau')
		\biggr\},
	\end{align}
where we disregard the source independent prefactor, all $\tau$-integrations run from $0$ to $\beta$, whereas the matrix $\Bom$ and the source $\bj$, introduced in (\ref{eq:dm_coord}) take the following particular form
	\begin{align}\label{eq:dm_equcl_pars}
		\hspace{-0.15cm}\Bom &\equiv
		\begin{bmatrix*}[r]
		-\overright{W}_{\!\eu}  G_D(\beta,\beta) \overleft{W}_{\!\eu} 
        & \overright{W}_{\!\eu}  G_D(\beta,0) \overleft{W}_{\!\eu} \,\\ 
		\overright{W}_{\!\eu}  G_D(0,\beta)
    \overleft{W}_{\!\eu}  & -\overright{W}_{\!\eu}  G_D(0,0) \overleft{W}_{\!\eu} \,
	\end{bmatrix*},\\
		\hspace{-0.1cm}\bj^T  &= \int_0^\beta d\tau \, J_\eu(\tau) \,
        \bw_\eu^T(\tau).
		\end{align}
Now, one can substitute the density matrix (\ref{eq:dm_eucl}), defined by the parameters (\ref{eq:dm_equcl_pars}) to the general expression for the generating functional (\ref{gen_fun_dir}). This leads to
	\begin{align} \label{Zfinal1}
	Z[\bJ, J_\eu]&= {\rm const}\times
    \exp \Biggl\{ -\frac{i}2 \int dt\,dt' \, \bJ^T\!(t) \bG(t,t') \bJ(t')\nonumber\\
	&-\int dt \, d\tau \,\bJ^T\!(t) \bG(t,0)
    \,\overright{W}_{\!\eu}  \bg_D(\tau) \, J_\eu(\tau) \nonumber\\
    &+ \frac12\int d\tau\,d\tau' \, J_\eu(\tau)\,G_\eu (\tau,\tau')\, J_\eu(\tau') \,\Biggr\}.
	\end{align}
Note that the kernel of the third integral here is the periodic Euclidean Green's function
    \begin{align} \label{G_\eu 1}
    G_\eu (\tau,\tau')=G_D(\tau,\tau') + i\,\bw_\eu^T(\tau)\,\bG(0,0)\, \bw_\eu (\tau')
    \end{align}
corresponding to the fact that with the Lorentzian sources switched off the functional $Z[0, J_\eu]$ represents the Euclidean path integral over periodic fields $\phi(\tau)$ on the time interval with the identified boundary points $\tau_\pm$. The expression for this Green's function seemingly dependent via $\bG(0,0)$ on Lorentzian objects is in fact independent of them. This property is based on the relation (\ref{G00}) and derived in Appendix~\ref{sec:app_G00_der}.

\subsection{Analytic continuation and KMS condition}\label{subs:analytic_kms}

The further transformation of the generating functional, which allows one to reveal its new analyticity properties, can be performed due to two assumptions. The first assumption is that the Euclidean action (\ref{eq:action_eucl}) is obtained by analytic continuation of the Lorentzian one (\ref{eq:action_lor}), namely
	\begin{equation} \label{eq:action_wick_rot}
		i S[\phi]\bigr|_{t=-i\tau} = -S_\eu[\phi]
	\end{equation}
This implies the following form of the Euclidean action coefficient functions
	\begin{align}
		&A_\eu(\tau) = A(-i\tau), \quad
		B_\eu(\tau) = -i B(-i\tau), \nonumber\\
		&C_\eu(\tau) = -C(-i\tau).
	\end{align}
Though this requirement sounds rather restrictive, it can be based on the assumptions discussed in Introduction about the properties of the Euclidean background underlying the quadratic action and sandwiched between the two (identified) turning points at which the analytic match between the Euclidean and Lorentzian branches can be done. Another assumption which we use in what follows is the possibility to make a special choice of the Neumann basis functions, derived above.
	
The first step is to rewrite the second and the third terms in the exponential of the generating functional (\ref{Zfinal1}) in terms of the Euclidean Neumann Green's function $G_N(\tau,\tau')$ instead of the Dirichlet one, i.e.\ $(W_\eu + \omega)G_N(\beta, \tau') = (W_\eu - \omega^*)G_N(0, \tau') = 0$ where $\omega$ is the same as in (\ref{eq:bv_eq})--(\ref{bom_diag}). This is done using the relations (\ref{green_dir_neum_1})--(\ref{green_dir_neum_2}) (after the replacement $\bom \mapsto -i \bom$ associated with the transition to the Euclidean version of Dirichlet and Neumann Green's functions) and the derivation in Appendix~\ref{sec:app_G00_der}. The result reads as the expression (\ref{Zfinal1}) with the kernel of the Lorentzian-Euclidean term $-\bG(t,0) \, \bw_\eu(\tau)$ replaced by $\bG(t,0)\,(\bom + \Bom)\, \bg_N(\tau)$ and the new form of the periodic Green's function $G_\eu (\tau,\tau')$ in the Euclidean-Euclidean block
	\begin{align}\label{gen_fun_neum}
		G_\eu (\tau,\tau')&=G_N(\tau, \tau')\nonumber\\
         &+ \bg_N^T(\tau)
        \sqrt{2\bom_\re} (\bnu^* + \bX ) \sqrt{2\bom_\re} \bg_N(\tau'),
	\end{align}
where $\bg_N(\tau)$ is the Euclidean version of the definition (\ref{GNrow}) for the Neumann Green's function.

To proceed further we have to derive several important properties of Euclidean Neumann Green's function which is the part of (\ref{gen_fun_neum}), specific to the choice (\ref{omega_bod}) of $\omega$. In terms of the Euclidean basis functions it reads as
	\begin{align}
		G_N(\tau, \tau') = &- u_+(\tau) (\Delta^N_{-+})^{-1} u_-(\tau')
        \, \theta(\tau-\tau')\nonumber\\
        &+ u_-(\tau) (\Delta^N_{+-})^{-1} u_+(\tau') \, \theta(\tau'-\tau),
	\end{align}	
Here $u_+$, $u_-$ are the basis functions obeying Neumann boundary conditions
	\begin{equation} \label{green_eucl_neum_bdy}
		(W_\eu + \omega)u_+ |_{\tau=\beta} = 0, \quad
		(W_\eu - \omega)u_- |_{\tau=0} = 0
	\end{equation}
and, as usual,
	\begin{equation} \label{deltas}
	\Delta_{+-}^N\!=\! u_+^T W_\eu u_-\! -\! (W_\eu u_+)^T u_-,
    \; \Delta_{-+}^N\!=\! - (\Delta_{+-}^N)^T.
	\end{equation}
Note that the boundary conditions on $u_\pm$ above are exactly the analytic continuation $t\mapsto -i\tau$ of the boundary conditions (\ref{eq:green_lor_neum_bdy}) on $v$, $v^*$.
	
Now, consider in detail the matrix of boundary values of the Euclidean Neumann Green's function at $\tau_+=\beta$ and $\tau_-=0$
	\begin{equation}
		\bG_N\|\!
		=\! \begin{bmatrix}
			\,u_-(\beta) (\Delta_{+-}^N)^{-1} u_+^T(\beta) & -u_+(\beta) (\Delta_{-+}^N)^{-1} u_-^T(0)\, \\
			\,u_-(0) (\Delta_{+-}^N)^{-1} u_+^T(\beta) & -u_+(0) (\Delta_{-+}^N)^{-1} u_-^T(0)\,
		\end{bmatrix}
	\end{equation}
(double vertical bar denotes here the restruction of the two Green's function arguments to two boundary surfaces thus forming the $2\times 2$ block matrix). Using the Euclidean version of the relation (\ref{green_dir_neum_2}), we find the alternative form of this matrix
	\begin{equation}
		\bG_N\| = (\bom + \Bom)^{-1} = \frac1{\sqrt{2\omega}}
		\begin{bmatrix}
			I & \frac{\nu}{\nu+I}\, \\
			\,\frac{\nu}{\nu+I} & I
		\end{bmatrix}
		\frac1{\sqrt{2\omega}}, \label{green_neum_bdy_bdy}
	\end{equation}
where we use the explicit form (\ref{nu_bar_bod}) of $\bnu$, corresponding to the particular choice of basis functions described in the previous subsection.
	
Equating these two expressions for $\bG_N \|$, with due regard to the structure of $\Delta^N_{+-}$ in (\ref{deltas}), we find the two sets of equalities. The first set follows from the diagonal blocks
	\begin{equation} \label{eq:basis_fun_bdy_add}
		(W_\eu + \omega)u_+ |_{\tau=0} = 0, \qquad
		(W_\eu - \omega)u_- |_{\tau=\beta} = 0,
	\end{equation}
	and means that the basis functions $u_+$,~$u_-$ obey the same Neumann boundary conditions at both boundary values of the Euclidean time (cf. Eq.~(\ref{green_eucl_neum_bdy})). This also implies the following explicit form of the matrices $\Delta_{+-}^N$,~$\Delta_{-+}^N$
	\begin{equation}
		\Delta_{+-}^N = -(\Delta_{-+}^N)^T = 2 u_+^T \, \omega \, u_-,
	\end{equation}
where the basis functions $u_+$,~$u_-$ are evaluated either at $\tau=0$ or $\tau=\beta$. Similarly, from the off-diagonal blocks of (\ref{green_neum_bdy_bdy}), one gets the formulas, relating the boundary values of the basis functions
	\begin{equation} \label{basis_fun_periodicity}
    \begin{aligned}
	&u_-(\beta) = \frac1{\sqrt{2\omega}} \frac{\nu+I}{\nu}\sqrt{2\omega}
    \, u_-(0), \\
	&u_+(\beta) = \frac1{\sqrt{2\omega}} \frac{\nu}{\nu+I}\sqrt{2\omega}
    \, u_+(0).
    \end{aligned}
	\end{equation}
		
It is useful to continue the Euclidean equations of motion
beyond the interval $0 < \tau < \beta $ with the period $\beta$ (which is again possible because $\tau=0$ and $\tau=\beta)$ are the turning points),
	\begin{equation}
    \begin{aligned}
		&A_\eu(\tau+\beta) = A_\eu(\tau), \quad B_\eu(\tau+\beta) = B_\eu(\tau),\\
        &C_\eu(\tau+\beta) = C_\eu(\tau).
        \end{aligned}
	\end{equation}
Together with (\ref{eq:eucl_action_herm_cond}) it also implies
	\begin{equation} \label{eq:coef_fun_eucl_refl}
    \begin{aligned}
		&A_\eu(\tau) = A_\eu(-\tau), \quad B_\eu(\tau) = -B_\eu(-\tau),\\
        &C_\eu(\tau) = C_\eu(-\tau).
        \end{aligned}
	\end{equation}
	
Once the functions $u_\pm(\tau)$ satisfy the same homogeneous boundary conditions for both $\tau = 0$ and $\tau=\beta$ (cf.\ (\ref{green_eucl_neum_bdy}) and (\ref{eq:basis_fun_bdy_add})), being translated by the period they can only differ by the multiplication with some non-singular matrices $L_\pm$, $u_\pm(\tau+\beta) = u_\pm(\tau) L_\pm$. From (\ref{basis_fun_periodicity}) we obtain their explicit form
	\begin{align}
    \begin{aligned}
		&\hspace{-0.3cm}u_-(\tau+\beta)= u_-(\tau) \, u_-^{-1}(0) \frac1{\sqrt{2\omega}} \frac{\nu+I}{\nu}\sqrt{2\omega} \, u_-(0), \\
		&\hspace{-0.3cm}u_+(\tau+\beta)= u_+(\tau) \, u_+^{-1}(0) \frac1{\sqrt{2\omega}} \frac{\nu}{\nu+I}\sqrt{2\omega} \, u_+(0).
        \end{aligned}
	\end{align}
With the normalization
	\begin{equation} \label{eq:eucl_bf_norm}
		u_+(0) = u_-(0) = \frac1{\sqrt{2\omega}},
	\end{equation}
this monodromy simplifies to
	\begin{align}
    \begin{aligned}
		u_-(\tau+\beta) = u_-(\tau) \, \frac{\nu+I}{\nu},\\
		u_+(\tau+\beta) = u_+(\tau) \, \frac{\nu}{\nu+I}.
    \end{aligned}
	\end{align}

Similarly, in view of the reflection symmetry (\ref{eq:coef_fun_eucl_refl}) of the operator $F_\eu$ the functions $u_+(\tau)$ and $u_-(-\tau)$ can differ at most by some non-degenerate matrix $L$, $u_+(\tau) = u_-(-\tau) \, L$. For the normalization (\ref{eq:eucl_bf_norm}) this implies
	\begin{equation}
		u_+(\tau) = u_-(-\tau).
	\end{equation}
	
For the choice (\ref{eq:eucl_bf_norm}) we have $\Delta_{+-}^N = -\Delta_{-+}^N = I$, so that the blocks of the Euclidean and Lorentzian-Euclidean Green's function in (\ref{gen_fun_neum}) read
	\begin{align}
		&\begin{aligned}
			G_\eu(\tau,\tau') = &\, u_+(\tau) \,  u_-^T(\tau')
        \,\theta(\tau-\tau')\\
        &+ u_-(\tau) \, u_+^T(\tau') \, \theta(\tau'-\tau) \\
        &+ \, u_+(\tau) \, \nu \, u_-^T(\tau') + u_-(\tau) \, \nu \, u_+^T(\tau'),
		\end{aligned} \\
		& \begin{bmatrix}
			G_{LE}^1(t,\tau) \\ G_{LE}^2(t,\tau)
		\end{bmatrix}
		= \bG(t,0) (\bom + \Bom) \bg_N(\tau)\nonumber\\
        & =
    \begin{bmatrix}
			&I& \\ &I&
		\end{bmatrix}
		\Bigl(v(t) \, \nu \, u_-^T(\tau) + v^*(t) \, (\nu + I)
        \, u_+^T(\tau)\Bigr).
	\end{align}
This finally leads us to the expression for the generating functional (\ref{Z_FINAL}) with the total block-matrix Green's function given by Eqs.(\ref{totalG})--(\ref{GLE1}), which was advocated in Section~\ref{sec:summary}.

If one introduces the Euclidean Wightmann Green's functions
	\begin{equation}
    \begin{aligned}
	&\hspace{-0.2cm}G_{E}^>(\tau, \tau')\! =\! u_+(\tau)(\nu + I)
    u_-^T(\tau')\! +\! u_-(\tau)\,\nu\, u_+^T(\tau'), \\
    &\hspace{-0.2cm}G_{E}^<(\tau,\tau') = \bigl[G_\eu^>(\tau', \tau)\bigl]^T,
    \end{aligned}
	\end{equation}
then $G_\eu(\tau, \tau')$ can be expressed as
	\begin{equation}
		G_\eu(\tau, \tau') = G_{E}^>(\tau, \tau') \, \theta(\tau-\tau')
        + G_{E}^<(\tau,\tau') \, \theta(\tau' - \tau),
	\end{equation}
and the Lorenzian Wighmann Green's function (\ref{lor_green_wightmann}) is an analytic continuation of $G_{E}^>(\tau)$.
	
	Now, it is time to connect the Euclidean basis functions $u_\pm$ and the Lorentzian ones $v$,~$v^*$. Specifically, let us show that both sets of functions can be obtained from a single function $V(z)$ of the complex time $z = t -i\tau$, obeying complexified equations of motion\ (\ref{eq:phi_eq})
	\begin{equation}
		\biggl[-\frac{d}{dz} A(z) \frac{d}{dz} - \frac{d}{dz} B(z)
    + B^T(z)\frac{d}{dz} + C(z) \biggr] V(z) = 0.
	\end{equation}
	This equation reduces to the Lorentzian e.o.m.\ for $z=t$ and to the Euclidean ones for $z=-i\tau$. Under the assumption that coefficient functions $A(t)$,~$B(t)$, and $C(t)$ are real, together with the reflection symmetry (\ref{eq:coef_fun_eucl_refl}), one can find that $V^*(z) \equiv (V(z^*))^*$ obeys the same equation. Moreover, the initial conditions (\ref{eq:green_lor_neum_bdy}) for $v$,~$v^*$ are connected with those  (\ref{green_eucl_neum_bdy}) for $u_\pm$ via analytic continuation $t\mapsto -i\tau$. This motivates us to impose the boundary condition on $V$ as follows
	\begin{equation} \label{eq:basis_V_in_cond}
		\bigl[i W_{\mathbb{C}} - \omega\,\bigr] V(z)\bigr|_{z=0}=0, \qquad W_{\mathbb{C}} \equiv A(z)\frac{d}{dz} + B(z)
	\end{equation}
	which reduces to those for $v$ or $u_+$ after the substitution $z=t$ or $z=-i\tau$, respectively. Supplementing the latter condition with the normalization
	\begin{equation}
		V(0) = \frac1{\sqrt{2\omega}},
	\end{equation}	
	one finds
	\begin{equation}
		v(t) = V(t), \quad u_+(\tau) = V(-i\tau),
	\end{equation}
	i.e.\ $v$ and $u_+$ are analytic continuation of each other. Similarly, complex conjugation of (\ref{eq:basis_V_in_cond}) and the same assumptions of coefficient functions reality and its reflection symmetry, we find that $V^*$ obeys the following boundary condition
	\begin{equation}
		\bigl[i W_{\mathbb{C}} + \omega\bigr]V^*(z)\bigr|_{z=0}=0,
	\end{equation}
	so that $v^*$ and $u_+$ can be obtained from $V^*$ as
	\begin{equation}
		v^*(t) = V^*(t), \quad u_-(\tau) = V^*(-i\tau).
	\end{equation}
	
	Thus, assuming that the complexified basis function $V(z)$, $z=t-i\tau$ is analytic on $0 \le t \le T$,~$0 \le \tau < \beta$, we have the following transformation law of the basis functions
	\begin{equation}
		v(t-i\beta) = v(t) \, \frac{\nu}{\nu + I}, \;
		v^*(t-i\beta) = v^*(t)\, \frac{\nu+I}{\nu}.
	\end{equation}
	Substituting to (\ref{lor_green_wightmann}), one has the following condition on Wightmann Green's function
	\begin{equation}
		G_>(t-i\beta, t') = G_<(t, t'),
	\end{equation}
	which is nothing but KMS condition advocated in Section~\ref{subs:analytic_kms}.

\section{Simple applications\label{Examples}}
	\subsection{Harmonic oscillator}
	In this section we consider harmonic oscillator as the simplest instructive example, which demonstrates the main concepts and quantities, introduced above, together with the convenience of the special choice of the basis functions $v$,~$v^*$. The corresponding action reads
	\begin{equation}
		S[\phi] = \frac12 \int dt \, \bigl(\dot \phi^2 - \omega_0^2 \phi^2\bigr),
	\end{equation}
	where $\phi$ is one-component field, defining the coordinate of oscillator, and $\omega_0$ is its frequency.

	We will consider the system in the state, defined by the Euclidean path integral (\ref{eq:density_m_eucl}), where the Euclidean action is an analytic continuation (\ref{eq:action_wick_rot}) of the Lorentzian one
	\begin{equation}
		S_\eu[\phi_\eu] = \frac12 \int d\tau \, \bigl(\dot \phi_\eu^2 + \omega_0^2 \phi_\eu^2\bigr).
	\end{equation}
Note that for $J_\eu=0$ density matrix (\ref{eq:density_m_eucl}) coincides with the thermal density matrix of the inverse temperature $\beta$. The corresponding differential operator defining the Euclidean equation of motion $F_\eu \phi_\eu=0$ and the Wronskian read
	\begin{equation}
		F_\eu = -\frac{d^2}{d\tau^2} + \omega_0^2,
        \qquad W_\eu = \frac{d}{d\tau}.
	\end{equation}
To exploit the answer (\ref{eq:dm_eucl}), one should first calculate the Dirichlet Green's function, which can be constructed out of corresponding basis functions $u_\pm^D(\tau)$ satisfying
	\begin{equation}
		F_\eu u_\pm^D(\tau) = 0, \quad u_+^D(\beta) = u_-^D(0) = 0.
	\end{equation}
	These basis function can be chosen as
	\begin{equation}
		u_+^D(\tau) = \sinh \omega_0(\tau - \beta), \quad u_-^D(\tau) = \sinh \omega_0 \tau
	\end{equation}
so that Dirichlet Green's function has the following form
	\begin{align}
	&G_D(\tau, \tau') = \frac1{\Delta_{+-}^D}
    \bigl[u_+^D(\tau)\, u_-^D(\tau') \, \theta(\tau-\tau')\nonumber\\
    &\qquad\qquad\qquad+ u_-^D(\tau) \, u_+^D(\tau') \, \theta(\tau'-\tau)\, \bigr],\\
	&\Delta_{+-}^D = -\sinh \beta\omega_0.
	\end{align}
	Substituting the Green's function obtained to (\ref{eq:dm_equcl_pars}), one finds the explicit form of the density matrix constituents
	\begin{align}
		\bOm&
        =\frac{\omega_0}{\sinh \beta\omega_0} \begin{bmatrix}
			\,\cosh \beta\omega_0  & -1 \\
			-1 & \cosh \beta\omega_0
		\end{bmatrix}                           \label{eq:osc_bom}\\
		\bj &
\if{= -\int_0^\beta d\tau \begin{bmatrix}
			u_-^D(\tau) / u_-^D(\beta) \\u_+^D(\tau) / u_+^D(0)
		\end{bmatrix} J_\eu(\tau)\nonumber\\
&}\fi
        =  \frac1{\sinh \beta\omega_0}
        \int_0^\beta d\tau \, \begin{bmatrix}
			- \sinh \omega_0 \tau \\ \,\sinh \omega_0(\tau-\beta)\,
		\end{bmatrix} J_\eu(\tau).
	\end{align}
	
	The basis functions, satisfying (\ref{eq:green_lor_neum_bdy}) are the linear combinations of $e^{\pm i\omega_0 t}$ which are the solutions of e.o.m.\, and read (cf.\ (\ref{eq:bogol_mode_inv}) and (\ref{eq:bogol_coef}))
	\begin{equation} \label{eq:v_osc}
		v(t) = \frac1{2\sqrt{2\omega}} \Bigl[\frac{\omega_0+\omega}{\omega_0} e^{-i\omega_0 t} + \frac{\omega_0-\omega}{\omega_0} e^{i\omega_0 t}\, \Bigr],
	\end{equation}
	where we assume $\omega$ to be real for the simplicity.
	
	The remaining component of the Lorenzian Green's function (\ref{lor_green}) is the matrix $\bnu$, defined in (\ref{nu_bar_def}). Substituting $\bOm$ defined in (\ref{eq:osc_bom}), one obtains
	\begin{gather} \label{bnu_osc}
    \begin{aligned}
		&\bnu =
		\begin{bmatrix}
			\,\,\kappa && \nu \,\\
            \,\,\nu && \kappa\,\,
		\end{bmatrix}, \quad
		\nu = \frac{\omega^2+\omega_0^2}{4 \omega \omega_0} \coth\frac{\beta\omega_0}{2}  - \frac12,\\
        &\kappa = \frac{\omega^2 - \omega_0^2}{4 \omega \omega_0} \coth\frac{\beta\omega_0}{2}
    \end{aligned}
	\end{gather}
	that makes the Green's function (\ref{lor_green}) rather cumbersome even for harmonic oscillator (see (\ref{eq:gf_wightman_kappa}) for Wightman function). Obviously, for the choice $\omega = \omega_0$ diagonal component $\kappa$ of $\bnu$ vanishes, that leads to significant simplifications. Let us show that this choice follows from the construction of Section~\ref{subs:omega_choice}. Extracting $R$ and $S$ from (\ref{eq:osc_bom}) as
	\begin{equation}
    \begin{aligned}
	&R =  \frac{\omega_0\cosh \beta\omega_0}{\sinh\beta\omega_0},
    \quad S = - \frac{\omega_0}{\sinh \beta\omega_0},\\
    &\sigma \equiv \frac{S}{R} = -\frac1{\cosh\beta\omega_0}.
    \end{aligned}
	\end{equation}
Substitution to (\ref{eq:omega_bod}) gives $\omega=\omega_0$ as expected. This immediately leads to vanishing $\kappa$ and
	\begin{equation}
	\bnu = \begin{bmatrix} \;0 && \nu\;\\ \;\nu && 0\; \end{bmatrix} ,
    \qquad \nu = \nu_0 \equiv \frac1{e^{\beta \omega_0} - 1}, \label{nu_bar_harm_osc}
	\end{equation}
	where one recognizes Bose-Einstein average occupation number in the expression for $\nu$ obtained. Basis function $v(t)$ takes the form of positive frequency basis function
	\begin{equation}
		v_0(t) \equiv v(t)\bigr|_{\omega=\omega_0} = \frac1{\sqrt{2\omega_0}} e^{-i\omega_0 t}.
	\end{equation} 	
	Thus, from (\ref{lor_green}) we obtain well-known expression for Wightman Green's function
	\begin{align}
		&G_>(t,t')=
		(\nu_0+1) \, v_0(t) v_0^*(t') +  \nu_0 \, v_0(t) v_0^*(t')  \nonumber\\ &\qquad=
		 \frac1{2\omega_0} \bigl( (\nu_0+1) \,e^{-i\omega_0(t-t')} + \nu_0 \, e^{i\omega_0(t-t')} \bigr),
	\end{align}
	and in terms of which the corresponding Feynman and anti-Feynman Green's functions can be expressed in a straightforward way. Note that (\ref{eq:gf_wightman_kappa}) with (\ref{bnu_osc})--(\ref{eq:v_osc}) substituted gives exactly the same answer, but in much more cumbersome form.
	
	\subsection{General one-dimensional system}
	Now, let us consider a more general case in which the field $\phi$ is one-component, i.e.\ defines a coordinate of some non-equilibrium mechanical system, and the assumptions of Section~\ref{subs:analytic_kms} are fulfilled. In this case the Euclidean basis functions defined in (\ref{green_eucl_neum_bdy}) are also one-component.
	Thus, from (\ref{eq:bf_monodromy}), we conclude that under a shift of the argument by the period the basis functions $u_\pm(\tau)$ simply acquire a numerical factor. According to Floquet theory of periodic differential equations (Euclidean equation of motion following from (\ref{eq:action_eucl}) belongs to exactly such a class of equations) this means that the basis functions $u_\pm(\tau)$ are close to the notion of Bloch functions (eigenfunctions of the translation-by-period operation). This fact motivates us to apply the Floquet theory \cite{Floquet}, which is especially powerful in one-dimensional case.
	
	In one-dimensional case the Euclidean equation of motion reads
	\begin{equation} \label{eq:eucl_eom_1d}
		\biggl[-\frac{d}{d\tau} A_\eu \frac{d}{d\tau} - \dot{B}_\eu(\tau) + C_\eu(\tau)\biggr] \phi_\eu(\tau) = 0.
	\end{equation}
	where the $A_\eu(\tau)$,~$B_\eu(\tau)$ and $C_\eu(\tau)$ become simply the functions ($1\times1$ matrices). Assuming that kinetic term is positive, i.e.\	$A_\eu(\tau) > 0$ one can define a new variable
	\begin{equation} \label{eq:y_u_canonical}
		y(\tau) = \sqrt{A_\eu(\tau)} \phi_\eu(\tau).
	\end{equation}
so that the e.o.m.\ acquires the canonical form
	\begin{equation} \label{eq:hills_eq}
		\biggl[\frac{d^2}{d\tau^2} + Q(\tau)\biggr] y(\tau) = 0,
	\end{equation}
where
	\begin{multline}
		Q(\tau) = -\frac12 \frac{d^2}{d\tau^2} \log A_\eu(\tau) - \frac14 \left(\frac{d}{d\tau} \log A_\eu(\tau)\right)^2 \\+ \frac1{A_\eu(\tau)}\left(\dot B_\eu(\tau) - C_\eu(\tau)\right)
	\end{multline}
and $Q(\tau)$ is periodic and reflection symmetric
	\begin{equation}
		Q(\tau + \beta) = Q(\tau), \qquad Q(\tau) = Q(-\tau).
	\end{equation}
	The equation (\ref{eq:hills_eq}) with periodic $Q(\tau)$ is usually referred as Hill's equation \cite{Hill}.
	
	Floquet theory guarantees that if the equation (\ref{eq:hills_eq}) has no periodic and doubly periodic solution then there exists the basis $y_\pm(\tau)$ of solutions such that
	\begin{equation}
		y_\pm(\tau+\beta) = e^{\mp \beta \varepsilon} y_\pm(\tau)
	\end{equation}
	where the parameter $\varepsilon$ is either real or imaginary, and functionally depends on $Q(\tau)$. Without loss of generality we set $\varepsilon>0$ in the real case and $\varepsilon = -iq$,~$0<q<\pi/\beta$ in the imaginary one.
	The basis function has the following important properties, depending on whether $\varepsilon$ is imaginary or real. Real $\varepsilon$ leads to real $y_\pm(\tau)$, whereas imaginary $\varepsilon$ implies $(y_\pm(\tau))^* = y_\mp(\tau)$. Reflection symmetry $Q(\tau)=Q(-\tau)$ leads to additional property $y_\pm(\tau) =  y_\mp(-\tau)$ so that $(y_\pm(\tau))^* = y_\pm(-\tau)$ for imaginary $\varepsilon$.
	
	Now, we can return to the original equation (\ref{eq:eucl_eom_1d}). Using (\ref{eq:y_u_canonical}), one can obtain the basis of its solutions $u_\pm(\tau)$ out of $y_\pm(\tau)$ as
	\begin{equation}
		u_\pm(\tau) = \frac1{\sqrt{A_\eu(\tau)}} y_\pm(\tau).
	\end{equation}
	This basis inherits the properties of $y_\pm(\tau)$ under translation by period, reflection and complex conjugation. In particular
	\begin{align}
		u_\pm(\tau+\beta) = e^{\mp\beta \varepsilon} \, u_\pm(\tau).
	\end{align}
	Comparing with (\ref{eq:bf_monodromy}) one concludes that the parameter $\varepsilon$ is connected to $\nu$ as
	\begin{equation} \label{eq:nu_varepsilon}
		\nu = \frac{1}{e^{\beta \varepsilon} - 1}.
	\end{equation}
The basis functions $u_\pm(\tau)$ have significantly different frequency properties depending on whether $\varepsilon$ is real or imaginary. Thus, real $\varepsilon$ implies
	\begin{equation} \label{eq:floq_omega}
		(W_\eu \pm \omega) u_\pm\bigr|_{\tau=0,\beta} = 0
	\end{equation}
where $\omega$ is a real number, which coincides with those defined in (\ref{eq:omega_bod}), as will be described below. In contrast, imaginary $\varepsilon$ leads to the property $(u_\pm(\tau))^* = u_\pm(-\tau)$, so that the fraction $W_\eu u_\pm(0) / u_\pm(0) = W_\eu u_\pm(\beta) / u_\pm(\beta)$ is imaginary\footnote{In deriving this property we use that $\dot{A}_\eu(0) = B_\eu(0)=0$, following from $A_\eu(\tau) = A_\eu(-\tau)$ and $B_\eu(\tau) = -B_\eu(-\tau)$.}, and one can write
	\begin{equation} \label{eq:floq_omega_prime}
		(iW_\eu \pm \omega') u_\pm\bigr|_{\tau=0,\beta} = 0,
	\end{equation}
	where the number $\omega' = i\omega$ is real.
	
	Let us calculate the density matrix (\ref{eq:density_m_eucl}) and examine its properties. To use the answer (\ref{eq:dm_eucl}), one should first construct the Dirichlet Green's function. The corresponding basis functions $u_\pm^D(\tau)$ obeying $u_-^D(0)=u_+^D(\beta)=0$ can be constructed as linear combinations of $u_\pm(\tau)$. Namely, one defines $u_-^D(\tau)$ as
	\begin{equation} \label{eq:floq_u_m}
		u_-^D(\tau) = \frac12\bigl(u_+(\tau) - u_-(\tau)\bigr),
	\end{equation}
so that $u_-^D(0) = 0$ due to $u_-(\tau) = u_+(-\tau)$. Due to reflection symmetry of (\ref{eq:eucl_eom_1d}) one can obtain $u_+^D(\tau)$ from $u_-^D(\tau)$ as
	\begin{equation} \label{eq:floq_u_p}
		u_+^D(\tau) \equiv u_-^D(\beta - \tau) = \frac12\bigl(e^{-\beta\varepsilon}u_-(\tau) - e^{\beta\varepsilon}u_+(\tau)\bigr).
	\end{equation}
	The corresponding Wronskian of $u_+^D$ and $u_-^D$ reads
	\begin{align}
		\Delta_{+-}^D &= u_+^D \, (W_\eu u_-^D) - (W_\eu u_+^D) \, u_-^D \nonumber\\&= - \sinh \beta\varepsilon \, u_+(0) \, W_\eu u_+(0)
	\end{align}
	where we use the relations (\ref{eq:floq_u_m})--(\ref{eq:floq_u_p}) between Dirichlet basis functions and $u_\pm(\tau)$, and its derivatives at the boundary points
	\begin{equation}
	\begin{aligned}
		W_\eu u_-^D(0) &= -W_\eu u_+^D(\beta) = W_\eu u_+(0), \\
		W_\eu u_-^D(\beta) &= -W_\eu u_+^D(0) = \cosh \beta \varepsilon \, W_\eu u_+(0).
	\end{aligned}
	\end{equation}
Substitution of the corresponding Dirichlet Green's function to (\ref{eq:dm_equcl_pars}) gives
	\begin{equation}
		\bOm = \frac{\omega}{\sinh \beta\varepsilon} \begin{bmatrix}
			\cosh \beta\varepsilon & -1 \\
			-1 & \cosh \beta\varepsilon
		\end{bmatrix},
	\end{equation}
where $\omega$ is defined in (\ref{eq:floq_omega}). Note that for real $\varepsilon$ this coincides with (\ref{eq:bom_nu}), with (\ref{eq:nu_varepsilon}) substituted. For imaginary $\varepsilon$ we express it as $\varepsilon = i q$, so $\bOm$ has the form
	\begin{equation}
		\bOm = \frac{\omega'}{\sin \beta q} \begin{bmatrix}
			\cos \beta q & -1 \\
			-1 & \cos \beta q
		\end{bmatrix},
	\end{equation}
	where $\omega'$ is defined in (\ref{eq:floq_omega_prime}).
	
	Following Appendix~\ref{sec:dm_def}, let us examine the properties of the underlying density matrix, defined by the obtained $\bOm$. For real $\varepsilon$ we have $R=\omega \coth \beta\varepsilon$ and $S=-\omega/\sinh \beta\varepsilon$, so that $R$,~$R+S$ and $R-S$ have the same sign, and we conclude that the density matrix is bounded, normalizable and positive-definite for $\omega > 0$. If it is the case, $\sigma \equiv S/R = -1/\cosh\beta\varepsilon$, so the definition (\ref{eq:floq_omega}) is consistent with (\ref{eq:omega_bod}), and particle interpretation is allowed.
	In contrast, for imaginary $\varepsilon$ we have $R=\omega' \cot \beta q$,~$S=-\omega'/\sin\beta q$, so $R+S$ and $R-S$ have different signs, so even if the density matrix is normalizable, the particle interpretation is not available.

\subsection{The case of a pure state: vacuum no-boundary wavefunction}
As we have shown above, the Euclidean density matrix prescription in a rather nontrivial way suggests a distinguished choice of the particle interpretation. In context of the pure Hartle-Hawking state this fact is well known and takes place in a much simpler way. Let us briefly discuss this here along with a general demonstration how the transition from a mixed state to the pure one proceeds via the change of spacetime topology of the underlying Euclidean instanton from Fig.~\ref{Fig.1} to Fig.~\ref{Fig.3}.

The no-boundary state defined by the path integral over the fields on the Euclidean ``hemisphere'' $D_+^4$ of Fig.~\ref{Fig.3} (and its reflection dual on $D_-^4$ considered as a factor in the factorizable pure density matrix of Fig.~\ref{Fig.3}) is the vacuum wavefunction (\ref{eq:vac_coord_repr}) with the real frequency (\ref{omega_in_vac}), $\omega= [iWv(t)][v(t)]^{-1}|_{t=0}$. The relevants positive-frequency basis function $v(t)$, similarly to (\ref{v_vs_u_+}), can be regarded as the analytic continuation of a special Euclidean basis function $u(\tau)$, $v(t)=u(\tau_++it)$. This basis function is selected by the requirement that it is regular everywhere inside $D_+^4$, including its pole which we label by $\tau=0$ \cite{Laflamme,reduction}.

To show this one should repeat the calculation of Section~\ref{subs:eucl_dm} on $D_+^4$ --- the support of the Euclidean action $S_\eu(\varphi)$ evaluated at the regular solution of equations of motion $F_\eu\phi(\tau)=0$ with the boundary value $\varphi=\phi(\tau_+)$ at the single boundary $\varSigma_+=\partial D_+^4$. This regular solution is given by the expression proportional to the regular basis function $u(\tau)$ of $F_\eu$ on $D_+^4$,
    \begin{equation}
    \phi(\tau)=u(\tau)[u(\tau_+)]^{-1}\varphi,
    \end{equation}
because the contribution of the complementary basis function dual to the regular $u(\tau)$ should be excluded in view of its singularity at $\tau=0$.\footnote{The point $\tau=0$ is an internal regular point of a smooth manifold $D_+^4$, so that this point with $\tau$ treated as a radial coordinate turns out to be a regular singularity of the equation $F_\eu\phi(\tau)=0$. Its two linearly independent solutions $u_\mp(\tau)$ have the asymptotic behavior $u_\mp\propto\tau^{\mu_\mp}$ with $\mu_->0>\mu_+$, so that only $u_-(\tau)\equiv u(\tau)$ is the regular one, while the contribution of the singular $u_+(\tau)\to\infty$, $\tau\to 0$, should be discarded from the solution $\phi(\tau)$ \cite{reduction}.} After the substitution into the expression for the action (\ref{eq:action_eucl}) its on-shell value reduces to the contribution of the single surface term at $\varSigma_+$, $S_\eu(\varphi)=\tfrac12\phi^T(W_\eu\phi)|_{\varSigma_+}$. As a result $S_\eu(\varphi)=\tfrac12\varphi^T\omega\varphi$, and the Hartle-Hawking wavefunction $\varPsi_{HH}(\varphi)\propto e^{-S_\eu(\varphi)}$ becomes the vacuum state (\ref{eq:vac_coord_repr}) with
    \begin{align}
    \omega&=-[W_\eu u(\tau_+)][u(\tau_+)]^{-1}\nonumber\\
    &\qquad=[iWv(t)][v(t)]^{-1}\bigr|_{t=0},
    \end{align}
where the second equality follows from the analytic continuation rule $v(t)=u(\tau_++it)$. Thus, the Hartle-Hawking no-boundary wavefunction of the linearized field modes is the vacuum of particles uniquely defined by a particular choice of positive-frequency basis functions $v(t)$ which in their turn are the analytic continuation of the \emph{regular} Euclidean basis functions $u(\tau)$, $v(t)=u(\tau_++it)$.\footnote{The set $u(\tau)$ is of course defined only up to a linear transformation with some constant matrix $L$, $u(\tau)\mapsto u(\tau)L$, $v(t)\mapsto v(t)L$, but this Bogoliubov transformation does not mix frequencies and therfore does not change particle interpretation.} This is a well-known fact \cite{Laflamme,reduction} which in the case of de Sitter cosmology corresponds to the Euclidean de Sitter invariant vacuum \cite{Mottola, Allen}.

It is known that vacuum in-in formalism in equilibrium models can be reached by taking the zero temperature limit $\beta\to\infty$. It is not quite clear how this limit can be obtained in generic non-equilibrium situations, but it is likely that the transition from mixed Euclidean density matrix to a pure state is always associated with ripping the Euclidean domain into two disjoint manifolds $D_+^4$ and $D_-^4$ depicted in Fig.~\ref{Fig.3}. To show this consider generic situation of the mixed state with the Euclidean density matrix of Fig.~\ref{Fig.1}. This density matrix has a Gaussian form (\ref{eq:dm_coord0})--(\ref{Omega}) with the matrix $\Bom$ given by Eq.~(\ref{Bomega}) with the Dirichlet Green's function which can be represented in terms of two sets of Dirichlet basis functions $u^D_\pm(\tau)$, $u^D_\pm(\tau_\pm)=0$,
	\begin{align}
	\hspace{-0.2cm}&G_D(\tau, \tau')=
    -u_+^D(\tau)\,(\Delta_{-+}^D)^{-1} [u_-^D(\tau')]^T \, \theta(\tau-\tau')\nonumber\\
    \hspace{-0.3cm}&\!\qquad\qquad+ u_-^D(\tau) \,(\Delta_{+-}^D)^{-1} [u_+^D(\tau')]^T\theta(\tau'-\tau).
	\end{align}
Now consider the case of a pure state, when the density matrix factorizes into the product of two wavefunctions, or the situation of $\varOmega_{+-}\equiv S=0$. This off-diagonal block of $\Bom$ reads as
	\begin{align}
	S&=\overright{W}_{\!\eu}  G_D(\tau_+,\tau_-)
    \overleft{W}_{\!\eu} \nonumber\\
    &=[W_\eu u^D_+(\tau_+)]\,[u^D_+(\tau_-)]^{-1},
	\end{align}
where we used the fact that
    \begin{align}\Delta_{-+}=[u^D_-(\tau_+)]^T  W_\eu u^D_+(\tau_+)=-[W_\eu u^D_-(\tau_-)]^Tu^D_+(\tau_-)\nonumber
    \end{align}
in view of boundary conditions on $u^D_\pm(\tau)$.
Therefore, the requirement of $S=0$ implies singularity of $u^D_+(\tau_-)$ which is impossible, because the Green's function $G_D(\tau,\tau')$ can have a singularity only at the coincidence point of its arguments $\tau=\tau'$. This means that no Dirichlet Green's function on a smooth connected Euclidean manifold of the topology $[\tau_-,\tau_+]\times S^3$ can generate the density matrix of a pure state. The only remaining option is ripping the bridge between $\varSigma_+$ and $\varSigma_-$ into the union of two disjoint parts $D^4_\pm$ by shrinking the middle time slice at $\bar\tau\equiv\tfrac{\tau_++\tau_-}2$ to a point.

In context of the cosmological model driven by the set of Weyl invariant quantum fields \cite{slih,why,SLIH_review} this option also matches with the interpretation of zero temperature limit $\beta\to\infty$, because the inverse temperature of the gas of conformal particles in this model is given by the instanton period in units of the  conformal time $\beta=2\int_{\bar\tau}^{\tau_+} d\tau/a(\tau)\to\infty$, which diverges because the cosmological scale factor (the size of the spatial $S^3$-section) $a(\tau)\to 0$ at $\tau\to\bar\tau$.

\section{Discussion and conclusions}\label{sec:Discussion}

Generality of the above formalism allows one to apply it to a wide scope of problems ranging from condensed matter physics to quantum gravity and cosmology. Our goal in future work will be its use in the calculation of the primordial CMB spectrum of cosmological perturbations in the model of microcanonical initial conditions for inflationary cosmology \cite{slih,why,hill-top}, which was briefly discussed as a motivation for this research.  Quasi-thermal nature of this setup was associated in these papers with the fact that the model was based on local Weyl invariant (conformal) matter which, on the one hand, generates the Friedmann background providing the necessary reflection symmetry and, on the other hand, turns out to be effectively in equilibrium, because in the comoving frame it describes a static situation.

Our results show, however, that thermal properties, including particle interpretation with the distinguished positive/negative frequency decomposition, are valid in much more general case. Specifically, the corresponding frequency matrix $\omega$ in the initial conditions problem for basis functions (\ref{eq:green_lor_neum_bdy0}) is shown to be determined by the parameters of Gaussian type density matrix (\ref{eq:omega_bod}), and the occupation number matrix $\nu$ reads as (\ref{nu})--(\ref{nu_bar_bod0}). In this setup, the Euclidean density matrix, which incorporates the reflection symmetry property guaranteed by (\ref{eq:eucl_action_herm_cond}), plays the role of the particular case. If in addition the Lorentzian action is related to the Euclidean action via the analytic continuation at the turning points of the bounce background (which, of course, respects its reflection symmetry), important analytic properties of correlation functions, including the KMS condition, begin to hold. These are the main results of the paper. They allow one to derive the full set of Lorentzian domain, Euclidean domain and mixed, Lorentzian-Euclidean, Green's functions of the in-in formalism and reveal its rich analytic structure. In particular, the results of Section~\ref{subs:in-in} significantly extend those of \cite{Hackl}, where the nonequilibrium evolution of Gaussian type density matrices was examined. The discussion of simple application examples of Section~\ref{Examples} shows the relation of the obtained formalism to the stability properties of dynamical systems in Floquet theory and the theory of Bloch functions. These properties, in their turn, are related to the eigenmode properties of the wave operator $F_E$ subject to periodic boundary conditions on the bounce instanton within Euclidean time $[\,\tau_-,\tau_+]$-range and deserve further studies.

Prospective nature of rich analytic structure of the
Euclidean-Lorentzian in-in formalism consists in the hope
that quantum equivalence of purely Euclidean calculations of loop effects with those of the Lorentzian calculations can be extended to generic bounce type backgrounds. This equivalence was proven in \cite{Higuchi,Korai} for the
vacuum case of the flat chart of the de Sitter spacetime
vs its Euclidean version --- $S^4$ instanton. A similar but
much simpler equivalence at the one-loop order was observed within covariant curvature expansion in asymptotically flat spacetime for systems with the Poincare-invariant vacuum which is prescribed as the initial condition at asymptotic past infnity \cite{Barvinsky_Beyound}. This equivalence is realized via a special type of analytic continuation
from Euclidean to Lorentzian spacetime, which guarantees unitarity and causality of relevant nonlocal form
factors.

Further applications of the in-in formalism in quantum cosmology require its extension to models with local gauge and diffeomorphism invariance (see also \cite{Dunne} for related problem in the context of quantum electrodynamics). What have been built thus far is the formalism in the physical sector of the theory for explicitly disentangled physical degrees of freedom. In cosmological models subject to time parametrization invariance time is hidden among the full set of metric and matter field variables, and disentangling time is a part of the Hamiltonian reduction to the physical sector. This reduction shows that the cosmological background can be devoid of physical degrees of freedom (just like Friedmann equation in FRW-metric models does not involve any physical degree of freedom in the metric sector of the system). This might play a major role in handling a zero mode of the wave operator $F_\eu $, which necessarily arises on the bounce type background \cite{Kolganov} and comprises in the cosmological context one of the aspects of the problem of time in quantum gravity \cite{Unitarity}. This and the other problems of cosmological applications of the in-in formalism go beyond the scope of this paper and will be the subject of future research.

\section*{Acknowledgements}
The authors are grateful to A.A.~Radovskaya, A.G.~Semenov and D.A.~Trunin for useful discussions. A.O.B. is also grateful for fruitful enjoyable conversations with Philip Stamp, Richard Woodard and especially grateful to Alexander Kamenshchik for long term collaboration on the problem of quantum initial conditions in cosmology. The work was supported by the Russian Science Foundation grant No 23-12-00051.

\appendix
\renewcommand{\thesection}{\Alph{section}}
\renewcommand{\theequation}{\Alph{section}.\arabic{equation}}

\section{Inversion of matrices} \label{sec:m_inv}
	Suppose we want to invert the following even-dimensional matrices of the form
	\begin{eqnarray} \label{M_to_inv}
		\boldsymbol M_1 = \bid - \boldsymbol P_\pm \boldsymbol A, \quad
		\boldsymbol M_2 = \bid - \boldsymbol A \boldsymbol P_\pm,
	\end{eqnarray}
where
	\begin{equation}
		\boldsymbol P_\pm = \bid \pm \bX, \qquad \bX = \begin{bmatrix}
			\;0 && I \\ I && 0\;
		\end{bmatrix},
	\end{equation}
and the matrix $\boldsymbol A$ satisfies the following property
	\begin{equation}
	\bX \, \boldsymbol A \, \bX = \boldsymbol A^*.
	\end{equation}
	In terms of the block-matrix representation of $A$ this simply means that $\boldsymbol A$ has the following form
	\begin{equation}
		\boldsymbol A = \begin{bmatrix*}[l]
			\;B & C \\ \;C^* & B^*
		\end{bmatrix*}.
	\end{equation}
Next, we formally expand (\ref{M_to_inv}) in Taylor series and obtain
	\begin{equation}
		(\boldsymbol M_1)^{-1} = \sum_{n=0}^\infty (\boldsymbol P_\pm \boldsymbol A)^n, \quad
		(\boldsymbol M_2)^{-1} = \sum_{n=0}^\infty (\boldsymbol A \boldsymbol P_\pm)^n
	\end{equation}
Observing that
		 $(\boldsymbol P_\pm \boldsymbol A)^n= \boldsymbol P_\pm (\boldsymbol A+\boldsymbol A^*)^{n-1} \boldsymbol A$, $(\boldsymbol A \boldsymbol P_\pm)^n= \boldsymbol A (\boldsymbol A+\boldsymbol A^*)^{n-1} \boldsymbol P_\pm$,
we immediately find the needed inversion formulae
	\begin{align}
		(\boldsymbol M_1)^{-1} &= \bid + \sum_{n=0}^\infty \boldsymbol P_\pm (\boldsymbol A+\boldsymbol A^*)^{n} \boldsymbol A \nonumber\\
                & = \bid +  \boldsymbol P_\pm (\bid - \boldsymbol A - \boldsymbol A^*)^{-1} \boldsymbol A, \label{eq:M1_inv}\\
		(\boldsymbol M_2)^{-1} &= \bid + \boldsymbol A  \sum_{n=0}^\infty (\boldsymbol A+\boldsymbol A^*)^{n}\boldsymbol P_\pm
\nonumber\\
                &= \bid +  \boldsymbol A  (\bid - \boldsymbol A - \boldsymbol A^*)^{-1} \boldsymbol P_\pm. \label{eq:M2_inv}
	\end{align}

\section{Derivation of Eq.~(\ref{nu_bar_def})\label{B}}
To obtain the expression (\ref{lor_green}) for the Green's function $\bG(t,t')$ it is sufficient to derive the expression (\ref{Green_1st_part}) for its part $\bv_+(t) (i \bDel_{-+} )^{-1} \bv_-^T(t')$. For this purpose we first write down the explicit form of $\bv_-^T$, substituting (\ref{eq:bv_bog_coef}) into (\ref{eq:bv_bog_tr}) which gives
	\begin{equation}
		\bv_-^T(t') = (\bom+\Bom)\frac1{\sqrt{2\bom_\re}}\bv^\dagger(t') + (\bom^*-\Bom)\frac1{\sqrt{2\bom_\re}}\bv^T(t').
	\end{equation}
Next, by adding and subtracting the expression
    $[-\bX\bom+\Bom\bX]\, (2\bom_\re)^{-1/2} \bv^\dagger(t') \nonumber$
we artificially disentangle the expression featuring in the square brackets of (\ref{eq:Delta_Green_full_expl}), so that we get
	\begin{align}
		\bv_-^T(t') =&
		 \Bigl[(\bid - \bX) \, \bom
    + \Bom \, (\bid + \bX)\,\Bigr]
    \frac1{\sqrt{2\bom_\re}}\bv^\dagger(t')\nonumber\\
     &+ (\bom^*-\Bom)\frac1{\sqrt{2\bom_\re}}\bv_+^T(t'),
	\end{align}	
where $\bv^T$ was complemented to $\bv_+^T$ in accordance with Eq.~(\ref{v+_vs_v}). Note that $\bom_\re$ and $\bX$ commute with each other. Further, by noting that $\bX\bom\bX=\bom^*$ let us rewrite the difference $\bom^*-\Bom$ so that we again disentangle the same expression as in square brackets above
	\begin{equation}
		\bom^*-\Bom = (\bom+\Bom) \bX - \Bigl[  (\bid - \bX) \, \bom    + \Bom \, (\bid + \bX)\Bigr] \bX.
	\end{equation}
As a result $\bv_-^T$ takes the form
	\begin{align}
		\bv_-^T(t')& =
		\Bigl[  (\bid - \bX) \, \bom
    + \Bom \, (\bid + \bX)\Bigr]\nonumber\\ &\qquad\times
    \frac1{\sqrt{2\bom_\re}}\bigl(\bv^\dagger(t')
    - \bX \bv_+^T(t') \bigr)\nonumber\\
    &+ (\bom+\Bom)\frac1{\sqrt{2\bom_\re}}\bv_+^T(t').
	\end{align}
Substitution to $\bv_+(t) (i \bDel_{-+} )^{-1} \bv_-^T(t')$ gives
	\begin{align} \label{B5}
	\bv_+(t) &(i \bDel_{-+} )^{-1} \bv_-^T(t') =
    \bv_+(t) \bv^\dagger(t') - \bv_+(t) \bX \bv_+^T(t')\nonumber\\
    &+ \bv_+(t) \sqrt{2\bom_\re} \Bigl[  (\bid - \bX) \, \bom  + \Bom \, (\bid + \bX)\Bigr]^{-1}\nonumber\\ &\qquad\qquad\qquad\times(\bom+\Bom)\bX\frac1{\sqrt{2\bom_\re}}\bv_+^T(t'),
	\end{align}
where the expression in the square brackets can be rearranged as follows by using the fact that the matrices $\bom_\re$ and $\bX$ commute and $\bX^2=\bid$
	\begin{align}
		&\hspace{-0.3cm}(\bid - \bX) \, \bom  + \Bom \, (\bid + \bX) =
		(\bom+\Bom)\bX\frac1{\sqrt{2\bom_\re}}\nonumber\\
    &\times\Bigl[\bid + \bX - \sqrt{2\bom_\re}\bX(\bom+\Bom)^{-1}\bX\sqrt{2\bom_\re}\Bigr]\sqrt{2\bom_\re}.
	\end{align}
Substituting this expression to (\ref{B5}) we get the desired result (\ref{Green_1st_part}) with the matrix $\bnu$ given by (\ref{nu_bar_def}).

\section{Derivation of Eq.~(\ref{gen_fun_neum})} \label{sec:app_G00_der}
The aim of this appendix is twofold. First of all, we derive the simple form (\ref{G00}) of $\bG(0,0)$, showing that the Euclidean Green's function of Eq.~(\ref{Zfinal1})
	\begin{multline} \label{eq:gf_eucl_dir}
		G_\eu(\tau,\tau') = G_D(\tau,\tau') \\ + i \,
    \bg_D \overleft{W}_{\!\eu}(\tau) \, \bG(0,0) \,
    \overright{W}_{\!\eu} \bg_D (\tau')
	\end{multline}
is indeed independent of Lorentzian quantities. Next, we express the generating functional (\ref{Zfinal1}) in terms of Neumann Green's function rather than Dirichlet one by using the relations (\ref{green_dir_neum_1})--(\ref{green_dir_neum_2}) and thus derive another form of the periodic Green's function (\ref{gen_fun_neum}).

Let us write down the explicit form of $i\bG(0,0)$ taken from equations (\ref{eq:Green_full}) and (\ref{eq:Delta_Green_full_expl})
	\begin{equation}
		i\bG(0,0) = (\bid + \bX) \Bigl[  (\bid - \bX) \,
    \bom+ \Bom \, (\bid + \bX)\Bigr]^{-1}.
	\end{equation}
Next, we identically add and subtract $\bOm^*$ inside the square brackets and extract the factor $\bOm + \bOm^*=2\bOm_\re$ out of the brackets. So, we obtain
	\begin{gather}
		i \bG(0, 0) = \frac{\bid + \bX}{2\bOm_\re}\bigl[\bid - (\bid-\bX)\bA\bigr]^{-1}, \\
		\bA \equiv (\Bom^* - \bom)(\Bom + \Bom^*)^{-1}, \nonumber
	\end{gather}
where we used the fact that $\bX\Bom^*\bX=\Bom$ and $\bX^2=\bid$ and also noted the the fraction is unambiguous since $\bid+\bX$ and $\bOm$ commute with each other. Now, the expression in the square brackets can be inverted with the use of (\ref{eq:M1_inv}). The result of this inversion is the identity matrix $\bid$ plus a second term, having $\bid - \bX$ as a left multiplier. Observing that $(\bid + \bX)(\bid-\bX)=0$, we conclude that only $\bid$ survives, hence the result reads
	\begin{equation} \label{eq:G00_expl}
		i\bG(0,0) = \frac{\bid + \bX}{2\bOm_\re}.
	\end{equation}
Thus, we see that the Euclidean Green's function (\ref{eq:gf_eucl_dir}) is independent of any Lorentzian quantities, $\bom$ in particular.

	Now, let us rewrite the Eucliden Green's function (\ref{eq:gf_eucl_dir}) following from the generating functional (\ref{Zfinal1}) in a different form, namely express Dirichlet Green's function in terms of Neumann one, satisfying $(W_\eu + \omega)G_N(\beta, \tau') = (W_\eu - \omega^*)G_N(0, \tau') = 0$, where $\omega$ is the same as in (\ref{eq:bv_eq})--(\ref{bom_diag}). For this purpose, we use the relations (\ref{green_dir_neum_2}) and (\ref{green_dir_neum_1}) with the substitution $\bom \mapsto -i \bom$ reflecting the Euclidean nature of the Neumann Green's function. Applying these relations together with (\ref{eq:G00_expl}) to the Euclidean Green's function (\ref{eq:gf_eucl_dir}), we obtain
	\begin{multline} \label{eq:gf_eucl_neum_pre}
		G_\eu(\tau,\tau') = G_N(\tau, \tau') \\+
		\bg_N^T(\tau) \biggl[(\bom+\bOm)\frac{\bid+\bX}{2\bOm_\re}(\bom+\bOm) \\ - (\bom+\bOm)\biggr] \bg_N(\tau').
	\end{multline}
	It turns out that this expression can be significantly simplified, and directly related to the matrix $\bnu$, defined in (\ref{nu_bar_def}). To show this rewrite $\bnu$ in a different form by defining the matrix
	\begin{equation}
		\boldsymbol{B}^{-1} \equiv \sqrt{2\bom_\re} \,\bX \,
		(\bom\! +\!\Bom)^{-1} \bX \sqrt{2\bom_\re}
	\end{equation}
and extracting it out of the square brackets in (\ref{nu_bar_def}). So we get
	\begin{equation}
		\bnu + \bX = -\Bigl[\bid - \boldsymbol{B}(\bid+\bX)\Bigr]^{-1} \boldsymbol{B},
	\end{equation}
where we moved $\bX$ for further convenience to the left hand side. Now, we explicitly invert the expression in the square brackets above, using (\ref{eq:M2_inv}). After straightforward rearrangements, the result reads
	\begin{multline}
		\bnu + \bX = \bX \frac1{\sqrt{2\bom_\re}} \biggl[(\bom+\bOm)\frac{\bid+\bX}{2\bOm_\re}(\bom+\bOm) \\ - (\bom+\bOm)\biggr] \frac1{\sqrt{2\bom_\re}} \bX.
	\end{multline}
	Thus, the comparison to (\ref{eq:gf_eucl_neum_pre}) gives
	\begin{multline} \label{eq:gf_eucl_neum}
		G_\eu(\tau,\tau') = G_N(\tau, \tau') \\+
		\bg_N^T(\tau) \sqrt{2\bom_\re} (\bnu^* + \bX) \sqrt{2\bom_\re}\, \bg_N(\tau'),
	\end{multline}
where we use the fact that $\bX \bnu \bX = \bnu^*$.

\section{Properties of Gaussian density matrices} \label{sec:dm_def}
	Suppose we have the Gaussian density matrix (\ref{eq:dm_coord}) which we rewrite here for the convenience
	\begin{equation}
		\rho(\varphi_+, \varphi_-) =
        \frac1Z \exp\left\{ -\frac12 \bvphi^T \bOm \, \bvphi + \bj^T \! \bvphi  \right\}, \;
        \bvphi = \begin{bmatrix}
			\;\varphi_+\; \\\varphi_-
		\end{bmatrix}
	\end{equation}
where
	\begin{equation}
		\bj = \begin{bmatrix*}[l]
			\;j \\ \;j^*
		\end{bmatrix*}, \;\bOm =
		\begin{bmatrix*}[l]
			\;\,R & S \\ \;\,S^* & R^*
		\end{bmatrix*}, \; R = R^T, \; S = S^\dagger.
	\end{equation}
and examine the following properties of it, namely
	\begin{enumerate}
		\item normalizability, i.e.\ finiteness of $\tr \hat \rho$,
		\item boundedness, i.e.\ finiteness of $\|\hat \rho |\psi\rangle\|$ for arbitrary normalizable state $|\psi\rangle$,
		\item positive definiteness, i.e.\ positivity of the eigenvalues of $\hat\rho$.
	\end{enumerate}

Normalizability is equivalent to the existence of the integral
	\begin{equation}
		\tr\hat\rho\! =\! \int d\varphi \, \rho(\varphi, \varphi)
    = \bigl[\det (R+S+R^*+S^*)\,\bigr]^{-1/2},
	\end{equation}
which is equivalent to positive definiteness of the real part of $R+S$. Boundedness of $\hat\rho$ is equivalent to existence of $\hat \rho^2$ whose coordinate form reads $\la\varphi_1| \mathinner{\hat \rho^2} |\varphi_2\ra \propto \bigl[\det (R+R^*)\,\bigr]^{-1/2}$, so that we should demand the positive definiteness of the real part of $R$.
	
Positive definiteness requires additional attention, namely the analysis of the eigenvalues and eigenvectors of $\hat\rho$. We will focus on the case in which $R$ and $S$ are real, and $\bj=0$. All the results will also hold for nonvanishing $\bj$ but its derivation will be more cumbersome. We will also assume that normalizability and boundedness of the density matrix are enforced, i.e.\ both $R$ and $R+S$ are positive definite. Let us consider a matrix element $\la \varphi |\mathinner{\hat \rho} |\alpha\ra$, where $|\alpha\ra$ is the coherent state defined by Eq.~(\ref{eq:coh_state}). Inserting a partition of unity in the coordinate representation, we have
	\begin{align}
		\la\varphi|\mathinner{\hat \rho}| \alpha \ra
    &= \int d\varphi' \, \rho(\varphi, \varphi') \, \la\varphi' | \alpha\ra \nonumber\\
    &\begin{aligned}=\frac1Z \exp \Bigl[& -\frac12 \varphi^T \bigl(R-S(R+\omega)^{-1}S\bigr) \varphi\\
    &- \alpha^T \sqrt{2\omega} (R+\omega)^{-1} S \varphi\\
     &- \frac12 \alpha^T \bigl(I-\sqrt{2\omega}(R+S)^{-1}\sqrt{2\omega}\bigr)\alpha \Bigr].\end{aligned}
	\end{align}
Now, let us assume $R-S$ is positive definite, i.e.\ the choice (\ref{omega_bod}) of $\omega$ can be made. After some calculations one can rewrite the matrix element above as
	\begin{multline}
		\la \varphi | \mathinner{\hat \rho} |\alpha \ra  =
		\frac1Z \exp \biggl[ -\frac12 \varphi^T \omega \, \varphi + \alpha^T \frac{\nu}{\nu+I} \sqrt{2\omega} \, \varphi \\
		- \frac12 \alpha^T \left(\frac{\nu}{\nu+I}\right)^2 \alpha \biggr],
	\end{multline}
where $\nu$ defined in terms of $R$ and $S$ in (\ref{nu_bar_bod}). Comparing the right hand side to (\ref{eq:coh_state}) one concludes that
	\begin{equation}
		\hat\rho \bigl| \alpha\bigr\ra
    = \bigl|\frac{\nu}{\nu+I}\alpha\bigr\ra.
	\end{equation}
Taking derivatives of the both sides of this equality with respect to $\alpha$, substituting $\alpha=0$ afterwards, and comparing to (\ref{eq:fock_gen}), one observes that eigenvalues of $\hat\rho$ are arbitrary products of eigenvalues of $\frac{\nu}{\nu+I}$. Thus, the latter matrix should be positive definite for $\hat \rho$ to be positive definite too. Using the expression (\ref{nu_bar_bod}) for $\nu$, one finds that positive definiteness is satisfied for a negative definite matrix $\sigma$ and consequently for a negative definite $S=R^{1/2}\sigma R^{1/2}$.
	
Summarizing, we found that the Gaussian density matrix (\ref{eq:dm_coord}) is normalizable if the real part of the sum $R+S$ is positive definite. Density matrix is bounded if a real part of $R$ is positive definite. If, in addition, the difference $R-S$ is positive definite, which is motivated by the necessity of particle interpretation, presented in Section~\ref{subs:omega_choice}, one concludes that the density matrix is positive definite if $S$ is negative definite.

\bibliography{main_arxiv}

\begin{thebibliography}{53}%
\makeatletter
\providecommand \@ifxundefined [1]{%
 \@ifx{#1\undefined}
}%
\providecommand \@ifnum [1]{%
 \ifnum #1\expandafter \@firstoftwo
 \else \expandafter \@secondoftwo
 \fi
}%
\providecommand \@ifx [1]{%
 \ifx #1\expandafter \@firstoftwo
 \else \expandafter \@secondoftwo
 \fi
}%
\providecommand \natexlab [1]{#1}%
\providecommand \enquote  [1]{``#1''}%
\providecommand \bibnamefont  [1]{#1}%
\providecommand \bibfnamefont [1]{#1}%
\providecommand \citenamefont [1]{#1}%
\providecommand \href@noop [0]{\@secondoftwo}%
\providecommand \href [0]{\begingroup \@sanitize@url \@href}%
\providecommand \@href[1]{\@@startlink{#1}\@@href}%
\providecommand \@@href[1]{\endgroup#1\@@endlink}%
\providecommand \@sanitize@url [0]{\catcode `\\12\catcode `\$12\catcode
  `\&12\catcode `\#12\catcode `\^12\catcode `\_12\catcode `\%12\relax}%
\providecommand \@@startlink[1]{}%
\providecommand \@@endlink[0]{}%
\providecommand \url  [0]{\begingroup\@sanitize@url \@url }%
\providecommand \@url [1]{\endgroup\@href {#1}{\urlprefix }}%
\providecommand \urlprefix  [0]{URL }%
\providecommand \Eprint [0]{\href }%
\providecommand \doibase [0]{https://doi.org/}%
\providecommand \selectlanguage [0]{\@gobble}%
\providecommand \bibinfo  [0]{\@secondoftwo}%
\providecommand \bibfield  [0]{\@secondoftwo}%
\providecommand \translation [1]{[#1]}%
\providecommand \BibitemOpen [0]{}%
\providecommand \bibitemStop [0]{}%
\providecommand \bibitemNoStop [0]{.\EOS\space}%
\providecommand \EOS [0]{\spacefactor3000\relax}%
\providecommand \BibitemShut  [1]{\csname bibitem#1\endcsname}%
\let\auto@bib@innerbib\@empty
\bibitem [{\citenamefont {Schwinger}(1961)}]{Schwinger}%
  \BibitemOpen
  \bibfield  {author} {\bibinfo {author} {\bibfnamefont {J.~S.}\ \bibnamefont
  {Schwinger}},\ }\bibfield  {title} {\bibinfo {title} {{Brownian motion of a
  quantum oscillator}},\ }\href {https://doi.org/10.1063/1.1703727} {\bibfield
  {journal} {\bibinfo  {journal} {J. Math. Phys.}\ }\textbf {\bibinfo {volume}
  {2}},\ \bibinfo {pages} {407} (\bibinfo {year} {1961})}\BibitemShut {NoStop}%
\bibitem [{\citenamefont {Keldysh}(1964)}]{Keldysh}%
  \BibitemOpen
  \bibfield  {author} {\bibinfo {author} {\bibfnamefont {L.~V.}\ \bibnamefont
  {Keldysh}},\ }\bibfield  {title} {\bibinfo {title} {{Diagram technique for
  nonequilibrium processes}},\ }\href@noop {} {\bibfield  {journal} {\bibinfo
  {journal} {Zh. Eksp. Teor. Fiz.}\ }\textbf {\bibinfo {volume} {47}},\
  \bibinfo {pages} {1515} (\bibinfo {year} {1964})}\BibitemShut {NoStop}%
\bibitem [{\citenamefont {Hartle}\ and\ \citenamefont {Hawking}(1983)}]{HH}%
  \BibitemOpen
  \bibfield  {author} {\bibinfo {author} {\bibfnamefont {J.~B.}\ \bibnamefont
  {Hartle}}\ and\ \bibinfo {author} {\bibfnamefont {S.~W.}\ \bibnamefont
  {Hawking}},\ }\bibfield  {title} {\bibinfo {title} {{Wave Function of the
  Universe}},\ }\href {https://doi.org/10.1103/PhysRevD.28.2960} {\bibfield
  {journal} {\bibinfo  {journal} {Phys. Rev. D}\ }\textbf {\bibinfo {volume}
  {28}},\ \bibinfo {pages} {2960} (\bibinfo {year} {1983})}\BibitemShut
  {NoStop}%
\bibitem [{\citenamefont {Hawking}(1984)}]{H}%
  \BibitemOpen
  \bibfield  {author} {\bibinfo {author} {\bibfnamefont {S.~W.}\ \bibnamefont
  {Hawking}},\ }\bibfield  {title} {\bibinfo {title} {{The Quantum State of the
  Universe}},\ }\href {https://doi.org/10.1016/0550-3213(84)90093-2} {\bibfield
   {journal} {\bibinfo  {journal} {Nucl. Phys. B}\ }\textbf {\bibinfo {volume}
  {239}},\ \bibinfo {pages} {257} (\bibinfo {year} {1984})}\BibitemShut
  {NoStop}%
\bibitem [{\citenamefont {Starobinsky}(1979)}]{Starobinsky}%
  \BibitemOpen
  \bibfield  {author} {\bibinfo {author} {\bibfnamefont {A.~A.}\ \bibnamefont
  {Starobinsky}},\ }\bibfield  {title} {\bibinfo {title} {Relict gravitation
  radiation spectrum and initial state of the universe},\ }\href@noop {}
  {\bibfield  {journal} {\bibinfo  {journal} {JETP lett}\ }\textbf {\bibinfo
  {volume} {30}},\ \bibinfo {pages} {131} (\bibinfo {year} {1979})}\BibitemShut
  {NoStop}%
\bibitem [{\citenamefont {Mukhanov}\ and\ \citenamefont
  {Chibisov}(1981)}]{Mukhanov-Chibisov}%
  \BibitemOpen
  \bibfield  {author} {\bibinfo {author} {\bibfnamefont {V.~F.}\ \bibnamefont
  {Mukhanov}}\ and\ \bibinfo {author} {\bibfnamefont {G.~V.}\ \bibnamefont
  {Chibisov}},\ }\bibfield  {title} {\bibinfo {title} {{Quantum Fluctuations
  and a Nonsingular Universe}},\ }\href@noop {} {\bibfield  {journal} {\bibinfo
   {journal} {JETP Lett.}\ }\textbf {\bibinfo {volume} {33}},\ \bibinfo {pages}
  {532} (\bibinfo {year} {1981})}\BibitemShut {NoStop}%
\bibitem [{\citenamefont {Ade}\ \emph {et~al.}(2014{\natexlab{a}})\citenamefont
  {Ade} \emph {et~al.}}]{Planck1}%
  \BibitemOpen
  \bibfield  {author} {\bibinfo {author} {\bibfnamefont {P.~A.~R.}\
  \bibnamefont {Ade}} \emph {et~al.} (\bibinfo {collaboration} {Planck}),\
  }\bibfield  {title} {\bibinfo {title} {{Planck 2013 results. XVI.
  Cosmological parameters}},\ }\href
  {https://doi.org/10.1051/0004-6361/201321591} {\bibfield  {journal} {\bibinfo
   {journal} {Astron. Astrophys.}\ }\textbf {\bibinfo {volume} {571}},\
  \bibinfo {pages} {A16} (\bibinfo {year} {2014}{\natexlab{a}})},\ \Eprint
  {https://arxiv.org/abs/1303.5076} {arXiv:1303.5076 [astro-ph.CO]}
  \BibitemShut {NoStop}%
\bibitem [{\citenamefont {Ade}\ \emph {et~al.}(2014{\natexlab{b}})\citenamefont
  {Ade} \emph {et~al.}}]{Planck2}%
  \BibitemOpen
  \bibfield  {author} {\bibinfo {author} {\bibfnamefont {P.~A.~R.}\
  \bibnamefont {Ade}} \emph {et~al.} (\bibinfo {collaboration} {Planck}),\
  }\bibfield  {title} {\bibinfo {title} {{Planck 2013 results. XXII.
  Constraints on inflation}},\ }\href
  {https://doi.org/10.1051/0004-6361/201321569} {\bibfield  {journal} {\bibinfo
   {journal} {Astron. Astrophys.}\ }\textbf {\bibinfo {volume} {571}},\
  \bibinfo {pages} {A22} (\bibinfo {year} {2014}{\natexlab{b}})},\ \Eprint
  {https://arxiv.org/abs/1303.5082} {arXiv:1303.5082 [astro-ph.CO]}
  \BibitemShut {NoStop}%
\bibitem [{\citenamefont {DeWitt}(1967)}]{DeWitt}%
  \BibitemOpen
  \bibfield  {author} {\bibinfo {author} {\bibfnamefont {B.~S.}\ \bibnamefont
  {DeWitt}},\ }\bibfield  {title} {\bibinfo {title} {{Quantum Theory of
  Gravity. 1. The Canonical Theory}},\ }\href
  {https://doi.org/10.1103/PhysRev.160.1113} {\bibfield  {journal} {\bibinfo
  {journal} {Phys. Rev.}\ }\textbf {\bibinfo {volume} {160}},\ \bibinfo {pages}
  {1113} (\bibinfo {year} {1967})}\BibitemShut {NoStop}%
\bibitem [{\citenamefont {Barvinsky}(2007)}]{why}%
  \BibitemOpen
  \bibfield  {author} {\bibinfo {author} {\bibfnamefont {A.~O.}\ \bibnamefont
  {Barvinsky}},\ }\bibfield  {title} {\bibinfo {title} {{Why there is something
  rather than nothing (out of everything)?}},\ }\href
  {https://doi.org/10.1103/PhysRevLett.99.071301} {\bibfield  {journal}
  {\bibinfo  {journal} {Phys. Rev. Lett.}\ }\textbf {\bibinfo {volume} {99}},\
  \bibinfo {pages} {071301} (\bibinfo {year} {2007})},\ \Eprint
  {https://arxiv.org/abs/0704.0083} {arXiv:0704.0083 [hep-th]} \BibitemShut
  {NoStop}%
\bibitem [{\citenamefont {Barvinsky}(2013{\natexlab{a}})}]{BFV}%
  \BibitemOpen
  \bibfield  {author} {\bibinfo {author} {\bibfnamefont {A.~O.}\ \bibnamefont
  {Barvinsky}},\ }\bibfield  {title} {\bibinfo {title} {{BRST technique for the
  cosmological density matrix}},\ }\href
  {https://doi.org/10.1007/JHEP10(2013)051} {\bibfield  {journal} {\bibinfo
  {journal} {JHEP}\ }\textbf {\bibinfo {volume} {10}},\ \bibinfo {pages}
  {051}},\ \Eprint {https://arxiv.org/abs/1308.3270} {arXiv:1308.3270 [hep-th]}
  \BibitemShut {NoStop}%
\bibitem [{\citenamefont {Barvinsky}(1993)}]{Unitarity}%
  \BibitemOpen
  \bibfield  {author} {\bibinfo {author} {\bibfnamefont {A.~O.}\ \bibnamefont
  {Barvinsky}},\ }\bibfield  {title} {\bibinfo {title} {{Unitarity approach to
  quantum cosmology}},\ }\href {https://doi.org/10.1016/0370-1573(93)90032-9}
  {\bibfield  {journal} {\bibinfo  {journal} {Phys. Rept.}\ }\textbf {\bibinfo
  {volume} {230}},\ \bibinfo {pages} {237} (\bibinfo {year}
  {1993})}\BibitemShut {NoStop}%
\bibitem [{\citenamefont {Laflamme}(1987)}]{Laflamme}%
  \BibitemOpen
  \bibfield  {author} {\bibinfo {author} {\bibfnamefont {R.}~\bibnamefont
  {Laflamme}},\ }\bibfield  {title} {\bibinfo {title} {{The Euclidean Vacuum:
  Justification From Quantum Cosmology}},\ }\href
  {https://doi.org/10.1016/0370-2693(87)91488-2} {\bibfield  {journal}
  {\bibinfo  {journal} {Phys. Lett. B}\ }\textbf {\bibinfo {volume} {198}},\
  \bibinfo {pages} {156} (\bibinfo {year} {1987})}\BibitemShut {NoStop}%
\bibitem [{\citenamefont {Mottola}(1985)}]{Mottola}%
  \BibitemOpen
  \bibfield  {author} {\bibinfo {author} {\bibfnamefont {E.}~\bibnamefont
  {Mottola}},\ }\bibfield  {title} {\bibinfo {title} {{Particle Creation in de
  Sitter Space}},\ }\href {https://doi.org/10.1103/PhysRevD.31.754} {\bibfield
  {journal} {\bibinfo  {journal} {Phys. Rev. D}\ }\textbf {\bibinfo {volume}
  {31}},\ \bibinfo {pages} {754} (\bibinfo {year} {1985})}\BibitemShut
  {NoStop}%
\bibitem [{\citenamefont {Allen}(1985)}]{Allen}%
  \BibitemOpen
  \bibfield  {author} {\bibinfo {author} {\bibfnamefont {B.}~\bibnamefont
  {Allen}},\ }\bibfield  {title} {\bibinfo {title} {{Vacuum States in de Sitter
  Space}},\ }\href {https://doi.org/10.1103/PhysRevD.32.3136} {\bibfield
  {journal} {\bibinfo  {journal} {Phys. Rev. D}\ }\textbf {\bibinfo {volume}
  {32}},\ \bibinfo {pages} {3136} (\bibinfo {year} {1985})}\BibitemShut
  {NoStop}%
\bibitem [{\citenamefont {Page}(1986)}]{Page}%
  \BibitemOpen
  \bibfield  {author} {\bibinfo {author} {\bibfnamefont {D.~N.}\ \bibnamefont
  {Page}},\ }\bibfield  {title} {\bibinfo {title} {{Density Matrix of the
  Universe}},\ }\href {https://doi.org/10.1103/PhysRevD.34.2267} {\bibfield
  {journal} {\bibinfo  {journal} {Phys. Rev. D}\ }\textbf {\bibinfo {volume}
  {34}},\ \bibinfo {pages} {2267} (\bibinfo {year} {1986})}\BibitemShut
  {NoStop}%
\bibitem [{\citenamefont {Barvinsky}\ and\ \citenamefont
  {Kamenshchik}(2006)}]{slih}%
  \BibitemOpen
  \bibfield  {author} {\bibinfo {author} {\bibfnamefont {A.~O.}\ \bibnamefont
  {Barvinsky}}\ and\ \bibinfo {author} {\bibfnamefont {A.~Y.}\ \bibnamefont
  {Kamenshchik}},\ }\bibfield  {title} {\bibinfo {title} {{Cosmological
  landscape from nothing: Some like it hot}},\ }\href
  {https://doi.org/10.1088/1475-7516/2006/09/014} {\bibfield  {journal}
  {\bibinfo  {journal} {JCAP}\ }\textbf {\bibinfo {volume} {09}},\ \bibinfo
  {pages} {014}},\ \Eprint {https://arxiv.org/abs/hep-th/0605132}
  {arXiv:hep-th/0605132} \BibitemShut {NoStop}%
\bibitem [{\citenamefont {Starobinsky}(1980)}]{Starobinsky_R^2}%
  \BibitemOpen
  \bibfield  {author} {\bibinfo {author} {\bibfnamefont {A.~A.}\ \bibnamefont
  {Starobinsky}},\ }\bibfield  {title} {\bibinfo {title} {{A New Type of
  Isotropic Cosmological Models Without Singularity}},\ }\href
  {https://doi.org/10.1016/0370-2693(80)90670-X} {\bibfield  {journal}
  {\bibinfo  {journal} {Phys. Lett. B}\ }\textbf {\bibinfo {volume} {91}},\
  \bibinfo {pages} {99} (\bibinfo {year} {1980})}\BibitemShut {NoStop}%
\bibitem [{\citenamefont {Barvinsky}\ \emph {et~al.}(2015)\citenamefont
  {Barvinsky}, \citenamefont {Kamenshchik},\ and\ \citenamefont
  {Nesterov}}]{slih_R^2}%
  \BibitemOpen
  \bibfield  {author} {\bibinfo {author} {\bibfnamefont {A.~O.}\ \bibnamefont
  {Barvinsky}}, \bibinfo {author} {\bibfnamefont {A.~Y.}\ \bibnamefont
  {Kamenshchik}},\ and\ \bibinfo {author} {\bibfnamefont {D.~V.}\ \bibnamefont
  {Nesterov}},\ }\bibfield  {title} {\bibinfo {title} {{Origin of inflation in
  CFT driven cosmology: $R^2$-gravity and non-minimally coupled inflaton
  models}},\ }\href {https://doi.org/10.1140/epjc/s10052-015-3817-7} {\bibfield
   {journal} {\bibinfo  {journal} {Eur. Phys. J. C}\ }\textbf {\bibinfo
  {volume} {75}},\ \bibinfo {pages} {584} (\bibinfo {year} {2015})},\ \Eprint
  {https://arxiv.org/abs/1510.06858} {arXiv:1510.06858 [hep-th]} \BibitemShut
  {NoStop}%
\bibitem [{\citenamefont {Barvinsky}(2012)}]{suppression}%
  \BibitemOpen
  \bibfield  {author} {\bibinfo {author} {\bibfnamefont {A.~O.}\ \bibnamefont
  {Barvinsky}},\ }\bibfield  {title} {\bibinfo {title} {{On suppression of
  topological transitions in quantum gravity}},\ }\href
  {https://doi.org/10.1088/1475-7516/2012/09/033} {\bibfield  {journal}
  {\bibinfo  {journal} {JCAP}\ }\textbf {\bibinfo {volume} {09}},\ \bibinfo
  {pages} {033}},\ \Eprint {https://arxiv.org/abs/1208.0838} {arXiv:1208.0838
  [hep-th]} \BibitemShut {NoStop}%
\bibitem [{\citenamefont {Barvinsky}\ \emph {et~al.}(2016)\citenamefont
  {Barvinsky}, \citenamefont {Kamenshchik},\ and\ \citenamefont
  {Nesterov}}]{hill-top}%
  \BibitemOpen
  \bibfield  {author} {\bibinfo {author} {\bibfnamefont {A.~O.}\ \bibnamefont
  {Barvinsky}}, \bibinfo {author} {\bibfnamefont {A.~Y.}\ \bibnamefont
  {Kamenshchik}},\ and\ \bibinfo {author} {\bibfnamefont {D.~V.}\ \bibnamefont
  {Nesterov}},\ }\bibfield  {title} {\bibinfo {title} {{New type of hill-top
  inflation}},\ }\href {https://doi.org/10.1088/1475-7516/2016/01/036}
  {\bibfield  {journal} {\bibinfo  {journal} {JCAP}\ }\textbf {\bibinfo
  {volume} {01}},\ \bibinfo {pages} {036}},\ \Eprint
  {https://arxiv.org/abs/1509.07270} {arXiv:1509.07270 [hep-th]} \BibitemShut
  {NoStop}%
\bibitem [{\citenamefont {Barvinsky}(2016)}]{CHS}%
  \BibitemOpen
  \bibfield  {author} {\bibinfo {author} {\bibfnamefont {A.~O.}\ \bibnamefont
  {Barvinsky}},\ }\bibfield  {title} {\bibinfo {title} {{CFT driven cosmology
  and conformal higher spin fields}},\ }\href
  {https://doi.org/10.1103/PhysRevD.93.103530} {\bibfield  {journal} {\bibinfo
  {journal} {Phys. Rev. D}\ }\textbf {\bibinfo {volume} {93}},\ \bibinfo
  {pages} {103530} (\bibinfo {year} {2016})},\ \Eprint
  {https://arxiv.org/abs/1511.07625} {arXiv:1511.07625 [hep-th]} \BibitemShut
  {NoStop}%
\bibitem [{\citenamefont {Barvinsky}\ and\ \citenamefont
  {Kamenshchik}(2023)}]{SLIH_review}%
  \BibitemOpen
  \bibfield  {author} {\bibinfo {author} {\bibfnamefont {A.~O.}\ \bibnamefont
  {Barvinsky}}\ and\ \bibinfo {author} {\bibfnamefont {A.~Y.}\ \bibnamefont
  {Kamenshchik}},\ }\bibinfo {title} {Nonminimal higgs inflation and initial
  conditions in cosmology},\ in\ \href
  {https://doi.org/10.1007/978-981-19-3079-9_13-1} {\emph {\bibinfo {booktitle}
  {Handbook of Quantum Gravity}}},\ \bibinfo {editor} {edited by\ \bibinfo
  {editor} {\bibfnamefont {C.}~\bibnamefont {Bambi}}, \bibinfo {editor}
  {\bibfnamefont {L.}~\bibnamefont {Modesto}},\ and\ \bibinfo {editor}
  {\bibfnamefont {I.}~\bibnamefont {Shapiro}}}\ (\bibinfo  {publisher}
  {Springer Nature Singapore},\ \bibinfo {address} {Singapore},\ \bibinfo
  {year} {2023})\ pp.\ \bibinfo {pages} {1--47}\BibitemShut {NoStop}%
\bibitem [{\citenamefont {Barvinsky}(2013{\natexlab{b}})}]{thermal}%
  \BibitemOpen
  \bibfield  {author} {\bibinfo {author} {\bibfnamefont {A.~O.}\ \bibnamefont
  {Barvinsky}},\ }\bibfield  {title} {\bibinfo {title} {{Thermal power spectrum
  in the CFT driven cosmology}},\ }\href
  {https://doi.org/10.1088/1475-7516/2013/10/059} {\bibfield  {journal}
  {\bibinfo  {journal} {JCAP}\ }\textbf {\bibinfo {volume} {10}},\ \bibinfo
  {pages} {059}},\ \Eprint {https://arxiv.org/abs/1308.4451} {arXiv:1308.4451
  [hep-th]} \BibitemShut {NoStop}%
\bibitem [{\citenamefont {Jordan}(1986)}]{Jordan}%
  \BibitemOpen
  \bibfield  {author} {\bibinfo {author} {\bibfnamefont {R.~D.}\ \bibnamefont
  {Jordan}},\ }\bibfield  {title} {\bibinfo {title} {{Effective Field Equations
  for Expectation Values}},\ }\href {https://doi.org/10.1103/PhysRevD.33.444}
  {\bibfield  {journal} {\bibinfo  {journal} {Phys. Rev. D}\ }\textbf {\bibinfo
  {volume} {33}},\ \bibinfo {pages} {444} (\bibinfo {year} {1986})}\BibitemShut
  {NoStop}%
\bibitem [{\citenamefont {Calzetta}\ and\ \citenamefont
  {Hu}(1987)}]{Calzetta-Hu}%
  \BibitemOpen
  \bibfield  {author} {\bibinfo {author} {\bibfnamefont {E.}~\bibnamefont
  {Calzetta}}\ and\ \bibinfo {author} {\bibfnamefont {B.~L.}\ \bibnamefont
  {Hu}},\ }\bibfield  {title} {\bibinfo {title} {{Closed Time Path Functional
  Formalism in Curved Space-Time: Application to Cosmological Back Reaction
  Problems}},\ }\href {https://doi.org/10.1103/PhysRevD.35.495} {\bibfield
  {journal} {\bibinfo  {journal} {Phys. Rev. D}\ }\textbf {\bibinfo {volume}
  {35}},\ \bibinfo {pages} {495} (\bibinfo {year} {1987})}\BibitemShut
  {NoStop}%
\bibitem [{\citenamefont {Onemli}\ and\ \citenamefont
  {Woodard}(2002)}]{Onemli-Woodard}%
  \BibitemOpen
  \bibfield  {author} {\bibinfo {author} {\bibfnamefont {V.~K.}\ \bibnamefont
  {Onemli}}\ and\ \bibinfo {author} {\bibfnamefont {R.~P.}\ \bibnamefont
  {Woodard}},\ }\bibfield  {title} {\bibinfo {title} {{Superacceleration from
  massless, minimally coupled phi**4}},\ }\href
  {https://doi.org/10.1088/0264-9381/19/17/311} {\bibfield  {journal} {\bibinfo
   {journal} {Class. Quant. Grav.}\ }\textbf {\bibinfo {volume} {19}},\
  \bibinfo {pages} {4607} (\bibinfo {year} {2002})},\ \Eprint
  {https://arxiv.org/abs/gr-qc/0204065} {arXiv:gr-qc/0204065} \BibitemShut
  {NoStop}%
\bibitem [{\citenamefont {Maldacena}(2003)}]{Maldacena}%
  \BibitemOpen
  \bibfield  {author} {\bibinfo {author} {\bibfnamefont {J.~M.}\ \bibnamefont
  {Maldacena}},\ }\bibfield  {title} {\bibinfo {title} {{Non-Gaussian features
  of primordial fluctuations in single field inflationary models}},\ }\href
  {https://doi.org/10.1088/1126-6708/2003/05/013} {\bibfield  {journal}
  {\bibinfo  {journal} {JHEP}\ }\textbf {\bibinfo {volume} {05}},\ \bibinfo
  {pages} {013}},\ \Eprint {https://arxiv.org/abs/astro-ph/0210603}
  {arXiv:astro-ph/0210603} \BibitemShut {NoStop}%
\bibitem [{\citenamefont {Weinberg}(2005)}]{Weinberg}%
  \BibitemOpen
  \bibfield  {author} {\bibinfo {author} {\bibfnamefont {S.}~\bibnamefont
  {Weinberg}},\ }\bibfield  {title} {\bibinfo {title} {{Quantum contributions
  to cosmological correlations}},\ }\href
  {https://doi.org/10.1103/PhysRevD.72.043514} {\bibfield  {journal} {\bibinfo
  {journal} {Phys. Rev. D}\ }\textbf {\bibinfo {volume} {72}},\ \bibinfo
  {pages} {043514} (\bibinfo {year} {2005})},\ \Eprint
  {https://arxiv.org/abs/hep-th/0506236} {arXiv:hep-th/0506236} \BibitemShut
  {NoStop}%
\bibitem [{\citenamefont {Ford}\ and\ \citenamefont
  {Woodard}(2005)}]{Ford-Woodard}%
  \BibitemOpen
  \bibfield  {author} {\bibinfo {author} {\bibfnamefont {L.~H.}\ \bibnamefont
  {Ford}}\ and\ \bibinfo {author} {\bibfnamefont {R.~P.}\ \bibnamefont
  {Woodard}},\ }\bibfield  {title} {\bibinfo {title} {{Stress tensor
  correlators in the Schwinger-Keldysh formalism}},\ }\href
  {https://doi.org/10.1088/0264-9381/22/9/011} {\bibfield  {journal} {\bibinfo
  {journal} {Class. Quant. Grav.}\ }\textbf {\bibinfo {volume} {22}},\ \bibinfo
  {pages} {1637} (\bibinfo {year} {2005})},\ \Eprint
  {https://arxiv.org/abs/gr-qc/0411003} {arXiv:gr-qc/0411003} \BibitemShut
  {NoStop}%
\bibitem [{\citenamefont {Donoghue}\ and\ \citenamefont
  {El-Menoufi}(2014)}]{Donoghue-El-Menoufi}%
  \BibitemOpen
  \bibfield  {author} {\bibinfo {author} {\bibfnamefont {J.~F.}\ \bibnamefont
  {Donoghue}}\ and\ \bibinfo {author} {\bibfnamefont {B.~K.}\ \bibnamefont
  {El-Menoufi}},\ }\bibfield  {title} {\bibinfo {title} {{Nonlocal quantum
  effects in cosmology: Quantum memory, nonlocal FLRW equations, and
  singularity avoidance}},\ }\href {https://doi.org/10.1103/PhysRevD.89.104062}
  {\bibfield  {journal} {\bibinfo  {journal} {Phys. Rev. D}\ }\textbf {\bibinfo
  {volume} {89}},\ \bibinfo {pages} {104062} (\bibinfo {year} {2014})},\
  \Eprint {https://arxiv.org/abs/1402.3252} {arXiv:1402.3252 [gr-qc]}
  \BibitemShut {NoStop}%
\bibitem [{\citenamefont {Higuchi}\ \emph {et~al.}(2011)\citenamefont
  {Higuchi}, \citenamefont {Marolf},\ and\ \citenamefont {Morrison}}]{Higuchi}%
  \BibitemOpen
  \bibfield  {author} {\bibinfo {author} {\bibfnamefont {A.}~\bibnamefont
  {Higuchi}}, \bibinfo {author} {\bibfnamefont {D.}~\bibnamefont {Marolf}},\
  and\ \bibinfo {author} {\bibfnamefont {I.~A.}\ \bibnamefont {Morrison}},\
  }\bibfield  {title} {\bibinfo {title} {{On the Equivalence between Euclidean
  and In-In Formalisms in de Sitter QFT}},\ }\href
  {https://doi.org/10.1103/PhysRevD.83.084029} {\bibfield  {journal} {\bibinfo
  {journal} {Phys. Rev. D}\ }\textbf {\bibinfo {volume} {83}},\ \bibinfo
  {pages} {084029} (\bibinfo {year} {2011})},\ \Eprint
  {https://arxiv.org/abs/1012.3415} {arXiv:1012.3415 [gr-qc]} \BibitemShut
  {NoStop}%
\bibitem [{\citenamefont {Korai}\ and\ \citenamefont {Tanaka}(2013)}]{Korai}%
  \BibitemOpen
  \bibfield  {author} {\bibinfo {author} {\bibfnamefont {Y.}~\bibnamefont
  {Korai}}\ and\ \bibinfo {author} {\bibfnamefont {T.}~\bibnamefont {Tanaka}},\
  }\bibfield  {title} {\bibinfo {title} {{Quantum field theory in the flat
  chart of de Sitter space}},\ }\href
  {https://doi.org/10.1103/PhysRevD.87.024013} {\bibfield  {journal} {\bibinfo
  {journal} {Phys. Rev. D}\ }\textbf {\bibinfo {volume} {87}},\ \bibinfo
  {pages} {024013} (\bibinfo {year} {2013})},\ \Eprint
  {https://arxiv.org/abs/1210.6544} {arXiv:1210.6544 [gr-qc]} \BibitemShut
  {NoStop}%
\bibitem [{\citenamefont {Adshead}\ \emph {et~al.}(2009)\citenamefont
  {Adshead}, \citenamefont {Easther},\ and\ \citenamefont
  {Lim}}]{Adshead-Easther-Lim}%
  \BibitemOpen
  \bibfield  {author} {\bibinfo {author} {\bibfnamefont {P.}~\bibnamefont
  {Adshead}}, \bibinfo {author} {\bibfnamefont {R.}~\bibnamefont {Easther}},\
  and\ \bibinfo {author} {\bibfnamefont {E.~A.}\ \bibnamefont {Lim}},\
  }\bibfield  {title} {\bibinfo {title} {{The 'in-in' Formalism and
  Cosmological Perturbations}},\ }\href
  {https://doi.org/10.1103/PhysRevD.80.083521} {\bibfield  {journal} {\bibinfo
  {journal} {Phys. Rev. D}\ }\textbf {\bibinfo {volume} {80}},\ \bibinfo
  {pages} {083521} (\bibinfo {year} {2009})},\ \Eprint
  {https://arxiv.org/abs/0904.4207} {arXiv:0904.4207 [hep-th]} \BibitemShut
  {NoStop}%
\bibitem [{\citenamefont {Di~Pietro}\ \emph {et~al.}(2022)\citenamefont
  {Di~Pietro}, \citenamefont {Gorbenko},\ and\ \citenamefont
  {Komatsu}}]{Gorbenko_etal}%
  \BibitemOpen
  \bibfield  {author} {\bibinfo {author} {\bibfnamefont {L.}~\bibnamefont
  {Di~Pietro}}, \bibinfo {author} {\bibfnamefont {V.}~\bibnamefont
  {Gorbenko}},\ and\ \bibinfo {author} {\bibfnamefont {S.}~\bibnamefont
  {Komatsu}},\ }\bibfield  {title} {\bibinfo {title} {{Analyticity and
  unitarity for cosmological correlators}},\ }\href
  {https://doi.org/10.1007/JHEP03(2022)023} {\bibfield  {journal} {\bibinfo
  {journal} {JHEP}\ }\textbf {\bibinfo {volume} {03}},\ \bibinfo {pages}
  {023}},\ \Eprint {https://arxiv.org/abs/2108.01695} {arXiv:2108.01695
  [hep-th]} \BibitemShut {NoStop}%
\bibitem [{\citenamefont {Heckelbacher}\ \emph {et~al.}(2022)\citenamefont
  {Heckelbacher}, \citenamefont {Sachs}, \citenamefont {Skvortsov},\ and\
  \citenamefont {Vanhove}}]{Heckelbacher-Sachs-Skvortsov-Vanhove}%
  \BibitemOpen
  \bibfield  {author} {\bibinfo {author} {\bibfnamefont {T.}~\bibnamefont
  {Heckelbacher}}, \bibinfo {author} {\bibfnamefont {I.}~\bibnamefont {Sachs}},
  \bibinfo {author} {\bibfnamefont {E.}~\bibnamefont {Skvortsov}},\ and\
  \bibinfo {author} {\bibfnamefont {P.}~\bibnamefont {Vanhove}},\ }\bibfield
  {title} {\bibinfo {title} {{Analytical evaluation of cosmological correlation
  functions}},\ }\href {https://doi.org/10.1007/JHEP08(2022)139} {\bibfield
  {journal} {\bibinfo  {journal} {JHEP}\ }\textbf {\bibinfo {volume} {08}},\
  \bibinfo {pages} {139}},\ \Eprint {https://arxiv.org/abs/2204.07217}
  {arXiv:2204.07217 [hep-th]} \BibitemShut {NoStop}%
\bibitem [{\citenamefont {Arkani-Hamed}\ \emph {et~al.}(2020)\citenamefont
  {Arkani-Hamed}, \citenamefont {Baumann}, \citenamefont {Lee},\ and\
  \citenamefont {Pimentel}}]{Arkani-Hamed}%
  \BibitemOpen
  \bibfield  {author} {\bibinfo {author} {\bibfnamefont {N.}~\bibnamefont
  {Arkani-Hamed}}, \bibinfo {author} {\bibfnamefont {D.}~\bibnamefont
  {Baumann}}, \bibinfo {author} {\bibfnamefont {H.}~\bibnamefont {Lee}},\ and\
  \bibinfo {author} {\bibfnamefont {G.~L.}\ \bibnamefont {Pimentel}},\
  }\bibfield  {title} {\bibinfo {title} {{The Cosmological Bootstrap:
  Inflationary Correlators from Symmetries and Singularities}},\ }\href
  {https://doi.org/10.1007/JHEP04(2020)105} {\bibfield  {journal} {\bibinfo
  {journal} {JHEP}\ }\textbf {\bibinfo {volume} {04}},\ \bibinfo {pages}
  {105}},\ \Eprint {https://arxiv.org/abs/1811.00024} {arXiv:1811.00024
  [hep-th]} \BibitemShut {NoStop}%
\bibitem [{\citenamefont {Di~Valentino}\ \emph {et~al.}(2021)\citenamefont
  {Di~Valentino} \emph {et~al.}}]{Hubble_tension}%
  \BibitemOpen
  \bibfield  {author} {\bibinfo {author} {\bibfnamefont {E.}~\bibnamefont
  {Di~Valentino}} \emph {et~al.},\ }\bibfield  {title} {\bibinfo {title}
  {{Snowmass2021 - Letter of interest cosmology intertwined II: The hubble
  constant tension}},\ }\href
  {https://doi.org/10.1016/j.astropartphys.2021.102605} {\bibfield  {journal}
  {\bibinfo  {journal} {Astropart. Phys.}\ }\textbf {\bibinfo {volume} {131}},\
  \bibinfo {pages} {102605} (\bibinfo {year} {2021})},\ \Eprint
  {https://arxiv.org/abs/2008.11284} {arXiv:2008.11284 [astro-ph.CO]}
  \BibitemShut {NoStop}%
\bibitem [{\citenamefont {Di~Valentino}\ \emph {et~al.}(2019)\citenamefont
  {Di~Valentino}, \citenamefont {Melchiorri},\ and\ \citenamefont
  {Silk}}]{Silk1}%
  \BibitemOpen
  \bibfield  {author} {\bibinfo {author} {\bibfnamefont {E.}~\bibnamefont
  {Di~Valentino}}, \bibinfo {author} {\bibfnamefont {A.}~\bibnamefont
  {Melchiorri}},\ and\ \bibinfo {author} {\bibfnamefont {J.}~\bibnamefont
  {Silk}},\ }\bibfield  {title} {\bibinfo {title} {{Planck evidence for a
  closed Universe and a possible crisis for cosmology}},\ }\href
  {https://doi.org/10.1038/s41550-019-0906-9} {\bibfield  {journal} {\bibinfo
  {journal} {Nature Astron.}\ }\textbf {\bibinfo {volume} {4}},\ \bibinfo
  {pages} {196} (\bibinfo {year} {2019})},\ \Eprint
  {https://arxiv.org/abs/1911.02087} {arXiv:1911.02087 [astro-ph.CO]}
  \BibitemShut {NoStop}%
\bibitem [{\citenamefont {Yang}\ \emph {et~al.}(2023)\citenamefont {Yang},
  \citenamefont {Giar\`e}, \citenamefont {Pan}, \citenamefont {Di~Valentino},
  \citenamefont {Melchiorri},\ and\ \citenamefont {Silk}}]{Silk2}%
  \BibitemOpen
  \bibfield  {author} {\bibinfo {author} {\bibfnamefont {W.}~\bibnamefont
  {Yang}}, \bibinfo {author} {\bibfnamefont {W.}~\bibnamefont {Giar\`e}},
  \bibinfo {author} {\bibfnamefont {S.}~\bibnamefont {Pan}}, \bibinfo {author}
  {\bibfnamefont {E.}~\bibnamefont {Di~Valentino}}, \bibinfo {author}
  {\bibfnamefont {A.}~\bibnamefont {Melchiorri}},\ and\ \bibinfo {author}
  {\bibfnamefont {J.}~\bibnamefont {Silk}},\ }\bibfield  {title} {\bibinfo
  {title} {{Revealing the effects of curvature on the cosmological models}},\
  }\href {https://doi.org/10.1103/PhysRevD.107.063509} {\bibfield  {journal}
  {\bibinfo  {journal} {Phys. Rev. D}\ }\textbf {\bibinfo {volume} {107}},\
  \bibinfo {pages} {063509} (\bibinfo {year} {2023})},\ \Eprint
  {https://arxiv.org/abs/2210.09865} {arXiv:2210.09865 [astro-ph.CO]}
  \BibitemShut {NoStop}%
\bibitem [{\citenamefont {Matsubara}(1955)}]{Matsubara}%
  \BibitemOpen
  \bibfield  {author} {\bibinfo {author} {\bibfnamefont {T.}~\bibnamefont
  {Matsubara}},\ }\bibfield  {title} {\bibinfo {title} {{A New approach to
  quantum statistical mechanics}},\ }\href {https://doi.org/10.1143/PTP.14.351}
  {\bibfield  {journal} {\bibinfo  {journal} {Prog. Theor. Phys.}\ }\textbf
  {\bibinfo {volume} {14}},\ \bibinfo {pages} {351} (\bibinfo {year}
  {1955})}\BibitemShut {NoStop}%
\bibitem [{\citenamefont {Takahasi}\ and\ \citenamefont
  {Umezawa}(1975)}]{Umezawa}%
  \BibitemOpen
  \bibfield  {author} {\bibinfo {author} {\bibfnamefont {Y.}~\bibnamefont
  {Takahasi}}\ and\ \bibinfo {author} {\bibfnamefont {H.}~\bibnamefont
  {Umezawa}},\ }\bibfield  {title} {\bibinfo {title} {{Thermo field
  dynamics}},\ }\href@noop {} {\bibfield  {journal} {\bibinfo  {journal}
  {Collect. Phenom.}\ }\textbf {\bibinfo {volume} {2}},\ \bibinfo {pages} {55}
  (\bibinfo {year} {1975})}\BibitemShut {NoStop}%
\bibitem [{\citenamefont {Kubo}(1957)}]{KMS1}%
  \BibitemOpen
  \bibfield  {author} {\bibinfo {author} {\bibfnamefont {R.}~\bibnamefont
  {Kubo}},\ }\bibfield  {title} {\bibinfo {title} {Statistical-mechanical
  theory of irreversible processes. i. general theory and simple applications
  to magnetic and conduction problems},\ }\href@noop {} {\bibfield  {journal}
  {\bibinfo  {journal} {Journal of the physical society of Japan}\ }\textbf
  {\bibinfo {volume} {12}},\ \bibinfo {pages} {570} (\bibinfo {year}
  {1957})}\BibitemShut {NoStop}%
\bibitem [{\citenamefont {Martin}\ and\ \citenamefont
  {Schwinger}(1959)}]{KMS2}%
  \BibitemOpen
  \bibfield  {author} {\bibinfo {author} {\bibfnamefont {P.~C.}\ \bibnamefont
  {Martin}}\ and\ \bibinfo {author} {\bibfnamefont {J.~S.}\ \bibnamefont
  {Schwinger}},\ }\bibfield  {title} {\bibinfo {title} {{Theory of many
  particle systems. 1.}},\ }\href {https://doi.org/10.1103/PhysRev.115.1342}
  {\bibfield  {journal} {\bibinfo  {journal} {Phys. Rev.}\ }\textbf {\bibinfo
  {volume} {115}},\ \bibinfo {pages} {1342} (\bibinfo {year}
  {1959})}\BibitemShut {NoStop}%
\bibitem [{\citenamefont {Leonidov}\ and\ \citenamefont
  {Radovskaya}(2019)}]{Leonidov-Radovskaya}%
  \BibitemOpen
  \bibfield  {author} {\bibinfo {author} {\bibfnamefont {A.~V.}\ \bibnamefont
  {Leonidov}}\ and\ \bibinfo {author} {\bibfnamefont {A.~A.}\ \bibnamefont
  {Radovskaya}},\ }\bibfield  {title} {\bibinfo {title} {{Quantum corrections
  to the Classical Statistical Approximation for the expanding quantum
  field}},\ }\href {https://doi.org/10.1140/epjc/s10052-019-6586-x} {\bibfield
  {journal} {\bibinfo  {journal} {Eur. Phys. J. C}\ }\textbf {\bibinfo {volume}
  {79}},\ \bibinfo {pages} {55} (\bibinfo {year} {2019})},\ \Eprint
  {https://arxiv.org/abs/1809.06812} {arXiv:1809.06812 [hep-ph]} \BibitemShut
  {NoStop}%
\bibitem [{\citenamefont {Radovskaya}\ and\ \citenamefont
  {Semenov}(2021)}]{Radovskaya-Semenov}%
  \BibitemOpen
  \bibfield  {author} {\bibinfo {author} {\bibfnamefont {A.~A.}\ \bibnamefont
  {Radovskaya}}\ and\ \bibinfo {author} {\bibfnamefont {A.~G.}\ \bibnamefont
  {Semenov}},\ }\bibfield  {title} {\bibinfo {title} {{Semiclassical
  approximation meets Keldysh\textendash{}Schwinger diagrammatic technique:
  scalar $\varphi ^4$}},\ }\href
  {https://doi.org/10.1140/epjc/s10052-021-09382-4} {\bibfield  {journal}
  {\bibinfo  {journal} {Eur. Phys. J. C}\ }\textbf {\bibinfo {volume} {81}},\
  \bibinfo {pages} {704} (\bibinfo {year} {2021})},\ \Eprint
  {https://arxiv.org/abs/2003.06395} {arXiv:2003.06395 [hep-ph]} \BibitemShut
  {NoStop}%
\bibitem [{\citenamefont {Eastham}(1973)}]{Floquet}%
  \BibitemOpen
  \bibfield  {author} {\bibinfo {author} {\bibfnamefont {M.~S.~P.}\
  \bibnamefont {Eastham}},\ }\href@noop {} {\emph {\bibinfo {title} {The
  spectral theory of periodic differential equations}}}\ (\bibinfo  {publisher}
  {Scottish Academic Press},\ \bibinfo {year} {1973})\BibitemShut {NoStop}%
\bibitem [{\citenamefont {Magnus}\ and\ \citenamefont {Winkler}(2013)}]{Hill}%
  \BibitemOpen
  \bibfield  {author} {\bibinfo {author} {\bibfnamefont {W.}~\bibnamefont
  {Magnus}}\ and\ \bibinfo {author} {\bibfnamefont {S.}~\bibnamefont
  {Winkler}},\ }\href@noop {} {\emph {\bibinfo {title} {Hill's equation}}}\
  (\bibinfo  {publisher} {Courier Corporation},\ \bibinfo {year}
  {2013})\BibitemShut {NoStop}%
\bibitem [{\citenamefont {Barvinsky}(1994)}]{reduction}%
  \BibitemOpen
  \bibfield  {author} {\bibinfo {author} {\bibfnamefont {A.~O.}\ \bibnamefont
  {Barvinsky}},\ }\bibfield  {title} {\bibinfo {title} {{Tunneling geometries.
  2. Reduction methods for functional determinants}},\ }\href
  {https://doi.org/10.1103/PhysRevD.50.5115} {\bibfield  {journal} {\bibinfo
  {journal} {Phys. Rev. D}\ }\textbf {\bibinfo {volume} {50}},\ \bibinfo
  {pages} {5115} (\bibinfo {year} {1994})},\ \Eprint
  {https://arxiv.org/abs/gr-qc/9311023} {arXiv:gr-qc/9311023} \BibitemShut
  {NoStop}%
\bibitem [{\citenamefont {Hackl}\ and\ \citenamefont {Bianchi}(2021)}]{Hackl}%
  \BibitemOpen
  \bibfield  {author} {\bibinfo {author} {\bibfnamefont {L.}~\bibnamefont
  {Hackl}}\ and\ \bibinfo {author} {\bibfnamefont {E.}~\bibnamefont
  {Bianchi}},\ }\bibfield  {title} {\bibinfo {title} {{Bosonic and fermionic
  Gaussian states from K\"ahler structures}},\ }\href
  {https://doi.org/10.21468/SciPostPhysCore.4.3.025} {\bibfield  {journal}
  {\bibinfo  {journal} {SciPost Phys. Core}\ }\textbf {\bibinfo {volume} {4}},\
  \bibinfo {pages} {025} (\bibinfo {year} {2021})},\ \Eprint
  {https://arxiv.org/abs/2010.15518} {arXiv:2010.15518 [quant-ph]} \BibitemShut
  {NoStop}%
\bibitem [{\citenamefont {Barvinsky}\ and\ \citenamefont
  {Vilkovisky}(1987)}]{Barvinsky_Beyound}%
  \BibitemOpen
  \bibfield  {author} {\bibinfo {author} {\bibfnamefont {A.~O.}\ \bibnamefont
  {Barvinsky}}\ and\ \bibinfo {author} {\bibfnamefont {G.~A.}\ \bibnamefont
  {Vilkovisky}},\ }\bibfield  {title} {\bibinfo {title} {{Beyond the
  Schwinger-Dewitt Technique: Converting Loops Into Trees and In-In
  Currents}},\ }\href {https://doi.org/10.1016/0550-3213(87)90681-X} {\bibfield
   {journal} {\bibinfo  {journal} {Nucl. Phys. B}\ }\textbf {\bibinfo {volume}
  {282}},\ \bibinfo {pages} {163} (\bibinfo {year} {1987})}\BibitemShut
  {NoStop}%
\bibitem [{\citenamefont {Dunne}\ and\ \citenamefont {Schubert}(2005)}]{Dunne}%
  \BibitemOpen
  \bibfield  {author} {\bibinfo {author} {\bibfnamefont {G.~V.}\ \bibnamefont
  {Dunne}}\ and\ \bibinfo {author} {\bibfnamefont {C.}~\bibnamefont
  {Schubert}},\ }\bibfield  {title} {\bibinfo {title} {{Worldline instantons
  and pair production in inhomogeneous fields}},\ }\href
  {https://doi.org/10.1103/PhysRevD.72.105004} {\bibfield  {journal} {\bibinfo
  {journal} {Phys. Rev. D}\ }\textbf {\bibinfo {volume} {72}},\ \bibinfo
  {pages} {105004} (\bibinfo {year} {2005})},\ \Eprint
  {https://arxiv.org/abs/hep-th/0507174} {arXiv:hep-th/0507174} \BibitemShut
  {NoStop}%
\bibitem [{\citenamefont {Kolganov}(2022)}]{Kolganov}%
  \BibitemOpen
  \bibfield  {author} {\bibinfo {author} {\bibfnamefont {N.}~\bibnamefont
  {Kolganov}},\ }\bibfield  {title} {\bibinfo {title} {{Real-time diagram
  technique for instantonic systems}},\ }\href@noop {} {\  (\bibinfo {year}
  {2022})},\ \Eprint {https://arxiv.org/abs/2211.05746} {arXiv:2211.05746
  [hep-th]} \BibitemShut {NoStop}%
\end{thebibliography}%

\end{document}